\documentclass[aps,twocolumn,showpacs,superscriptaddress,preprintnumbers,amsmath,amssymb,pra,10pt,nofootinbib]{revtex4-2}
\usepackage{array}[=2016-10-06]
\usepackage{graphicx}
\usepackage{dcolumn}
\usepackage{amsmath}    
\usepackage{appendix}
\usepackage{bm}
\usepackage{bold-extra}
\usepackage[T1]{fontenc}
\usepackage[usenames,dvipsnames]{xcolor}
\usepackage{subfigure}
\usepackage{bbm}
\usepackage{enumerate}
\usepackage{float}
\usepackage{blochsphere}  
\usepackage{booktabs}
\usepackage{pifont}
\usepackage{subcaption}
\newcommand{\xmark}{\ding{55}}
\usepackage{tikz-3dplot} 
\definecolor{Cerulean}{rgb}{0.,0.59,0.835}
\definecolor{RubineRed}{rgb}{0.61,0.07,0.12}
\definecolor{myblue}{rgb}{0.2,0.2,0.8}
\usepackage[colorlinks=true,citecolor=myblue,linkcolor=RubineRed,urlcolor=Cerulean]{hyperref}
\usepackage{titlesec}
\usepackage{array}
\usepackage{dsfont}
\usepackage{physics}
\usepackage{mathtools}

\usepackage{amsmath}
\usepackage{caption}  
\usepackage{stfloats}
\usepackage{float}
\usepackage{amssymb}
\usepackage{physics}
\usepackage{graphicx}
\usepackage{dcolumn}
\usepackage{bm}
\usepackage{color}
\usepackage{quantikz} 
\usepackage{tabularx}
\usepackage{appendix}
\usepackage{capt-of}
\usepackage[normalem]{ulem}

\begin{document}
\preprint{AAPM/123-QED}

\title{Ultrafast quantum gate operations in a Kramers-Henneberger atom Qubit}

\author{A. Tasnim Aynul$^1$, C. Li$^1$, L. Cruz Rodriguez$^{1,2}$, C. Figueira de Morisson Faria}
\affiliation{Department of Physics and Astronomy, University College London, Gower Street, London WC1E 6BT, UK\\$^2$Department of Chemistry at the University of Warwick, Coventry
CV4 7AL, UK}

\date{\today}

\begin{abstract}
We propose and demonstrate the Kramers-Henneberger 
(KH) atom as a novel qubit platform for ultrafast 
single-qubit gate operations. In the KH frame, the 
time-averaged strong laser field engineers a 
double-well potential whose two lowest eigenstates 
define the qubit basis, so that the computational structure 
is created and maintained by the driving field itself. 
A weak resonant control field drives coherent gate 
operations: full time-dependent Schr\"{o}dinger 
equation simulations with the complete time-dependent 
KH potential confirm a Z gate and S gate of the order of femtoseconds, six orders of magnitude 
faster than laser-driven superconducting qubit gates. Decoherence 
characterisation gives longer decoherence times than those required for the gate operations. 
The complete six-gate 
single-qubit set is demonstrated in the 
time-averaged two-level limit, with strong agreement 
between the full and time-averaged descriptions 
confirming that fidelity is limited by structured 
leakage rather than stochastic decoherence. In principle, this is an 
error channel suppressible through pulse engineering. 
These results constitute the first demonstration of 
coherent single-qubit gates in a strong-field setting, 
with a clear pathway toward attosecond-scale 
operations.

\end{abstract}

\maketitle
\section{\label{sec:intro}Introduction}

Coherent superpositions occur when a quantum system occupies a linear combination of states while retaining well-defined relative phases. They underpin coherent quantum dynamics, in which the evolution of amplitudes and relative phases gives rise to population transfer, phase accumulation, and interference between competing pathways. Such dynamics arise across many fields, including ultrafast photoinduced charge transfer in molecular systems \cite{AndreaRozzi2013,Falke2014}, excitation-energy transfer in photosynthetic light-harvesting complexes \cite{Engel2007,Lee2007,Scholes2011,Cao2020}, coherent exciton and spin dynamics in solid-state systems \cite{Awschalom2013}, and the controlled preparation of entangled qubit states for quantum information processing \cite{Turchette1998,Horodecki2009}.

Coherent-state superpositions are also crucial to interferometric schemes in strong-field and attosecond science such as 
multichannel interference \cite{Smirnova2009,Augstein2012,Hassouneh2014}, Reconstruction of Attosecond Beating By Interference of Two-photon Transitions (RABITT) \cite{Paul2001,Dahlstrom2012,Agostini2024} and interferometric spectroscopy, laser-driven dynamics in
molecules \cite{Nisoli2017}, and multiphoton ionization in solids \cite{Ghimire2018,Kruchinin2018}. They also provide the basis for implementing quantum gates: precisely controlled unitary operations that transform quantum states within a chosen computational space. Roughly speaking, quantum gates behave like logic gates in a classical computer, except that they act on probability amplitudes and relative phases rather than only on binary values \cite{Deutsch1989,Barenco1995}. The common requirement is a controllable set of coherent states, together with fields or interactions capable of producing reproducible transformations within that state space. Quantum gates have been implemented in a wide range of physical systems, such as trapped ions \cite{Monroe1995,Schmidt-Kaler2003}, superconducting circuits \cite{Yamamoto2003,Plantenberg2007}, photons \cite{OBrien2003,Hacker2016}, neutral atoms \cite{Isenhower2010}, quantum dots \cite{Petta2005} and molecules \cite{Chuang1998}. The common requirement is a controllable set of coherent states, together with fields or interactions capable of producing reproducible transformations within that state space. Therefore, achieving precise control over such superpositions is a fundamental goal in both basic and applied quantum science. In conventional implementations, however, unwanted coupling to environmental degrees of freedom causes energy relaxation and dephasing, which respectively alter the state populations and randomise the relative phases of the superposition, thereby degrading gate fidelity \cite{Zurek2003,Schlosshauer2005}.

Light-induced potentials provide a versatile route for this control. By applying strong tailored fields to matter, one can modify the effective binding potentials, enabling selective manipulation of energy levels, couplings, and population distributions. This practice is widespread in attoscience, and examples range from field-dressed potential energy surfaces in molecules \cite{Ibrahim2018,Kubel2020,Abanador2020}, light-induced conical intersections, in which a topology is created by the laser field \cite{Demekhin2013,Csehi2018,Baekho2018}, laser-dressed bound-state structure \cite{Chini2012,Wu2013} and Floquet engineering \cite{Oka2017,Lucchini2022}, in which periodic fields serve as tools to create light-induced bound states and resonances. Light-induced potentials have also been explored in valleytronics \cite{Kim2014,Sie2015,Rana2023,Mrudul_2021,Mrudul2021,Tyulnev2024}, and valley polarisation
via skewed linearly polarised pulses has been
proposed as a route to valley-based quantum
devices \cite{Gopalan2026}, providing a more flexible resource than, for instance, doping \cite{Dixit2026_private}. 

Among these, the Kramers–Henneberger (KH) potential \cite{Henneberger1968} represents a paradigmatic example. Transforming into the oscillating electron frame gives an effective KH dichotomous potential, whose structure resembles a molecule \cite{Gavrila2002,Morales2011}. The KH potential exhibits  field-dressed bound states with tunable properties,  
making them ideal for studying dynamical stabilization \cite{Gavrila2002,Pont1990} high-order harmonic generation \cite{Reed1993,Madsen2021}, and  high-order Kerr nonlinearities \cite{Richter2013}. Over the past decade, renewed interest in the Kramers–Henneberger (KH) atom has led to proposals for IR–XUV pump–probe schemes \cite{Ivanov2022} and velocity-map imaging photoelectron spectroscopy \cite{Morales2011} as routes to image its field-dressed structure. The former schemes rely on coherent superpositions of states the interference between different quantum pathways to image the KH atom and highlight its differences from a diatomic molecule \cite{Ivanov2022}. Phase-space studies performed by some of us also shed light on features present in molecules and absent in the KH potential \cite{Aynul2025}. These studies also show that ionization is strongly suppressed if the system is prepared in a KH eigenstate, while coherent superpositions lead to a cyclic behavior in phase space. 

Furthermore, early studies show that two- or multicolor laser fields constitute a powerful tool for shaping the KH potential. For instance, linearly polarized bichromatic fields may considerably modify the KH potential wells, and introduce biases dependent on the time-delay between the two driving waves  \cite{Protopapas1994,Cheng1999,Potvliege1999}. Circularly polarized fields lead to a generalization of the well-known dichotomy to tri-, tetra-, and higher-order “-chotomies”, i.e.,  potentials with a larger number of minima \cite{Bauer2002}. 
This raises the question whether one may build and control coherent superpositions of KH eigenstates using multicolor laser fields, but without destroying or significantly altering the underlying structure of the KH potential. In principle, by introducing a secondary weaker field alongside a strong primary field, specific transitions between KH states can be driven, population distributions manipulated, and phase relationships engineered. 
This multicolor approach may provide a flexible framework for designing coherent operations in light-driven atomic system, with potential implications for quantum computing. 

In this paper, we explore this question using a strong field to create the light induced KH potential, and a much weaker, resonant field to couple KH eigenstates, thus creating an effective two-level atom. Individual KH bound states were chosen to serve as qubit states, with controlled transitions implementing single-qubit gates. Ultrafast gates may not only provide access to a parameter range for which the usual decoherence mechanisms may not develop, but also connect quantum information concepts with ultrafast electronic wave-packet control and field-dressed state engineering. Furthermore,the KH qubit basis is maintained by the strong field and therefore exists as a controllable dressed-state structure, which may pave the way for performing coherent logic within a transient strong-field-defined Hilbert space.

The work is meant as a proof of concept, in which we adapt the reduced-dimensionality model in \cite{Aynul2025} to a two-color field: a strong field, which creates the dichotomous KH potential, and a weak field, which is employed to perform quantum gate operations.  The weak field is resonant with the energy gap between the most deeply bound KH eigenstates, which are chosen to serve as the  qubit state. We construct single-qubit gates, and test them against the DiVicenzo criteria \cite{DiVincenzo2000} of being a well-characterised quantum system, reliable initialization, long coherence relative to gates,  universal control,  measurement, scalability and communication. Although not all criteria are satisfied, our model addresses several elementary prerequisites for quantum-information processing: preparation of a computational space, coherent phase manipulation, ultrafast gate operation, explicit characterisation of leakage and ionisation, and a proposed spatially resolved readout. In our computations, we consider both the time-dependent Schr\"odinger equation, and a time-averaged KH potential. 

This article is organized as follows. In Sec.~\ref{sec:backgd}, we provide the necessary theoretical background to understand our results and an outline of how our model is implemented. This includes the general expressions for the TDSE \ref{sec:tdse} and time-averaged \ref{sec:time-averaged} computations, and details about the model \ref{sec:model}. Subsequently, in Sec.~\ref{sec:ultrafastgates}, we explain how we build the KH gates, and also provide a brief summary of what to expect and of the observables. Our results are presented in Sec.~\ref{sec:results}, starting from qubit characterization, moving to assessing decoherence and the construction of gates applying additional pulses. Finally, in Sec.~\ref{sec:discussion} we state our conclusions. 

\section{\label{sec:backgd}Background}

\subsection{General expressions}
\label{sec:tdse}

Here, we briefly outline the equations employed in \cite{Aynul2025}, which have  been extended to a two-color field. We consider a one-dimensional model atom, which evolves according to the time-dependent Schr\"odinger equation (TDSE) 
\begin{equation}\label{eq:tdse}
    i\dfrac{\partial \psi_L(x,t)}{\partial t}=H_L(t)\psi_L(x,t),
\end{equation}
in the length-gauge and dipole approximation. In Eq.~\eqref{eq:tdse}, $\psi_L(x,t)$ is the time-dependent wave function, and \begin{equation}\label{eq:hamiltonian}
H_L(t)=\dfrac{\hat p^2}{2}+V(\hat{x})-\hat{x}[\varepsilon_s(t)+\varepsilon_w(t)]
\end{equation}
is the length-gauge Hamiltonian, where the hats denote operators, $\varepsilon_s(t)=\varepsilon_{0s}f_s(t)$ and $\varepsilon_w(t)=\varepsilon_{0w}f_w(t)$, where $f_s(t)$ and $f_w(t)$ are time-dependent functions, give the strong and the weak field, respectively ($\varepsilon_{0s}\gg \varepsilon_{0w}$), and $V(\hat{x})$ is the binding potential. We use atomic units throughout, for which $m=e=\hbar=1$. The operators are written in the position representation, so that $\hat p=-i\partial/\partial x$ and $\hat x=x$. We then apply the gauge transformation 
\begin{equation}\label{eq:unitary}
\mathcal{T}_{L\rightarrow KH} = \exp[i\int^t_0 A^2(\tau)d\tau]\exp[i\alpha(t)\hat{p}_x]\exp[-iA(t)\hat{x}]
\end{equation}
from the length gauge to the KH frame \cite{Henneberger1968,Faria1999,Richter2016},
where 
\begin{equation}
A(t)=A_s(t)+A_w(t)=-\int_0^t \varepsilon_s(\tau)d\tau - \int_0^t \varepsilon_w(\tau) d\tau
\label{eq:Afield}
\end{equation}
is the total vector potential, and 
\begin{equation}
\alpha(t) = \alpha_s(t) + \alpha_w(t) =\int_0^t A_s(\tau) d\tau+ \int_0^tA_w(\tau) d\tau
\label{eq:alpha}
\end{equation}
is the electron's quiver motion in the electromagnetic field. Definite integrals guarantee that the quiver motion  and momentum transfer vanish at the beginning and the end of the pulse and thus eliminate possible artifacts \cite{Fring1996,Faria1998b,Faria1998c}. The notation in Eqs.~\eqref{eq:Afield} and \eqref{eq:alpha} follow that of the field, i.e., the subscripts $s$ and $w$ refer to its strong and weak component, respectively.  

We consider that the time-dependence of the electric fields are a trigonometric function with an envelope or a ramp giving a turn-on and off, so that the vector potential and the classical displacement will be of the form 
\[A_i(t)=A_{0i}\tilde{f}_i(t)=\varepsilon_{0i}/\omega_i\tilde{f}_i(t)\]
and
$$\alpha_i(t)=\alpha_{0i}\tilde{\tilde{f}}_i(t)=\varepsilon_{0i}/\omega^2_i\tilde{\tilde{f}}_i(t),$$
where $i=s,w$ and, given $f_i(t)$, the functions $\tilde{f}_i(t)$ and $\tilde{\tilde{f}}_i(t)$ can be inferred from the integrals in Eqs.~\eqref{eq:Afield} and \eqref{eq:alpha}.

We obtain the full quantum dynamics by numerically solving the TDSE [Eq.~\eqref{eq:tdse}] in the length gauge. The propagation is performed
using the split-operator method on a spatial grid extending from
\(x=-1500\) to \(x=1500\)~a.u. To suppress artificial reflections from the
grid boundaries, an absorbing potential is introduced for
\(\lvert x\rvert>800\)~a.u. 
The analysis is restricted to the interval
\[
-60~\mathrm{a.u.} \leq x \leq 60~\mathrm{a.u.},
\]
corresponding to approximately six times the electron excursion amplitude.

The TDSE in the KH frame reads
\begin{equation}\label{eq:HKH}
i\dfrac{\partial \psi_{KH}(x_{KH},t)}{\partial t} =  H_\mathrm{KH}(x,t)\psi_{KH}(x_{KH},t),
\end{equation}
\noindent where 
\begin{equation}
    H_\mathrm{KH}(x,t) = \frac{\hat{p}^2}{2} + V\!\left(x + \alpha(t)\right),
    \label{eq:H_KH_full}
\end{equation}
denotes the KH-frame Hamiltonian, $\psi_{KH}(x_{KH},t)$ give the KH-frame wave function and we have used $\hat {p}=\hat {p}_{{KH}}$ and $x_{KH}\ =x+\alpha(t)$. We have omitted $\hat{p}$ in the arguments of $ H_\mathrm{KH}(x,t)$ as we work within the position representation, for which $\hat{p}=-i\partial/\partial x$. For the same reason, the hats have been dropped in the Hamiltonian.

All the time dependence in Eq.~\eqref{eq:HKH} is embedded in the argument of the binding potential, which, physically, implies that the system is now oscillating with the electron's quiver motion $\alpha(t)$. 
One should note that Eq.~\eqref{eq:HKH} is exact, but it brings additional physical intuition by framing the problem in a way that approximations can be performed in $V(x_{KH}+\alpha(t))$.

To obtain the time-dependent wave function in the KH frame, we apply the transformation \eqref{eq:unitary} in the length-gauge wave function obtained from the numerical solution of the TDSE. In practice, this  means that the KH wave function evaluated at the grid points $x_i$ in the length gauge has the form
\begin{multline}\label{eq:psiKH}
    \psi_{KH}(x_i,t) = \exp \biggl(\frac{i}{2} \int_0^{t}A^2(\tau)d\tau-iA(t)(x_i+\alpha(t))\biggr)\\
    \psi_L(x_i+\alpha(t),t),
\end{multline}
\noindent where the operators have been applied accordingly and $\psi_L(x_i,t)$ is obtained from the numerical solution of equation \eqref{eq:tdse}. A spatial translation is performed to obtain $\psi(x_i+\alpha(t),t)$.

An alternative way of approaching the problem is to perform a unitary transformation considering the strong-field only. This choice on one field component being much stronger than the other, and thus responsible, in practice, for creating a dichotomous potential.   Furthermore, the weak field is only applied in the stabilization regime, many cycles after the strong field has been switch on, and before it is switched off. This would translate into taking $\alpha_s(t)$ and $A_s(t)$ as defined in Eqs.~\eqref{eq:Afield} and \eqref{eq:alpha}, instead of the full fields in the unitary transformation \eqref{eq:unitary}. 

This gives the Hamiltonian
\begin{equation}\label{eq:mixedgauge}
    H^{(s)}_{KH}
    =
    \frac{\hat{p}^2}{2}
    + V(\hat{x}+\alpha_s(t))
    +(\hat{x}+\alpha_s(t))\varepsilon_w(t).
\end{equation}
In Eq.~\eqref{eq:mixedgauge}, the strong field shifts the argument of the
binding potential, while the weak field is associated with the dipole-type
coupling
\begin{equation}
    \widetilde{H}_{\mathrm{coupl}}(t)
    =
    (\hat{x}+\alpha_s(t))\varepsilon_w(t).
    \label{eq:Hmixed}
\end{equation}
One should note that this transformation leads to a mixed gauge and is not the
KH transformation in its conventional form. A similar approach has been used
in~\cite{Potvliege1998,Potvliege1999}. Eq.~\eqref{eq:mixedgauge} is also exact, but within a different physical picture: the strong field causes the potential to shift, while the weak field induces dipole-like transitions. Numerically, we have verified that using the full argument, as in \eqref{eq:HKH} or only the strong field, as in \eqref{eq:mixedgauge} will yield practically identical results in the parameter range of interest (not shown). Thus, we consider the total field when solving the TDSE, as stated above.  Nonetheless, the mixed gauge is useful when considering time-averaged models for the KH potential, as discussed below. 

\subsection{Time-averaged potential models}
\label{sec:time-averaged}

Next, we derive a time-averaged dichotomous potential using either the full KH Hamiltonian or Eq.~\eqref{eq:mixedgauge}. As a starting point, we expand the shifted potential in either Eq.~\eqref{eq:HKH} or \eqref{eq:mixedgauge} in a Fourier series. For the full KH potential, we consider an additional step, namely expanding the argument of the KH in series up to first order in $\alpha_w$. This  is valid for $\alpha_{0s}\gg \alpha_{0w}$ and gives 
\begin{equation}\label{eq:KH1storder}
    H_{KH}(t) \approx \frac{p^2}{2} + V(x_{KH} + \alpha_s(t)) + \left. \frac{dV}{dx} \right|_{x_{KH} + \alpha_s(t)} \hspace*{-1.5cm}\alpha_w(t),
\end{equation}
so that, to this approximation, the argument of the potential is dependent on the strong field via the quiver motion $\alpha_s(t)$. For the mixed-gauge Hamiltonian \eqref{eq:mixedgauge}, this step is not required as the argument of the potential contains only $\alpha_s(t)$. 

\begin{figure}[H]
   \centering
    \includegraphics[width=0.5\textwidth]{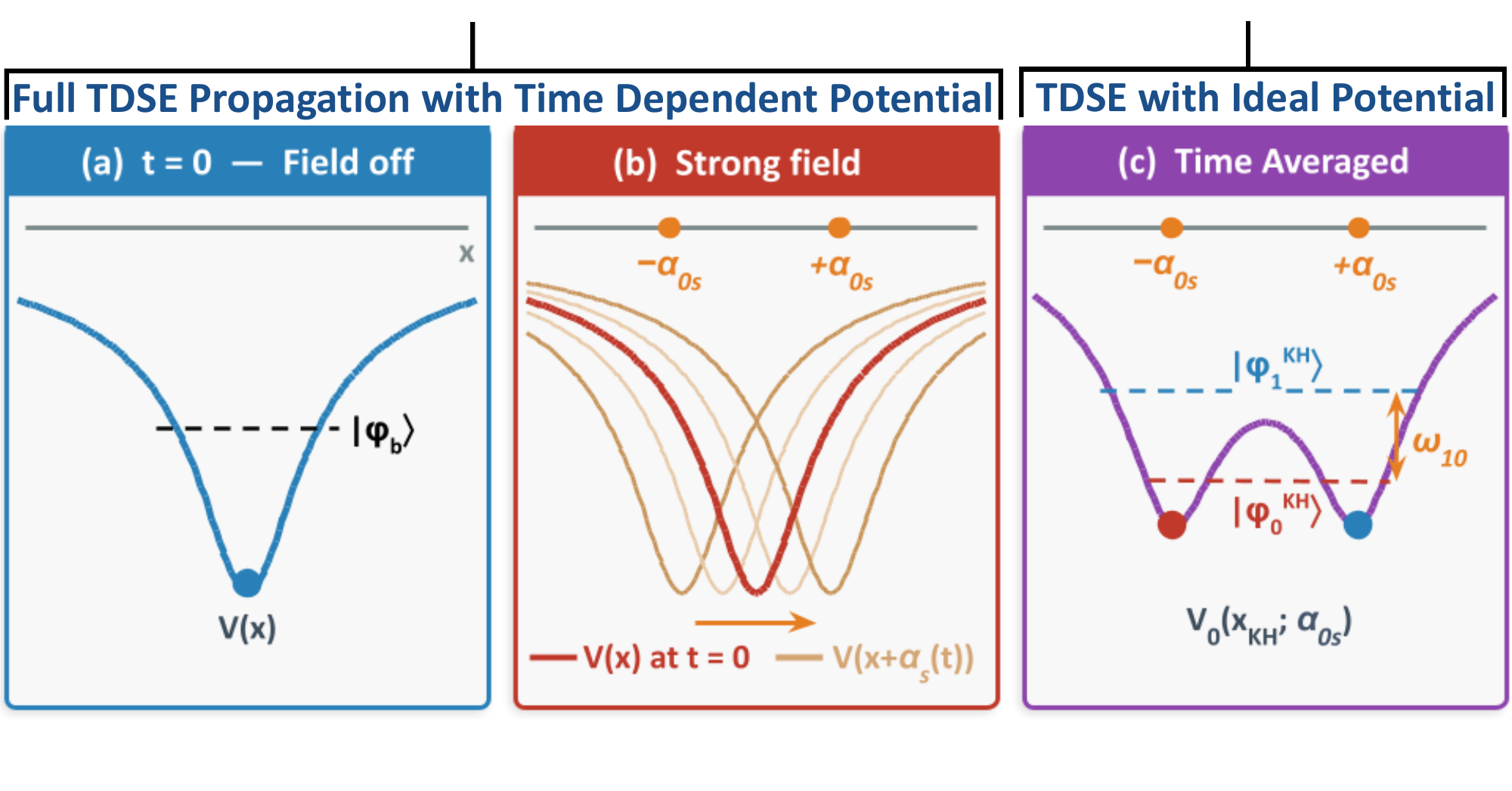}
   \captionof{figure}{Illustration of the Kramers-Henneberger 
approximation. \textbf{(a)} The field-free 
soft-core potential $V(x) = -1/\sqrt{x^2 + 1}$ 
at $t = 0$, showing its bound ground state 
$\ket{\phi_b}$, with energy $E_g = -0.669$ a.u. 
\textbf{(b)} The time-dependent KH potential 
$V(x + \alpha_s(t))$ during the strong laser 
pulse, showing the displaced soft-core well at 
successive instants within one optical cycle as 
the electron frame oscillates between the 
classical turning points $\pm\alpha_{0s}$ (orange 
markers), with quiver amplitude 
$\alpha_{0s} = \varepsilon_{0s}/{\omega_s}^2 = 10.2$ a.u. The 
solid red curve marks $t = 0$ and the arrow 
indicates the direction of the excursion. 
\textbf{(c)} The time-averaged KH 
potential \eqref{eq:fourierterm}, obtained by 
averaging panel (b) over one optical cycle, whose 
minima lie at the turning points $\pm\alpha_{0s}$. 
The strong-field dressing transforms the 
single-well atomic potential into a symmetric 
double-well structure supporting two low-lying 
bound eigenstates $\ket{\phi_0^{KH}}$ (symmetric) 
and $\ket{\phi_1^{KH}}$ (antisymmetric), separated 
by the energy gap $\omega_{10}$.}
    \label{fig:KHapproximation}
\end{figure}

The Fourier expansion gives 
\cite{Richter2016}
\begin{multline}\label{eq:fourier}
    V(x_{KH}+\alpha_s(t))=\sum_{n=-\infty}^{+\infty} V_n(x_{KH};\alpha_{0s})e^{in\omega_s t}\\=V_0(x_{KH};\alpha_{0s})+\sum_{n\ne 0}V_n(x_{KH};\alpha_0)e^{in\omega_s t}, 
\end{multline}
whose $n^{th}$ harmonic reads
\begin{equation}\label{eq:nfourierterm}
    V_n(x_{KH};\alpha_{0s})=\dfrac{1}{T_s}\int_0^{T_s} V(x_{KH}+\alpha_s(t))dt e^{-in\omega_s t},
\end{equation}
where $T_s$ and $\omega_s=2\pi/T_s$ are the period and frequency of the strong field, respectively. The zeroth order term 
\begin{equation}\label{eq:fourierterm}
    V_0(x_{KH};\alpha_{0s})=\dfrac{1}{T_s}\int_0^{T_s} V(x_{KH}+\alpha_s(t))dt
\end{equation}
gives the time-averaged, dichotomous KH potential, which dominates the system's dynamics. Within this approximation, the KH Hamiltonian is time independent, with
eigenstates $\phi^{KH}_n(x)$ obtained from the time-independent Schr\"odinger equation
\begin{equation}\label{eq:KHeigenstates}
      \hspace*{-0.3cm}\biggl(\dfrac{\hat p^2}{2}+V_0(x_{KH};\alpha_{0s})\biggr)\phi_n^{KH}(x_{KH})=E^{KH}_n\phi_n^{KH}(x_{KH}),
\end{equation}
where $E^{KH}_n$ is a generic eigenenergy associated with the $n^{th}$ KH eigenstate, such that $\braket{x}{\phi^{KH}_n}=\phi^{KH}_n(x)$ \cite{Gavrila1984,popov1999applicability}. Eq.~\eqref{eq:KHeigenstates} is employed here to determine the KH eigenstates and eigenenergies, which will be a cornerstone of the coherent superpositions and gates we intend to build.  Fig.~\ref{fig:KHapproximation} illustrates the potentials considered in the present work, i.e., the field-free potential, the full time-dependent potential obtained in the KH frame, and the time-averaged KH potential for the parameters used in the present article (left, middle and right panel, respectively). 

Inspecting Eq.~\eqref{eq:KH1storder} and considering the time-averaged potential shows us that the interaction with the field is given by 
\begin{equation}
    H_{\text{coupl}}(t)=\left. \frac{dV}{dx} \right|_{x + \alpha_s(t)} \alpha_w(t). \label{eq:couplingKH}
\end{equation}
$ H_{\text{coupl}}(t)$ couples different KH eigenstates. For the mixed-gauge Hamiltonian \eqref{eq:mixedgauge}, under the same time-averaged approximation, the coupling between different KH eigenstates will be given by Eq.~\eqref{eq:Hmixed}.

\subsection{Model and implementation}
\label{sec:model}

In our calculations, we use a soft-core potential 
\begin{equation}
    V(x) = -1/\sqrt{(x^2+1)}\label{eq:softcore}
\end{equation}
with ground state energy $E_g = -0.669$ a.u. and ground wavefunction, $\phi_b(x)$, as its initial condition, under the influence of the strong field 
\begin{equation}
    \varepsilon_s=\varepsilon_{0s}f_s(t)\sin[\omega_s t ],\label{eq:estrong}
\end{equation}
where
\begin{equation}\label{eq:envelope}
     f_s(t)=
     \begin{cases}
     t/(6T_s)      &  0\le t\le 6 T_s\\
     1           & 6T_s\le t\le 444 T_s,\\
     -1(t-T_{sf})/(6T_s) &  444T_s<t<T_{sf}
     \end{cases}
\end{equation}
is a trapezoidal envelope with a 6-cycle switch on and off, and a flat top extending for over 444 periods. The time $T_{sf}=450T_s$ marks the end of that pulse, and the time $T_s=2\pi/\omega_s$ gives the period of this field. Throughout, we take the field amplitude and frequency as  $\varepsilon_{0s} = 5.0$ a.u. and  $\omega_s = 0.7$ a.u., respectively. These parameters lead to a time-averaged KH potential supporting 13 bound states, whose minima are at $\alpha_{0s}=\pm10.2$ a.u. The KH eigenenergies are $E^{KH}_n$ ($n=0-12$) are given in  Table \ref{tab:energy_eigenvalues}. These parameters have been employed in \cite{Norman2015}, and deviate from the short-range potential used in our previous publication \cite{Aynul2025}. The present potential leads to more deeply bound states for the KH atom and an earlier onset of stabilization. 
\begin{table}[H]
\centering
\begin{tabular}{cc}
\hline
$n$ & Energy eigenvalue \\
\hline
0  & $-0.23832$ \\
1  & $-0.22805$ \\
2  & $-0.18838$ \\
3  & $-0.14145$ \\
4  & $-0.09251$ \\
5  & $-0.06044$ \\
6  & $-0.04480$ \\
7  & $-0.03245$ \\
8  & $-0.02574$ \\
9  & $-0.01993$ \\
10 & $-0.01527$ \\
11 & $-0.00952$ \\
12 & $-0.00321$ \\
\hline
\end{tabular}
\caption{Energy eigenvalues $E^{KH}_n$ (up to the fifth decimal point) associated with the eigenstates of the time-averaged KH potential \eqref{eq:fourierterm} for the soft-core potential \eqref{eq:softcore} in the strong field given by Eq.~\eqref{eq:estrong} and the above-stated parameters. }
\label{tab:energy_eigenvalues}
\end{table}

Using a weak field 
\begin{equation}
\varepsilon_w(t)=\varepsilon_{0w}f_w(t)
    \sin[\omega_w(t-T_{G}) + \varphi]\label{eq:weakpulse}
\end{equation}
where $\varepsilon_{0w}$ is the field amplitude, 
$\omega_w$ is the carrier frequency, $\varphi$ is the 
carrier phase, and $T_{G}$ is the pulse start time, 
with trapezoidal envelope
\begin{equation}\label{eq:envelopeweak}
     f_w(t)=
     \begin{cases}
     (t-T_{G})/\tau_2      &  0\le t-T_{G}\le \tau_2\\
     1           & \tau_2\le t-T_{G}\le T_{wf}-\tau_2,\\
     -1((t-T_{G})-T_{wf})/\tau_2 &  T_{wf}-\tau_2<t-T_{G}<T_{wf}\\
     0 & \text{otherwise}
     \end{cases}
\end{equation}
where $\tau_2$ is the ramp-up/ramp-down duration and 
$T_{wf}$ is the total pulse duration, of frequency 
$\omega_w=0.010270461$ a.u., with 
approximately half a cycle switch-on and half a 
cycle switch-off ($\tau_2 = 358$~a.u.), which is resonant with the 
transition $\omega_{10}=E^{KH}_1-E^{KH}_0$ 
coupling the two most deeply bound eigenstates. The carrier phase, $\varphi$, is set to $\pi$ unless stated otherwise. The 
parameters of the strong field are kept fixed, while 
those of the weak field are varied.

In the full TDSE, $T_{G}$ marks the time at which the 
weak field switches on relative to the strong-field 
turn-on. In the ideal two-level case, by contrast, 
there is no strong-field turn-on transient: the initial 
state is set directly and the weak pulse begins at 
$t=0$, so $T_{G}=0$ throughout.

For the specific atomic potential employed here, the coupling given by Eq.~\eqref{eq:couplingKH} reads 
\begin{equation}\label{eq:coupling-H}
H_{\text{coupl}} = \frac{(x_{KH} + \alpha_{0s} \sin \omega_s t)}{\left[ 1 + (x_{KH} + \alpha_{0s} \sin \omega_s t)^2 \right]^{3/2}} \alpha_{0w} \sin (\omega_w t+\varphi)
\end{equation}
and, if the mixed gauge is employed, Eq.~\eqref{eq:Hmixed} is given by 
\begin{equation}\label{eq:dodgy_hamiltonian}
    \widetilde{H}_{\text{coupl}} = (\hat{x} + \alpha_{0s} \sin \omega_s t) \varepsilon_{0w} \sin (\omega_w t+\varphi)
.\end{equation}
Both analytic expressions were computed considering the regions in time for which $f_s(t)=1$ and $f_w(t)=1$, and provide good insight in the selection rules expected for each coupling. Eq.~\eqref{eq:coupling-H} may couple even or odd eigenstates, while Eq.~\eqref{eq:dodgy_hamiltonian} only couples eigenstates of different parities. 

\section{Building ultrafast gates}
\label{sec:ultrafastgates}

Next, we aim to build ultrafast single-qubit gates in 
the KH potential. A single qubit's state can be 
visualised as a point on the Bloch sphere, and a 
single-qubit gate is a unitary operation that rotates 
this point, changing the amplitudes and relative phase 
of the basis states. In physical 
terms it is a precisely timed interaction that 
transforms one quantum state into another, analogous to 
a classical logic gate but acting on superpositions 
rather than binary values. This geometric picture, in 
which pure states lie on the surface of the Bloch 
sphere and gate operations correspond to rotations, is 
the natural language of conventional single-qubit gate 
theory~\cite{Nielsen2000}.

For that purpose, we will need a discrete basis with two quantum states, and a way to control coherent superpositions associated with them. To construct this basis, we will employ the strong field and to perform these operations, we will utilize the weak field. The KH atom supports two equivalent and physically meaningful qubit 
basis representations, illustrated in 
Fig.~\ref{fig:qubitbasis}. The choice of basis determines what gate can be implemented by a given pulse. 

\subsection{Qubit basis and initial conditions}
\label{sec:qubitbasis}

\begin{figure}[H]
  \hspace*{-0.5cm}  \includegraphics[width=0.52\textwidth]{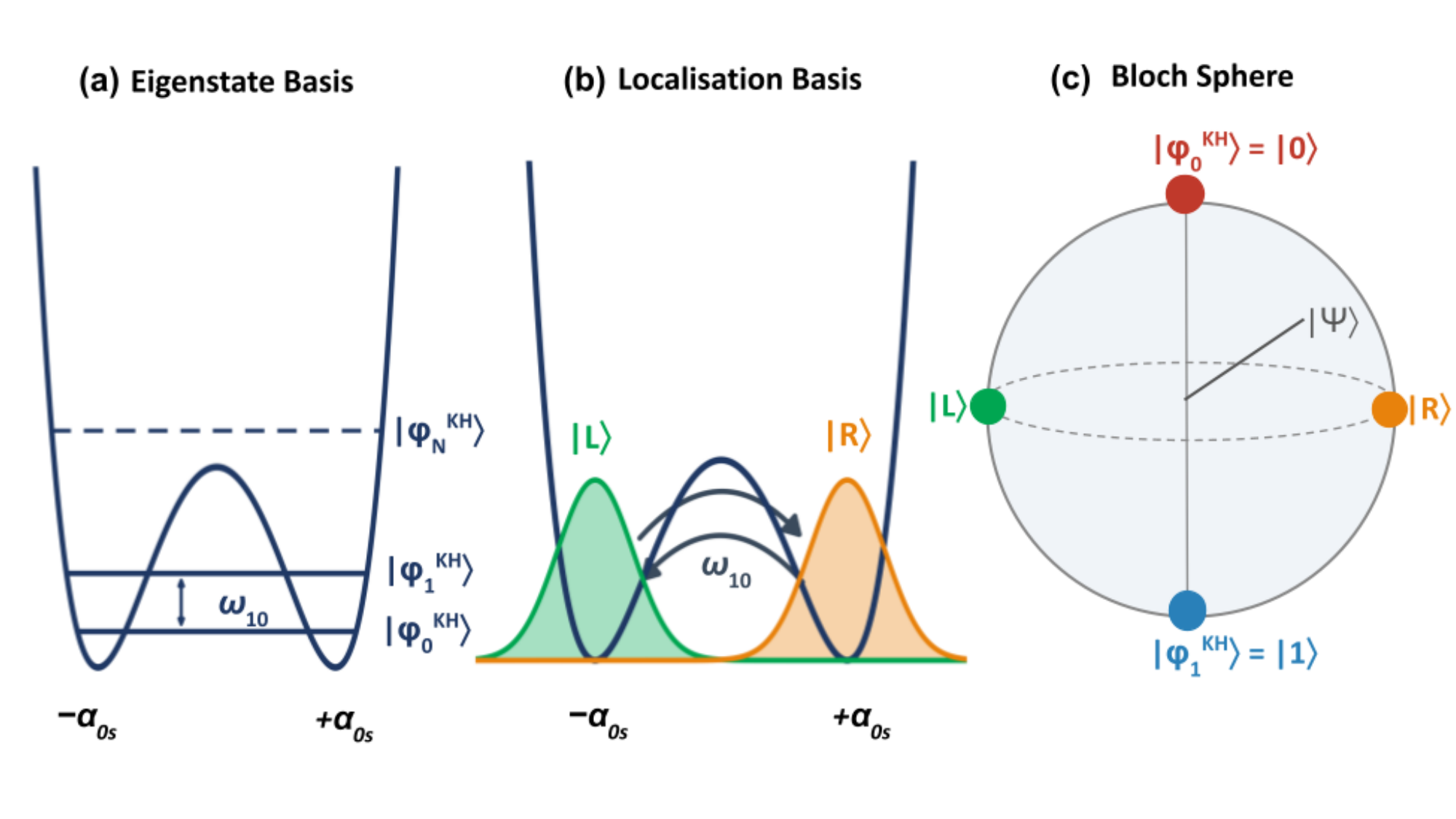}
    \caption{Two equivalent basis representations for the KH 
atom qubit [panels (a) and (b)], together with 
their Bloch sphere representation [panel (c)]. 
Panel (a) shows the energy level structure of 
the KH double-well potential: the two lowest KH 
eigenstates $\ket{\phi^{KH}_0}$ and 
$\ket{\phi^{KH}_1}$, separated by the energy 
gap $\omega_{10}$, form the eigenstate qubit 
basis, with higher-lying states 
$\ket{\phi^{KH}_N}$ indicated by the dashed 
threshold line. Panel (b) shows the equivalent 
localisation basis: the left- and right-localised 
states $\ket{L}$ and $\ket{R}$, given by 
Eqs.~\eqref{eq:left} and \eqref{eq:right}, are 
coherent superpositions of the KH eigenstates 
localised in the left ($-\alpha_{0s}$) and right 
($+\alpha_{0s}$) wells respectively, and undergo 
coherent oscillation at frequency $\omega_{10}$. 
Panel (c) shows the Bloch sphere representation: 
the KH eigenstates $\ket{\phi^{KH}_0}$ and 
$\ket{\phi^{KH}_1}$ occupy the north and south 
poles respectively, playing the roles of $\ket{0}$ 
and $\ket{1}$, while $\ket{L}$ and $\ket{R}$ are 
antipodal equatorial states. A general qubit state 
$\ket{\Psi}$ is shown as a near-equatorial state 
on the sphere. }
\label{fig:qubitbasis}
\end{figure}

One may, for instance, employ the two lowest  KH eigenstates $\{\ket{\phi^{KH}_0}, \ket{\phi^{KH}_1}\}$, which are resonantly coupled by the weak field.  A schematic representation is displayed in Fig.~\ref{fig:qubitbasis}(a), 
where $\ket{\phi^{KH}_0}$ is symmetric and has even parity, while $\ket{\phi^{KH}_1}$ is antisymmetric and has odd parity. On the Bloch sphere, displayed  in Fig.~\ref{fig:qubitbasis}(c), $\ket{\phi^{KH}_0}$ ($\ket{\phi^{KH}_1}$ ) is placed at the north (south) pole and is playing the role of 
$\ket{0}$ ($\ket{1}$).

An equivalent encoding  uses spatially localized states to the left or to the right well of the KH potential. These are given as coherent superpositions of KH eigenstates, i.e., 
\begin{equation}
    \ket{L} = (\ket{\phi^{KH}_0} + \ket{\phi^{KH}_1})/\sqrt{2}  
    \label{eq:left}
\end{equation}
and 
\begin{equation}
    \ket{R} = (\ket{\phi^{KH}_0} - \ket{\phi^{KH}_1})/\sqrt{2},
    \label{eq:right}
\end{equation}
localized on the left and the right KH potential well, respectively. These states are shown in Fig.~\ref{fig:qubitbasis}(b), and sit on the equator of the Bloch sphere 
(Fig.~\ref{fig:qubitbasis}(c)), at antipodal points $\ket{L}$ and 
$\ket{R}$. This encoding is the analogue of a charge qubit in a 
semiconductor double quantum dot, \cite{Hayashi2003}, where the 
computational basis states represent the electron occupying the left 
or right dot. In the quantum dot case, inter-dot charge transfer drives 
coherent oscillation between the two basis states; in the KH atom the 
analogous process is coherent population transfer between the two wells 
of the KH potential.

In the ideal time-averaged 
KH picture, the system is initialised in the ground 
eigenstate $\ket{\phi^{KH}_0}$ (the north pole of the 
Bloch sphere) since the time-averaged potential is 
time-independent from the outset and $\ket{\phi^{KH}_0}$ 
is a stationary state of this Hamiltonian, which in 
the absence of the weak control field remains 
unchanged in time. In the 
full TDSE, however, an inspection of the bound-state populations of the KH atom show that no single KH  eigenstate is uniquely populated.  For some specific times $T_{G}$, 
the strong-field turn-on evolves the initial 
field-free ground state onto an approximately equal superposition of 
the two lowest KH eigenstates,
\begin{equation}
    \ket{\Psi(T_{G})} \approx \frac{1}{\sqrt{2+\epsilon^2}}
    \bigl(\ket{\phi^{KH}_0} + e^{i\phi}\ket{\phi^{KH}_1}\bigr)
    + \epsilon\ket{\phi^{KH}_{2+}},\label{eq:hybrid}
\end{equation}
where $\epsilon$ represents the small but non-negligible leakage 
amplitude to higher KH states. The latter, denoted $\ket{\phi^{KH}_{2+}}$ here, are a coherent superposition of ${\ket{\phi^{KH}_{2}}..\ket{\phi^{KH}_{N}}}$.   Eq.~\eqref{eq:hybrid} gives a localized state to the left or the right KH well, which corresponds to a near-equatorial 
state on the Bloch sphere, as indicated in 
Fig.~\ref{fig:qubitbasis}(c). Gate operations 
are applied to this equatorial initial state, and the start time 
$T_{G}$ of the control pulse selects the phase $\phi$ at the moment 
the gate begins, thereby controlling the effective rotation axis on the 
Bloch sphere, as discussed in Section~\ref{sec:results}. In our previous publication \cite{Aynul2025}, we illustrate how these basis states behave in position and phase space, and also discuss the stability of the KH atom based on its initial conditions. 

\subsection{Coherent superpositions and gate operations}
\label{sec:gates}

In Table~\ref{tab:gates}, we summarise the set of  gates relevant to the present work, along with their Bloch sphere representations. These are standard single-qubit
operations that transform a quantum state on the Bloch sphere. 

The identity gate (first row) leaves the state unchanged. 
The Hadamard
gate (second row) creates coherent superpositions by mapping the computational basis states
as
$
\ket{\phi^{KH}_0}\rightarrow |R\rangle, \qquad
\ket{\phi^{KH}_1}\rightarrow |L\rangle,
$
 thereby connecting the \(z\)- and \(x\)-basis descriptions. In our problem, this means that this gate takes a system in $\ket{\phi^{KH}_0}$ ($\ket{\phi^{KH}_1}$ ) to a state localized in the right (left) well. 
The Pauli gates
correspond to \(\pi\)-rotations about the Bloch-sphere axes: The Pauli-\(X\)
gate (third row) performs a bit flip, exchanging $\ket{\phi^{KH}_0}$ and $\ket{\phi^{KH}_1}$. The
Pauli-\(Z\) gate (fourth row) performs a phase flip, changing the sign of $\ket{\phi^{KH}_1}$ 
while leaving $\ket{\phi^{KH}_0}$ unchanged. The Pauli-\(Y\) gate (fifth row) combines a bit
flip with a phase change, mapping
$
\ket{\phi^{KH}_0} \rightarrow i\ket{\phi^{KH}_1}, \qquad
\ket{\phi^{KH}_1} \rightarrow -i\ket{\phi^{KH}_0} .
$
The \(S\) gate (sixth row) is a
phase gate corresponding to a \(\pi/2\) rotation about the \(z\)-axis, adding a
phase \(i\) to the $\ket{\phi^{KH}_1}$ component while leaving $\ket{\phi^{KH}_0}$
unchanged. Finally, the \(T\) gate (bottom row) is likewise a
phase gate, corresponding to a \(\pi/4\) rotation about
the \(z\)-axis and adding a phase \(e^{i\pi/4}\) to the
$\ket{\phi^{KH}_1}$ component while leaving
$\ket{\phi^{KH}_0}$ unchanged.

In order to actualize these gates, one uses the fact that $\ket{\phi^{KH}_0}$ and $\ket{\phi^{KH}_1}$ are strongly coupled by the resonant weak field. This implies that one may define a two-level subspace using these two eigenstates, and that the population transfer between them will be governed by the matrix element   $\bra{\phi^{KH}_1}\hat{x} \varepsilon_{0w} \sin (\omega_{10} t+\varphi)\ket{\phi^{KH}_0}$, where $\omega_{10}$ is the energy difference between both KH eigenstates. The matrix element containing the $\alpha_s(t)$ term in Eq.~\eqref{eq:Hmixed} 
vanishes due to the orthogonality of the KH eigenstates confirming that the physically relevant coupling for 
the two-level description stems from the dipole term alone. This will lead to Rabi oscillations between these two eigenstates, characterised by the transition 
dipole matrix element,
\begin{equation}
    \mu = \langle\phi^{KH}_1|\hat{x}|\phi^{KH}_0\rangle 
    = 6.13754 \text{ a.u.},
    \label{eq:dipole}
\end{equation}
and the Rabi frequency
\begin{equation}
    \Omega_R = \varepsilon_{0w}\mu.
    \label{eq:rabifreq}
\end{equation}
The Rabi frequency will be used to determine the pulse areas needed for the gates we intend to apply. 

For an ideal two-level atom, the population of the excited state  as a function of the pulse duration $t$ is given by ~\cite{Allen1975}
\begin{equation}
    \mathcal{P}_{1}(t) = \sin^2\!\left(\frac{\Omega_R t}{2}\right),
    \label{eq:rabi}
\end{equation}
where  complete population transfer ($\mathcal{P}_1 = 1$) is 
achieved at
\newcommand{\blochscale}{0.55}      
\newcommand{\blochlab}{\scriptsize} 
\newcommand{\blochax}{\scriptsize}  

\begin{table}[h]
\centering
\small
\renewcommand{\arraystretch}{1.6}
\begin{tabularx}{\columnwidth}{|c|c|c|X|}
\hline
\textbf{Gate} & \textbf{Symbol} & \textbf{Bloch Sphere} & \textbf{Purpose} \\
\hline
Identity &
\begin{quantikz}[baseline=-0.5ex]
\lstick{} & \gate{I} & \qw
\end{quantikz} &
\begin{tikzpicture}[baseline=-0.5ex, scale=\blochscale]
  \draw (0,0) circle (1);
  \draw[dashed, gray] (0,0) ellipse (1 and 0.3);
  \draw[->] (0,-1.2) -- (0,1.2) node[above, font=\blochax] {$z$};
  \draw[->] (-1.2,0) -- (1.4,0) node[right, font=\blochax] {$x$};
  \draw[->, red, thick] (0,0) -- (0,1);
  \node[font=\blochlab, red, above left=-1pt] at (0,1) {$\ket{0}$};
\end{tikzpicture} &
Leaves state unchanged: $\ket{0} \to \ket{0}$. \\
\hline
Hadamard &
\begin{quantikz}[baseline=-0.5ex]
\lstick{} & \gate{H} & \qw
\end{quantikz} &
\begin{tikzpicture}[baseline=-0.5ex, scale=\blochscale]
  \draw (0,0) circle (1);
  \draw[dashed, gray] (0,0) ellipse (1 and 0.3);
  \draw[->] (0,-1.2) -- (0,1.2) node[above, font=\blochax] {$z$};
  \draw[->] (-1.2,0) -- (1.4,0) node[right, font=\blochax] {$x$};
  \draw[->, red!60, dashed, thick] (0,1) arc (90:0:1);
  \draw[->, red, thick] (0,0) -- (0,1);
  \draw[->, blue, thick] (0,0) -- (1,0);
  \node[font=\blochlab, red,  above left=-1pt]  at (0,1) {$\ket{0}$};
  \node[font=\blochlab, blue, below right=-1pt] at (1,0) {$\ket{R}$};
\end{tikzpicture} &
Creates a coherent superposition: $\ket{0} \to \ket{R}$, $\ket{1} \to \ket{L}$. \\
\hline
Pauli-X &
\begin{quantikz}[baseline=-0.5ex]
\lstick{} & \gate{X} & \qw
\end{quantikz} &
\begin{tikzpicture}[baseline=-0.5ex, scale=\blochscale]
  \draw (0,0) circle (1);
  \draw[dashed, gray] (0,0) ellipse (1 and 0.3);
  \draw[->] (0,-1.2) -- (0,1.2) node[above, font=\blochax] {$z$};
  \draw[->] (-1.2,0) -- (1.4,0) node[right, font=\blochax] {$x$};
  \draw[->, red!60, dashed, thick] (0,1) arc (90:-80:1);
  \draw[->, red, thick] (0,0) -- (0,1);
  \draw[->, blue, thick] (0,0) -- (0,-1);
  \node[font=\blochlab, red,  above left=-1pt] at (0,1)  {$\ket{0}$};
  \node[font=\blochlab, blue, below left=-1pt] at (0,-1) {$\ket{1}$};
\end{tikzpicture} &
Acts as quantum NOT: $\ket{0} \to \ket{1}$, $\ket{1} \to \ket{0}$. \\
\hline
Pauli-Z &
\begin{quantikz}[baseline=-0.5ex]
\lstick{} & \gate{Z} & \qw
\end{quantikz} &
\begin{tikzpicture}[baseline=-0.5ex, scale=\blochscale]
  \draw (0,0) circle (1);
  \draw[dashed, gray] (0,0) ellipse (1 and 0.3);
  \draw[->] (0,-1.2) -- (0,1.2) node[above, font=\blochax] {$z$};
  \draw[->] (-1.2,0) -- (1.4,0) node[right, font=\blochax] {$x$};
  \draw[->, red!60, dashed, thick] (1,0) arc (0:170:1 and 0.3);
  \draw[->, red,  thick] (0,0) -- (1,0);
  \draw[->, blue, thick] (0,0) -- (-1,0);
  \node[font=\blochlab, red,  below right=-1pt] at (1,0)  {$\ket{R}$};
  \node[font=\blochlab, blue, below left=-1pt]  at (-1,0) {$\ket{L}$};
\end{tikzpicture} &
Phase flip: $\ket{R} \to \ket{L}$, $\pi$ rotation about $z$-axis. \\
\hline
Pauli-Y &
\begin{quantikz}[baseline=-0.5ex]
\lstick{} & \gate{Y} & \qw
\end{quantikz} &
\begin{tikzpicture}[baseline=-0.5ex, scale=\blochscale]
  \draw (0,0) circle (1);
  \draw[dashed, gray] (0,0) ellipse (1 and 0.3);
  \draw[->] (0,-1.2) -- (0,1.2) node[above, font=\blochax] {$z$};
  \draw[->] (-1.2,0) -- (1.4,0) node[right, font=\blochax] {$x$};
  \draw[->, red!60, dashed, thick] (0,1) arc (90:260:1);
  \draw[->, red,  thick] (0,0) -- (0,1);
  \draw[->, blue, thick] (0,0) -- (0,-1);
  \node[font=\blochlab, red,  above right=-1pt] at (0,1)  {$\ket{0}$};
  \node[font=\blochlab, blue, below right=-1pt] at (0,-1) {$\ket{1}$};
\end{tikzpicture} &
Bit-flip and phase-flip: $\ket{0} \to i\ket{1}$, $\ket{1} \to -i\ket{0}$. \\
\hline
S &
\begin{quantikz}[baseline=-0.5ex]
\lstick{} & \gate{S} & \qw
\end{quantikz} &
\begin{tikzpicture}[baseline=-0.5ex, scale=\blochscale]
  \draw (0,0) circle (1);
  \draw[dashed, gray] (0,0) ellipse (1 and 0.3);
  \draw[->] (0,-1.2) -- (0,1.2) node[above, font=\blochax] {$z$};
  \draw[->] (-1.2,0) -- (1.4,0) node[right, font=\blochax] {$x$};
  \draw[->, red!60, dashed, thick] (1,0) arc (0:88:1 and 0.3);
  \draw[->, red,  thick] (0,0) -- (1,0);
  \draw[->, blue, thick] (0,0) -- (0,0.3);
  \node[font=\blochlab, red, below right=-1pt] at (1,0) {$\ket{R}$};
  \node[font=\blochlab] at (1.55,0.60) {$\pi/2$};
\end{tikzpicture} &
$\pi/2$ rotation about $z$-axis. \\
\hline
T &
\begin{quantikz}[baseline=-0.5ex]
\lstick{} & \gate{T} & \qw
\end{quantikz} &
\begin{tikzpicture}[baseline=-0.5ex, scale=\blochscale]
  \draw (0,0) circle (1);
  \draw[dashed, gray] (0,0) ellipse (1 and 0.3);
  \draw[->] (0,-1.2) -- (0,1.2) node[above, font=\blochax] {$z$};
  \draw[->] (-1.2,0) -- (1.4,0) node[right, font=\blochax] {$x$};
  \draw[->, red!60, dashed, thick] (1,0) arc (0:45:1 and 0.3);
  \draw[->, red,  thick] (0,0) -- (1,0);
  \draw[->, blue, thick] (0,0) -- (0.71,0.21);
  \node[font=\blochlab, red, below right=-1pt] at (1,0) {$\ket{R}$};
  \node[font=\blochlab] at (1.65,0.60) {$\pi/4$};
\end{tikzpicture} &
$\pi/4$ rotation about $z$-axis. \\
\hline
\end{tabularx}
\caption{Summary of standard single-qubit gates (first column), their circuit symbols (second column),
Bloch-sphere representations (third column), and action on basis states (fourth column). In the Bloch-sphere
diagrams, the Cartesian axes indicate the orientation of the sphere, with the
computational basis states located at the poles and superposition states on the
equator when relevant. The coloured arrows indicate how representative input
states are transformed by the gate, with the initial and final states represented in red and blue, respectively.}
\label{tab:gates}
\end{table}

\begin{equation}
    t_{\pi,r} = \frac{\pi}{\Omega_R} = \frac{\pi}{\varepsilon_{0w} \mu},
    \label{eq:tpi}
\end{equation}
denoted the $\pi$-pulse time. These expressions assume a rectangular pulse envelope, standard in quantum information gate 
formalisms~\cite{Nielsen2000}. For the purpose of this study, here we add the turn-on time $\tau_2$ to Eq.~\eqref{eq:tpi} to account for the additional pulse area accumulated 
during the ramp-up before the field, defined as a trapezoidal envelope in in 
Eq.~\eqref{eq:envelopeweak}, reaches full 
amplitude. This is what we refer to as $t_\pi$. Along the same 
line of thought, one may create an equal 
superposition of $\ket{\phi^{KH}_0}$ and 
$\ket{\phi^{KH}_1}$, rotating the Bloch vector from 
the north pole to the equator using a $\pi/2$ pulse 
of duration $t_{\pi/2} = t_\pi/2$~\cite{Nielsen2000}. 
More generally, the marked completion time for a 
$\pi/n$ rotation is $t_{\pi/n} = \tau_2 + 
\pi/(n\Omega_R)$, with $n=1$ for the $\pi$ pulse, $n=2$ for the $\pi/2$ pulse and 
$n=4$ for the $\pi/4$ pulse. Whether this coincides 
with the true gate completion time depends on how 
the flat-top is constructed.

The pulse area $\theta = \Omega_R t$ determines 
the gate operation rather than the peak intensity 
alone~\cite{Allen1975}, making gates robust to 
pulse shape provided the total area is preserved. For the 
trapezoidal envelope used here, the ramp-up and 
ramp-down periods contribute a fixed absolute excess 
pulse area, $\Omega_R\tau_2$, independent of the target 
rotation angle so this excess represents a 
progressively larger fraction of the target for 
smaller-angle pulses. We consider two 
constructions, both of which retain the full 
trapezoidal envelope but differ only in the length 
of the flat-top portion. In the shortened 
construction, the flat-top is trimmed so that the 
total pulse area, including both ramps,
delivers exactly the target rotation $\theta$ at the 
marked completion time $t_\pi$. In the unshortened 
construction, the flat-top retains its full length 
$\pi/\Omega_R$ and the ramps add their excess area 
on top, so that the target rotation is reached a 
time $\tau_2/2$ before $t_\pi$, since half of the 
ramp area has already been accumulated by the end of 
the flat-top. For the $\pi$ pulse this excess fraction is small 
enough that the qualitative gate signature is 
unaffected, so the unshortened flat-top is retained; 
fidelity as defined below is nevertheless evaluated 
at $t_\pi$ throughout, giving a conservative estimate 
of gate performance. The same reasoning applies to 
the $\pi/2$ pulse as well. 

Any unitary operation associated with a  single-qubit gate can be written as a product of rotations. Therefore, having the 
ability to perform arbitrary rotations, even just about 
two axes, such as $x$ and $z$, is enough for 
single-qubit universality~\cite{Nielsen2000}. In the KH 
qubit, the carrier phase $\varphi$ sets the pulse's 
rotation axis anywhere in the $xy$-plane of the Bloch 
sphere, giving direct 
access to $R_x(\theta)$ and $R_y(\theta)$ for any 
$\theta$ set by the pulse area. 
The $S$, $T$  and $Z$ gates are therefore implemented as 
composite three-pulse sequences,
\begin{equation}
    R_z(\theta) = R_x(\pi/2)\,R_y(\theta)\,R_x(-\pi/2),
    \label{eq:composite}
\end{equation}
with $\theta=\pi/2$ giving $S$ and $\theta=\pi/4$ 
giving $T$: the outer pulses are identical and only the 
middle pulse's area differs, so the measured phase 
shifts are expected in the ratio $2\!:\!1$. The same 
decomposition at $\theta=\pi$ reproduces the $Z$ gate. 

The above formulation is based on the ideal two-level 
limit, in which the composite $S$, $T$ and $Z$ sequences, 
and the Hadamard and Y gates, use the shortened flat-top 
described above so that the target rotation angle in 
Eq.~\eqref{eq:composite} is delivered exactly at the 
marked completion time; the single-pulse $X$, $Z$, and 
initialisation tests instead retain the unshortened-flat 
construction matched to the full TDSE, so that any 
discrepancy between ideal and full results isolates the 
effect of leakage rather than a difference in pulse 
construction. In the full TDSE, residual fast 
oscillations at $\omega_s$ persist, driving wavepacket 
excursions beyond the time-averaged double-well and 
coupling the qubit subspace to higher KH eigenstates. 
This is the primary source of decoherence, which is 
practically absent from the ideal time averaged KH 
case. The degree to which the full TDSE dynamics 
departs from this ideal picture is precisely what the 
population diagnostics and phase fidelity $F_\varphi$ 
quantify, as discussed below.

\subsection{Observables}
\label{sec:observables}

Next we outline the observables that will be employed in the subsequent studies: the conditional populations, the asymmetry parameter and the phase fidelity. 
 
The conditional populations $\mathcal{P}_0$, $\mathcal{P}_1$, 
and $\mathcal{P}_{2+}$, the projections onto the two relevant eigenstates $\ket{\phi^{KH}_0}$, $\ket{\phi^{KH}_1}$, 
and all higher KH eigenstates, respectively, provide direct evidence 
of gate quality. The conditional population of the $i$-th KH 
eigenstate is given by
\begin{equation}
    \mathcal{P}_i(t) = \frac{\left|\displaystyle\int 
    \psi_\mathrm{KH}(x,t)\,\phi_i^{*\,\mathrm{KH}}(x)\,
    \mathrm{d}x\right|^2}{\displaystyle\int 
    |\psi_\mathrm{KH}(x,t)|^2\,\mathrm{d}x},
    \label{eq:population_KH}
\end{equation}
where $\psi_\mathrm{KH}(x,t)$ is the wavefunction in the KH frame 
obtained by applying the KH transformation to the full TDSE 
solution, $\phi_i^\mathrm{KH}(x)$ are the wave functions associated with the KH eigenstates, obtained from the eigenvalue equation associated with the KH Hamiltonian [Eq.~\eqref{eq:KHeigenstates}], and the denominator 
$\int|\psi_\mathrm{KH}|^2\,\mathrm{d}x$ is the surviving bound 
norm at each time step, normalising out the overall ionisation 
loss so that $\sum_i \mathcal{P}_i(t) = 1$ at all times. This 
conditional normalisation isolates the qubit-relevant dynamics 
from the overall ionisation decay and enables direct comparison 
with the ideal two-level case where no ionisation occurs. In idealized models with time-averaged potential, and when the bound-state populations are first characterized, this normalisation is not employed.  

A clean gate within the qubit subspace requires 
$\mathcal{P}_0+\mathcal{P}_1$ to remain close to its strong-field-only value 
throughout the pulse, with $\mathcal{P}_{2+}$ at its 
strong-field-only baseline, confirming that the weak field 
drives intra-subspace rotation rather than additional coupling 
to higher states. Any increase in $\mathcal{P}_{2+}$ above the 
strong-field-only baseline is the direct measure of 
gate-induced leakage. Eq.~\eqref{eq:population_KH} can be modified to account for other populations by considering coherent superpositions of eigenstates. 

Localisation of the 
wavepacket to a single well constitutes an effective 
readout of the qubit in the localisation basis, 
motivating the following observable. A natural observable used to quantify wave-packet localization 
is the inter-well asymmetry ~\cite{Chelkowski2004}
\begin{equation}
    \mathcal{A}(t) = \frac{\mathcal{P}_L(t) - \mathcal{P}_R(t)}{\mathcal{P}_L(t) + \mathcal{P}_R(t)},
    \label{eq:asymmetry}
\end{equation}
where $\mathcal{P}_R(t)$ and $\mathcal{P}_L(t)$ are the conditional probabilities of 
finding the electron in the right and left wells respectively. They are computed by integrating the probability density $|\psi_{KH}(x,t)|^2$ over $x > 5$ and 
$x < -5$ a.u., excluding the central barrier region where it is shared between both wells and a well 
assignment is ambiguous. These limits were determined by overlaying the full TDSE outcome on the time-averaged potential. For a wave packet fully on the left (right) well, $\mathcal{A}=1$ ($\mathcal{A}=-1$).

Note that $\mathcal{A}(t)=0$ does not necessarily imply a pure 
eigenstate; it can occur for a superposition 
with any population split between $\ket{\phi_0^{KH}}$ 
and $\ket{\phi_1^{KH}}$, marking a moment of spatial 
balance between the wells rather than a return to either 
eigenstate. Our definition of $\mathcal{P}_L$ and 
$\mathcal{P}_R$ via spatial integration is an approximate 
proxy for the idealised localisation basis, but the 
resulting asymmetry still oscillates at $\omega_{10}$ as 
expected, consistent with the coherent-superposition 
case~\cite{Aynul2025}.

We have verified that, in  the strong-field-only case,  the asymmetry 
oscillates as $\mathcal{A}(t) \approx \mathcal{A}\cos(\omega_{10} t + 
\phi)$, where $\phi$ is a phase shift observed in our computations, arising 
from the strong-field turn-on transient. This will be shown subsequently, and reflects the coherent inter-well 
population transfer at the eigenfrequency $\omega_{10}$. 
Application of the weak control pulse modifies both the amplitude 
and phase of this oscillation and the phase shift 
$\delta = \phi_\mathrm{both} - \phi$ relative 
to the unperturbed case is the primary gate observable.

The phase shift $\delta$ is used to define the
phase fidelity $F_\varphi$, which  quantifies how closely the observed 
phase shift matches the target gate rotation angle. It is defined 
separately for $\pi$, $\pi/2$ and $\pi/4$ pulses as
\begin{equation}
    F_\varphi^{(\eta)} = 1 - \frac{|\delta - \eta|}{\eta},
    \qquad \eta \in \{\pi, \pi/2\,,  \pi/4\},
    \label{eq:fidelity}
\end{equation}
where $\delta$ is extracted by fitting 
$\mathcal{A}(t) = \mathcal{A}\cos(\omega_{10}t + \phi)$ independently to 
the strong-field-only and both-fields time evolution over a chosen 
time window, with the difference wrapped to $(-\pi, \pi]$. 
Two such windows are used: the \emph{late-pulse} 
window, taken over the later part of the pulse in 
the vicinity of the marked completion time $t_\pi$ 
(or $t_{\pi/n}$) while the field is still on, and 
the \emph{post-pulse} window, beginning once the 
field has switched off. Because the target rotation 
is completed slightly before $t_\pi$ (owing to the 
unshortened flat-top), the late-pulse window 
captures the gate's action around completion, before 
any ramp-down-induced leakage accumulates further, 
while the post-pulse window tests whether the phase 
shift persists once the drive is removed. Both are 
reported where the coherent oscillation remains 
cleanly extractable in the strong-field-only and 
both-fields signals; the specific window boundaries 
are given in the relevant figure captions. Where a 
different window, or the point $t_\pi$ itself, is 
used to determine fidelity, this is specified in the 
relevant section.
$F_\varphi = 1$ corresponds to a perfect gate and $F_\varphi = 0$ 
to no gate action. This measure is valid when the  oscillation 
is present in both the strong-field-only and 
both-fields asymmetry parameter over the chosen time window. In cases where the oscillation collapses post-pulse 
due to localisation locking, the mean asymmetry bias 
$\mathcal{B} = \langle \mathcal{A} \rangle_\mathrm{post} - 
\langle \mathcal{A} \rangle_\mathrm{strong}$ serves as a complementary 
observable, where $\langle\mathcal{A} \rangle_\mathrm{post}$ gives the asymmetry parameter once the field oscillations have collapsed. These are summarised in Table~\ref{tab:observables}.

In the present work, each gate is characterised by a 
specific combination of pulse parameters and expected 
signature in the observables. The identity gate 
corresponds to no measurable gate action: the 
populations and asymmetry track their 
strong-field-only values, giving $\delta \simeq 0$. The 
$X$ gate, applied to $\ket{\phi^{KH}_0}$, is a $\pi$ 
pulse producing complete population inversion, 
$\mathcal{P}_0 \to 0$ and $\mathcal{P}_1 \to 1$. The 
$Y$ gate is a $\pi$ pulse with carrier phase 
$\varphi = \pi/2$; since the rotation axis is set 
directly by $\varphi$, its population signature is 
identical to the $X$ gate, and the two are distinguished 
by the $\pi/2$ shift in the asymmetry oscillation that 
follows from their $\pi/2$ difference in carrier phase. 
The Hadamard gate is a $\pi/2$ pulse applied to 
$\ket{\phi^{KH}_0}$, driving the Bloch vector from the 
pole to the equator and producing an equal 
superposition, $\mathcal{P}_0 = \mathcal{P}_1 = 0.5$. For the phase gates ($Z$, $S$, and $T$), the 
populations $\mathcal{P}_0$ and $\mathcal{P}_1$ are by definition 
insensitive to the relative phase and are expected to 
remain unchanged; the asymmetry $\mathcal{A}(t)$ is 
therefore the clear indicator of completed gate 
action, with the phase-shift fidelity $F_\varphi$ being the 
relevant metric.
The $Z$ gate is characterised by the phase-shift 
observable $F_\varphi$ for a $\pi$-pulse, which tests 
whether the pulse produces a relative phase shift 
$\delta \simeq \pi$ while oscillations are 
present and populations return to their pre-pulse 
values. The $S$ and $T$ gates are 
characterised by the same observable, $F_\varphi$, but 
for the composite three-pulse sequences of 
Eq.~\eqref{eq:composite}, corresponding to target phase 
shifts $\delta \simeq \pi/2$ and $\delta \simeq \pi/4$ 
respectively in a non-collapsed post-pulse oscillation. 
By contrast, the localisation bias $\mathcal{B}$ 
characterises the localisation, or locking, operation by 
measuring the post-pulse well preference once the 
oscillation has collapsed.


\begin{table}[h]
\centering
\small
\caption{Observables used to characterise gate operations.}
\label{tab:observables}
\begin{tabular}{p{3.0cm}p{4.5cm}}
\hline\hline
Observable & Measures \\
\hline
 $\mathcal{P}_0$, $\mathcal{P}_1$, $\mathcal{P}_{2+}$ &
Eigenstate occupations and leakage to higher KH states \\[8pt]
$F_\varphi$ ($\pi$, $\pi/2$, $\pi/4$ pulse) &
Phase shift $\delta$  \\[8pt]
 $\mathcal{A}$ &
Time-dependent asymmetry parameter \\\\[8pt] $\mathcal{B}$  &
Well preference when oscillation collapses \\
\hline\hline
\\
\end{tabular}
\end{table}

\section{Results}
\label{sec:results}
\subsection{Characterising the qubit}
\label{sec:resultsinitialqubit}

\begin{figure}[H]
    \centering
   \hspace*{-0.5cm} \includegraphics[width=1.1\linewidth]{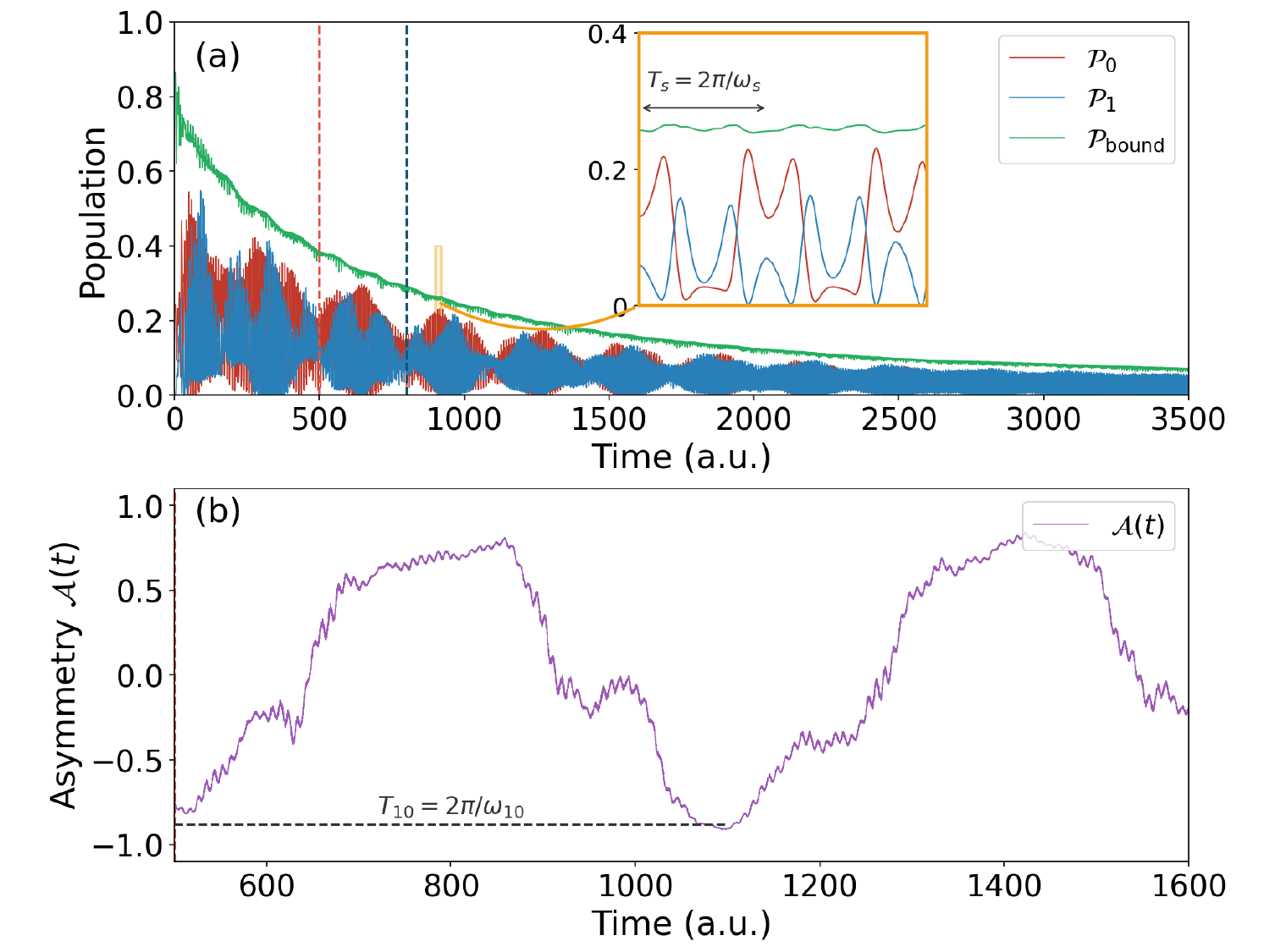}
    \caption{Unnormalised qubit populations and asymmetry under 
the strong field alone, for $\varepsilon_{0s} = 5$ 
a.u.\ and $\omega_s = 0.7$ a.u. 
(a)~The populations $\mathcal{P}_0$ and 
$\mathcal{P}_1$ of the two lowest KH eigenstates 
$\ket{\phi^{KH}_0}$ and $\ket{\phi^{KH}_1}$ are 
shown together with the total bound population 
$\mathcal{P}_\mathrm{bound}$. Following the initial 
ionisation burst during the strong-field turn-on, 
$\mathcal{P}_\mathrm{bound}$ decays slowly, 
retaining roughly 10--15\% of the initial 
population over the pulse. The dashed vertical 
lines mark two arbitrary times $T^{(1)}_{G} = 500$ 
a.u.\ and $T^{(2)}_{G} = 800$ a.u., after the onset 
of stabilisation at $T_{\mathrm{st}} = 470$ a.u.\ 
(see Appendix~\ref{app:stability}), for which weak pulses will be 
switched on in the subsequent analysis. The inset 
magnifies $900 \le t \le 920$ a.u., spanning about 
two cycles of the strong field $T_s = 2\pi/\omega_s 
\approx 9$ a.u. and shows the near-equal split of the surviving 
qubit population between $\mathcal{P}_0$ and 
$\mathcal{P}_1$ on average.
(b)~Well asymmetry $\mathcal{A}(t)$ over the 
interval $500 \le t \le 1600$ a.u., oscillating 
between the left-localised ($\mathcal{A} \to +1$) 
and right-localised ($\mathcal{A} \to -1$) 
configurations at the eigenstate beating period 
$T_{10} = 2\pi/\omega_{10} \approx 612$ a.u.\ 
(dashed line), confirming coherent population 
transfer between the wells at the splitting 
frequency $\omega_{10}$. At the initialisation 
time $T^{(1)}_{G} = 500$ a.u.\ the asymmetry is 
close to $-1$, corresponding to the right-localised 
state $\ket{R}$ -- which indicates an approximately equally weighted coherent superposition of the two KH eigenstates and translates as a near-equatorial state on the 
Bloch sphere as 
discussed in Section~\ref{sec:qubitbasis}.
}
\label{fig:resultinitialpop}
\end{figure}

Before discussing gate operations it is important to establish what 
fraction of the total population participates in the qubit dynamics 
and how the populations are interpreted. Fig.~\ref{fig:resultinitialpop}(a)
shows the unnormalised populations $\mathcal{P}_0$, $\mathcal{P}_1$, and $\mathcal{P}_\mathrm{bound}=\mathcal{P}_0+\mathcal{P}_1+\mathcal{P}_{2+}$ 
as  functions of time under the strong field only, prior to any weak control 
pulse. Here, unnormalised means that the denominator in Eq.~\eqref{eq:population_KH} is omitted in order to quantify irreversible ionization. The total bound population $\mathcal{P}_\mathrm{bound}$ decays continuously due 
to ionisation, falling to approximately 10--15\% of the initial 
population over the pulse duration.  Nonetheless, further tests show the onset of stabilization and that the electronic wave packet is trapped in a dichotomous potential. For details, see Figs.~\ref{fig:1appendix} and \ref{fig:2appendix} in Appendix \ref{app:stability}. These standard tests have been discussed elsewhere \cite{popov1999applicability, Gavrila2002, Aynul2025}.
Of this surviving bound population, 
approximately 60\% resides on average in the qubit subspace $\{|\phi^{KH}_0\rangle, 
|\phi^{KH}_1\rangle\}$, split roughly equally between $\mathcal{P}_0$ and $\mathcal{P}_1$, 
with the remaining $\sim$40\% in higher KH eigenstates $\mathcal{P}_{2+}$. The 
absolute qubit population is therefore only around 10--15\% of the 
original wavefunction, yet it undergoes well-defined coherent dynamics. 

The inset in Fig.~\ref{fig:resultinitialpop}(a), taken after stabilization has been reached, shows that the bound-state population oscillates with different phases and these dynamics are being driven by the strong field. For later times, the surviving wave 
packet settles into a 
coherent superposition of these two qubit basis states, with $\mathcal{P}_0+\mathcal{P}_1$ approaching the overall bound-state population. Superimposed on the fast residual oscillations at $\omega_s$, the main part of the figure shows a slow envelope modulation, whose frequency we determine next. 
For that purpose, we employ populations as given Eq.~\eqref{eq:population_KH} and the asymmetry parameter as given in Eq.~\eqref{eq:asymmetry}. The asymmetry is plotted in Fig.~\ref{fig:resultinitialpop}(b), for $500 \hspace{0.1cm}\mathrm{a.u.}\leq t \leq 1600$ a.u. The figure shows an oscillation $T_{10} = 2\pi/\omega_{10} \approx 612$ a.u., where  $\omega_{10}$ is the energy difference between the two KH eigenstates.  These oscillations demonstrate that the strong-field 
wavepacket is in a coherent superposition of both KH 
eigenstates, with a stable relative phase: the slow oscillations are associated with the time-averaged potential \eqref{eq:fourierterm} and the fast oscillations with the higher orders in the KH expansion \eqref{eq:nfourierterm}. 
Fig.~\ref{fig:resultinitialpop}(b) also shows that, at the chosen times $T_{G}=500$ a.u. and $T_{G}=800$ a.u. the system is in a roughly equally weighted coherent superposition of the KH eigenstates {$\ket{\phi^{KH}_0}$, $\ket{\phi^{KH}_1}$}. This confirms 
the near-equatorial initial state on the Bloch sphere discussed in 
Section~\ref{sec:qubitbasis} for those times. Thus, one may characterize a qubit using the time-averaged light-induced KH potential, and the frequency of interest is $\omega_{10}$.

All subsequent population plots use a smooth line 
representation that 
follows the slow envelope of the population dynamics 
and excludes the fast micromotion oscillations at 
$\omega_s$ visible in the raw data 
[see \ Fig.~\ref{fig:resultinitialpop}(a)], since it 
is the slow envelope that captures the qubit-relevant 
transitions between the KH eigenstates. Beyond this 
smoothing, all following population plots will use the same convention as in Fig.~\ref{fig:resultinitialpop}(b), for which the 
surviving bound population is normalised at each time step, showing 
conditional populations for the bound-state subspace. 
This 
normalisation serves two purposes: it isolates the qubit-relevant 
dynamics from the overall ionisation decay, and it enables direct 
comparison with the ideal two-level case where no ionisation occurs. 
The conditional populations therefore satisfy $\mathcal{P}_0 + \mathcal{P}_1 + \mathcal{P}_{2+} = 1$ 
at all times, and deviations of $\mathcal{P}_0 + \mathcal{P}_1$ from its strong-field-only 
baseline directly measure gate-induced leakage rather than ionisation 
loss.
Gate-induced enhancement of ionisation could in 
principle contribute to population loss without 
appearing in the conditional populations; this is 
addressed by operating in the KH stabilisation 
regime where ionisation is strongly suppressed, and 
by monitoring both the total bound population and 
total ionisation fraction directly, confirming that 
the weak field introduces no significant additional 
ionisation above the strong-field-only baseline at 
the field strengths used here.

\subsection{Leakage times and decoherence}
\label{sec:decoherence}

\begin{figure}[H]
    \centering
\hspace*{-0.5cm}\includegraphics[width=1.1\linewidth]{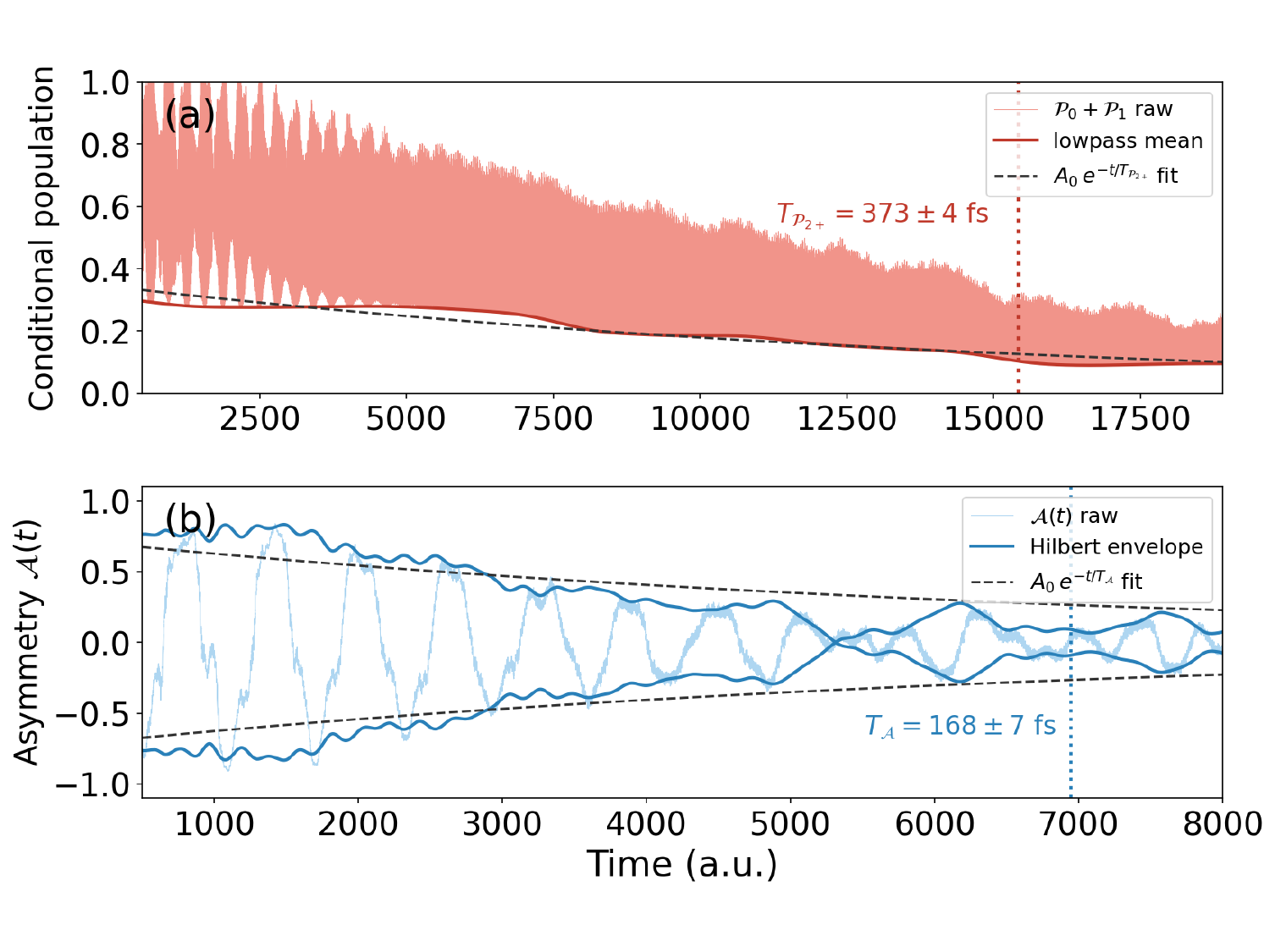} 
    \caption{Leakage times of the KH qubit under the strong 
field alone, for $\varepsilon_{0s} = 5$ a.u.\ and 
$\omega_s = 0.7$ a.u. 
\textbf{(a)} Population decay: the qubit-subspace 
conditional population $\mathcal{P}_0 + 
\mathcal{P}_1$ (raw, light red) is lowpass filtered 
to remove the fast $\omega_s$ oscillations and 
isolate the slow envelope (lowpass mean, solid red). 
An exponential fit (dashed) gives the time constant 
$T_{\mathcal{P}_{2+}} = 373 \pm 4$ fs. The decay 
reflects irreversible leakage to higher KH 
eigenstates $\mathcal{P}_{2+}$, driven by the 
residual fast oscillations at $\omega_s$, rather 
than coupling to an external bath. 
\textbf{(b)} Leakage time computed from the 
asymmetry parameter: the raw inter-well asymmetry 
$\mathcal{A}(t)$ (light blue) is bandpass filtered 
around $\omega_{10}$, and the amplitude of the 
resulting analytic signal gives the Hilbert 
envelope (solid blue). An exponential fit (dashed) 
gives $T_{\mathcal{A}} = 168 \pm 7$ fs, with the 
dashed vertical line marking $t = T_{\mathcal{A}}$ 
where the envelope has decayed to $1/e$ of its 
initial value. The loss arises from irreversible 
amplitude leakage out of the qubit subspace rather 
than reversible ensemble dephasing.
}\label{fig:resultdecoherence}
\end{figure}

Gate operations must be completed within the coherence 
time of the qubit; if the qubit loses its phase 
information before the gate finishes, the operation is 
meaningless. In this section, we will estimate for how long the gate can operate considering that our qubit is not perfect. In principle, there are two sources of imperfection. First, 
 the fast quiver motion at 
$\omega_s$ drives the electron wavepacket into regions 
beyond the time-averaged double-well on each optical 
cycle $T_s$ associated with the strong field. 
In the ideal time-averaged picture these 
excursions are removed by construction, but in the 
full TDSE they persist, repeatedly coupling the qubit 
subspace to higher KH eigenstates $\mathcal{P}_{2+}$ 
and returning population with a scrambled phase, which 
washes out the coherent inter-well oscillation. Second, even if a time-averaged KH potential is considered, the dichotomous potential obtained from the soft-core potential \eqref{eq:softcore} supports many bound states, as given in Table \ref{tab:energy_eigenvalues}. Therefore, there could be leakage also in this case, which would also affect the gate one is trying to build.

With that aim in mind, we calculate a time scale $T_{\mathcal{P}_{2+}}$, which is associated with how much population leaks from the qubit subspace into other KH eigenstates, using the TDSE computation. This is done by taking the joint conditional population $\mathcal{P}_0(t)+\mathcal{P}_1(t)$ as a function of time, whose envelop exhibits a slow 
decay, as shown in 
Fig.~\ref{fig:resultdecoherence}(a). This timescale is extracted by 
lowpass filtering the population to remove the fast 
oscillations at $\omega_s$ and isolate the slow envelope placed in the 
wide gap between the two frequencies of the problem: 
well below $\omega_s = 0.7$ a.u.\ so the fast quiver 
oscillation is rejected, with the large separation 
$\omega_{10} \ll \omega_s$ making the result 
insensitive to its precise value. This lowpass step 
removes both the fast $\omega_s$ oscillation and the 
$\omega_{10}$-scale inter-well oscillation, isolating 
the much slower underlying decay trend that is fitted 
for $T_{\mathcal{P}_{2+}}$.

Subsequently, we compute the mean of the filtered populations, for which an exponential fit 
$\mathcal{A}_0\exp(-t/ T_{\mathcal{P}_{2+}})$ gives
\begin{equation}
     T_{\mathcal{P}_{2+}} = 373 \pm 4 \text{ fs},
    \label{eq:T1}
\end{equation}
reflecting the slow irreversible loss of population 
from the qubit subspace through wavepacket excursions 
to higher KH eigenstates $\mathcal{P}_{2+}$ and 
eventual ionisation, and setting the upper bound on 
the gate operation window.

An alternative way of calculating leakage times is to use the asymmetry parameter and compute its decay time  $T_{\mathcal{A}}$. This is based on the argument that population transfer outside the qubit subspace {$\ket{\phi^{KH}_0}$, $\ket{\phi^{KH}_1}$} causes the inter-well oscillation 
amplitude to decay, as visible in 
Fig.~\ref{fig:resultdecoherence}(b). Unlike the lowpass 
filter used for $T_{\mathcal{P}_{2+}}$, here a bandpass 
filter around $\omega_{10}$ is required, since the 
quantity of interest is the oscillation itself rather 
than a slowly-decaying trend. The 
envelope is extracted by bandpass filtering the 
asymmetry around $\omega_{10}$ and taking the amplitude 
of the resulting analytic signal, shown as the solid 
blue curve in Fig.~\ref{fig:resultdecoherence}(b), with 
the dashed vertical line marking $t = T_{\mathcal{A}}$ where the 
fitted envelope has decayed to $1/e$ of its initial 
value. An exponential fit gives
\begin{equation}
    T_{\mathcal{A}} = 168 \pm 7 \text{ fs}.
    \label{eq:T2}
\end{equation}
Finally, one may also estimate the dephasing by comparing the actual asymmetry parameter with an idealized quantity $\mathcal{A}(t)=\mathcal{A}\cos(\omega_{10}t+\pi)$. 
 
The above timescales can be employed to estimate the number of  gate 
operations achievable in our system. For safety, we will consider the most conservative estimate $T_{\mathcal{A}}$. Using 
the $\pi$-pulse gate time $t_\pi = 2064$ a.u.\ $= 50$ fs 
at $\varepsilon_{0w} = 0.0003$ a.u., the number of gates 
within $T_{\mathcal{A}}$ is
\begin{equation}
    N_{\mathrm{gates}}= \frac{T_{\mathcal{A}}}{t_{\pi}} =   \frac{168}{50} = 3.4,
    \label{eq:Ngates}
\end{equation}
satisfying the DiVincenzo coherence criterion \cite{DiVincenzo2000}
$N_\mathrm{gates} > 1$. This gate budget can be improved either by 
increasing $\varepsilon_{0w}$ to reduce $t_\pi$, with 
ionisation essentially unchanged 
(Section~\ref{sec:ideal_comparison}), where at 
$\varepsilon_{0w} = 0.00065$ a.u.\ the gate time falls 
to $t_\pi = 1145$ a.u.\ $\approx 28$ fs giving 
$N_\mathrm{gates} = 6$, or by working at higher 
laser frequency where the KH stabilisation is stronger 
and wavepacket excursions to higher KH eigenstates are 
suppressed, increasing $T_{\mathcal{A}}$.

\subsection{Applying additional pulses}
Next, we apply additional pulses given by Eq.~\eqref{eq:weakpulse} to the light-induced system created by the strong field, in order to build ultrafast gates. Unless otherwise stated, we consider the full time dependent TDSE computation and initial times $T_{G}$ after the onset of stabilization.   As the main observables, we employ the asymmetry $\mathcal{A}(t)$, which gives us information about the phase and wavepacket localization, and the populations $\mathcal{P}_n$ associated with the KH eigenstates. We choose the driving-field intensities to be optimal according to the discussion in Sec.~\ref{sec:E1_results}, in which this is discussed in more detail. 

\subsubsection{$\pi$ pulse — Z gate}
\label{sec:pi_results}
\begin{widetext}
\begin{figure}[H]
    \centering
\includegraphics[width=0.9\textwidth]{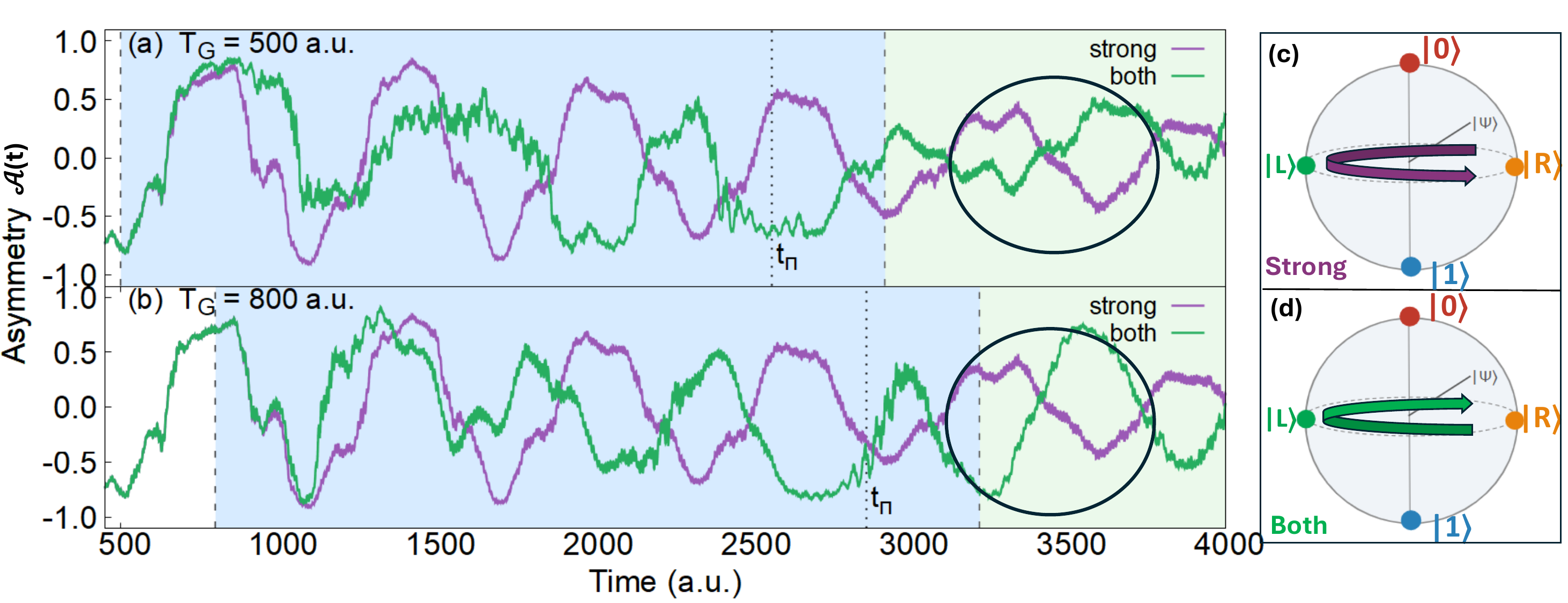}
\begin{minipage}{\textwidth}
    \caption{Asymmetry $\mathcal{A}(t)$ for the $\pi$ pulse at 
$\varepsilon_{0w} = 0.0003$ a.u. 
\textbf{(a)} $T_{G} = 500$ a.u.\ and 
\textbf{(b)} $T_{G} = 800$ a.u.: the both-fields 
result (green) is compared against the 
strong-field-only reference (purple). The 
background shading indicates the pulse timing: the 
white region before the first dashed line is the 
pre-pulse interval where only the strong field is 
present, the blue shaded region marks the weak 
pulse window (bounded by the dashed vertical lines 
at the start and end of the weak pulse), and the 
green region marks the post-pulse evolution under 
the strong field alone, after the weak pulse has 
switched off. The dotted vertical line marks the completion 
time $t_\pi = \tau_2 + \pi/\Omega_R$ (at $T_{G} + 
t_\pi$ on the absolute axis).
The black circles highlight the post-pulse regions 
where the both-fields oscillation is inverted 
relative to the strong-field-only reference --- the 
two oscillation curves are out of phase, the 
signature of coherent phase control by the weak 
pulse --- from which the phase shift is cleanly 
extractable. The phase shift is measured over a 
late-pulse and a post-pulse window: $[1700, 2200]$ 
and $[2922, 4000]$ a.u.\ for (a), and 
$[2200, 2864]$ and $[3222, 4000]$ a.u.\ for (b), 
chosen where the coherent oscillation is cleanly 
extractable in both signals. This yields post-pulse 
phase shifts $\delta = +0.937\pi$ 
($F_\varphi = 0.937$) for (a) and $\delta = -0.892\pi$ 
($F_\varphi = 0.892$) for (b), confirming a 
near-complete $Z$ gate. 
\textbf{(c, d)} Bloch sphere representations of the 
same effect: introducing the weak pulse flips the 
precession direction around the $z$-axis from 
anticlockwise for the strong field only 
[(c)] to clockwise for both fields [(d)], the 
$\pi$ phase inversion signature of a $Z$ gate and 
the manifestation of the same coherent phase control 
seen in the circled regions. The corresponding 
late-pulse fidelities are given in 
Table~\ref{tab:pi_pop_T2one}.}

    \label{fig:resultasymmetrypi}
   \end{minipage}
\end{figure}
\end{widetext}

We start from the $Z$ gate, which causes a phase flip in the equatorial states of the Bloch sphere (see Table \ref{tab:gates}). 
Fig.~\ref{fig:resultasymmetrypi} shows the asymmetry $\mathcal{A}(t)$ before and after the  $\pi$ pulse is applied. We start from the time for which there is the onset of stabilization and take  two pulse 
start times $T_{G} = 500$ a.u.\ [panel (a)] and 
$T_{G} = 800$ a.u.\ [panel (b)], each compared against 
the strong-field-only reference (purple).  For reference, the precession on the Bloch sphere is indicated on the right side [panels (c) and (d)] of the figure. The marked gate completion time $t_\pi = \tau_2 + 
\pi/\Omega_R$ is used as a fixed reference throughout. 
Since the full-TDSE time axis is absolute (measured 
from $t = 0$), this marker appears at $T_{G} + t_\pi$ 
in the figures, $t_\pi$ itself being measured from the 
pulse start $T_{G}$. 
In the strong-field-only case, the qubit state precesses 
anticlockwise around the $z$-axis on the Bloch sphere at frequency 
$\omega_{10}$. This precession manifests as the coherent inter-well oscillation of the asymmetry 
$\mathcal{A}(t) = \mathcal{A}\cos(\omega_{10}t + \phi) $. 
The introduction of the weak $\pi$ pulse inverts this 
precession direction, flipping the rotation from 
anticlockwise to clockwise around the $z$-axis and 
producing a $\pi$ phase shift in the asymmetry relative 
to the strong-field-only reference; the phase 
signature expected of a $Z$ gate. This inversion is visible in both panels as a 
phase shift of the both-fields oscillation relative to the 
strong-field-only reference, most clearly in the circled 
post-pulse regions where the two trajectories are out of phase.
At $T_{G} = 500$ a.u.\ the post-pulse phase shift is 
$\delta = +0.937\pi$, giving $F_\varphi = 0.937$, and at 
$T_{G} = 800$ a.u.\ the post-pulse phase shift is 
$\delta = -0.892\pi$, giving $F_\varphi = 0.892$. Both values 
confirm a near-complete $Z$ gate in the eigenstate basis.

\begin{widetext}
\begin{figure}[H]
    \centering
    \includegraphics[width=\textwidth]{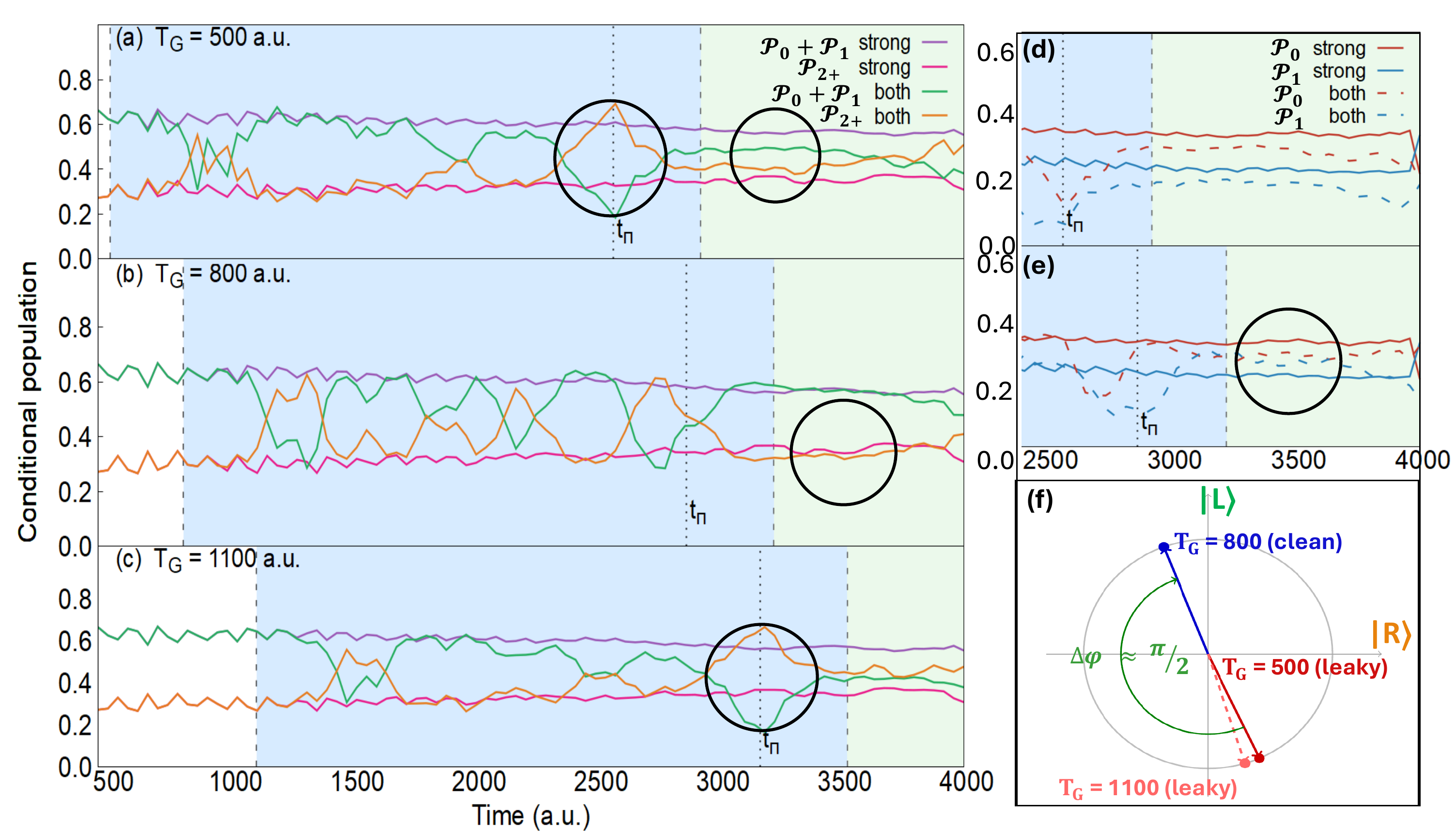}
    
    \vspace*{0.8cm}
    \begin{minipage}{\textwidth}
    \caption{Conditional populations for the $\pi$ pulse at three 
gate start times, $\varepsilon_{0w} = 0.0003$ a.u. 
\textbf{(a--c)} Qubit-subspace population 
$\mathcal{P}_0 + \mathcal{P}_1$ and higher-state 
leakage $\mathcal{P}_{2+}$ for both the 
strong-field-only reference (purple and magenta) and 
the both-fields case (green and orange), at 
$T_{G} = 500$ a.u.\ (a), $T_{G} = 800$ a.u.\ (b), 
and $T_{G} = 1100$ a.u.\ (c). The blue shaded region 
marks the weak pulse window, the green region the 
post-pulse evolution under the strong field alone, 
and the dotted vertical line marks the completion 
time $t_\pi = \tau_2 + \pi/\Omega_R$ (at $T_{G} + 
t_\pi$ on the absolute axis). In (b), the 
circled regions show that $\mathcal{P}_{2+}$ remains 
at the strong-field-only baseline (magenta) and 
$\mathcal{P}_0 + \mathcal{P}_1$ stays at its 
strong-field-only value (purple), with only 
redistribution between $\mathcal{P}_0$ and 
$\mathcal{P}_1$ [resolved in (e)] --- the signature 
of a clean intra-subspace gate. In (a) and (c), by 
contrast, the circled region at $t_\pi$ highlights 
transient leakage to $\mathcal{P}_{2+}$ above the 
strong-field-only baseline alongside the 
$\mathcal{P}_0$--$\mathcal{P}_1$ redistribution; the 
two cases share the same leakage pattern, correlated 
with the phase analysis in (f). 
\textbf{(d, e)} Individual eigenstate populations 
$\mathcal{P}_0$ (red) and $\mathcal{P}_1$ (blue) for 
the strong-field-only (solid) and both-fields 
(dashed) cases, for $T_{G} = 500$ a.u.\ (d) and 
$T_{G} = 800$ a.u.\ (e). The circled region in (e) 
shows the clean $\mathcal{P}_0 \leftrightarrow 
\mathcal{P}_1$ exchange --- $\mathcal{P}_0$ falls and 
$\mathcal{P}_1$ rises, a transfer between the two 
eigenstates driven by the weak field --- whereas in 
(d) both populations fall simultaneously, indicating 
leakage out of the qubit subspace. 
\textbf{(f)} Top-down view of the Bloch sphere, 
looking down the $z$-axis onto the equatorial plane 
(with $z$ directed out of the page), showing the 
qubit state at pulse start for the three cases: 
$T_{G} = 800$ a.u.\ (clean, blue) is separated from 
$T_{G} = 500$ a.u.\ (leaky, red) by a phase 
$\Delta\varphi = \omega_{10}(800-500) \approx \pi/2$, 
while $T_{G} = 1100$ a.u.\ (leaky, pink) --- one full 
inter-well oscillation period $T_{10} = 2\pi/\omega_{10} 
\approx 612$ a.u.\ after $T_{G} = 500$ a.u.\ --- 
returns to nearly the same phase point as 
$T_{G} = 500$ a.u., reproducing its leaky behaviour. %
Gate quality is thus governed by the oscillation phase 
$\varphi = \omega_{10} T_{G}$ at pulse start; see 
Tables~\ref{tab:pi_pop_T2one} and 
\ref{tab:T2one_axis_pi} for a summary.
}
    \label{fig:resultpopulationpi}        
    \end{minipage}
\end{figure} 
\end{widetext}

The population dynamics shed additional light on the gate quality, as illustrated in Fig.~\ref{fig:resultpopulationpi}. 
While both $T_{G}$ values produce comparable phase fidelities, 
the populations in Fig.~\ref{fig:resultpopulationpi} 
reveal a critical difference in gate quality. At 
$T_{G} = 800$ a.u.\ [panels (b) and (e)], all three conditions 
for a clean intra-subspace gate are met simultaneously at 
$t_\pi$: (i) $\mathcal{P}_1$ rises while $\mathcal{P}_0$ falls by the same amount, 
the clean $\mathcal{P}_0 \leftrightarrow \mathcal{P}_1$ exchange visible in 
panel (e); (ii) $\mathcal{P}_0 + \mathcal{P}_1$ remains conserved at its 
strong-field-only value in panel (b), and $\mathcal{P}_{2+}$ stays at 
the strong-field-only baseline throughout. (iii) Post-pulse, both 
$\mathcal{P}_0 + \mathcal{P}_1$ and $\mathcal{P}_{2+}$ return to and remain at the 
strong-field-only level, confirming a clean Z gate with no 
residual gate-induced leakage. We note that $\mathcal{P}_0$ returns close to its value 
at pulse-on ($T_G=800$ a.u.) by $t_\pi$, whereas 
$\mathcal{P}_1$'s return is less clean, with some 
genuine exchange between the two populations evident 
post-pulse. This partial exchange is difficult to 
resolve unambiguously from population data alone, since 
the near-equal $\mathcal{P}_0\approx\mathcal{P}_1$ at 
pulse-on makes a small swap and a return numerically 
similar; the phase evolution of the asymmetry, discussed 
above, remains the more reliable indicator of the gate's 
action. The clean post-pulse population exchange $\mathcal{P}_0 \leftrightarrow \mathcal{P}_1$, with 
$\mathcal{P}_{2+}$ remaining at the strong-field-only baseline, suggests that there is a transverse $R_x$ component alongside the 
$R_z(\pi)$ phase rotation. 

At $T_{G} = 500$ a.u.\ [panels (a) and (d)], by contrast, both 
$\mathcal{P}_0$ and $\mathcal{P}_1$ fall simultaneously at $t_\pi$, visible in 
panel (d), while $\mathcal{P}_{2+}$ rises transiently above the 
strong-field-only baseline in panel (a). The qubit subspace 
population $\mathcal{P}_0 + \mathcal{P}_1$ is not conserved at $t_\pi$, indicating 
that the weak field couples population out of the qubit subspace 
rather than rotating within it. Unlike the oscillatory fluctuations present at both $T_G$ 
values throughout the pulse window, this depletion at 
$T_G=500$ a.u.\ persists through the ramp-down and into 
the post-pulse region, with $\mathcal{P}_0+\mathcal{P}_1$ 
remaining below, and $\mathcal{P}_{2+}$ above, their 
respective strong-field-only baselines. Despite the comparable phase 
fidelity, this constitutes a leaky rotation rather than a pure 
intra-subspace Z gate. These results are summarised in 
Table~\ref{tab:pi_pop_T2one}. 

\begin{table}[H]
\small
\caption{$\pi$ pulse — gate characterisation, 
$\varepsilon_{0w} = 0.0003$ a.u., 
$\varepsilon_{0s}  = 5$ a.u., $\omega_{s} = 0.7$ a.u.}
\label{tab:pi_pop_T2one}
\begin{tabular}{p{2.2cm}p{2.8cm}p{2.8cm}}
\hline\hline
& $T_{G} = 500$ a.u. & $T_{G} = 800$ a.u. \\
\hline
$t_{\pi,r} = \pi/\Omega_R$ (a.u.) & 1706 & 1706 \\
$t_\pi = \tau_2 + \pi/\Omega_R$ (a.u.) & 2064 & 2064 \\
$t_\pi$ (fs) & 50 & 50 \\
$\delta$ (late pulse)    & $-0.945\pi$ & $+0.853\pi$ \\
$F_\varphi$ (late pulse) & 0.945       & 0.853       \\
$\delta$ (post pulse)    & $+0.937\pi$ & $-0.892\pi$ \\
$F_\varphi$ (post pulse) & 0.937       & 0.892       \\
$P_0+P_1$ at $t_\pi$ &
Depleted, both $P_0$ and $P_1$ fall &
Conserved, matches strong-only \\
$P_0 \leftrightarrow P_1$ exchange &
\xmark\ both fall simultaneously &
\checkmark\ $P_1$ rises, $P_0$ falls \\
$P_{2+}$ at $t_\pi$ &
Rises above strong-only &
Stays at strong-only \\
$P_{2+}$ post pulse &
Partially elevated &
Returns to strong-only \\
Overall assessment &
Leaky rotation &
Clean Z gate \\
\hline\hline
\end{tabular}
\end{table}

The contrast between both initialization times demonstrates 
that the pulse start time influences the subsequent dynamics, 
with $T_{G} = 800$ a.u.\ uniquely favouring a clean 
parity-selective intra-subspace rotation while $T_{G} = 500$ 
a.u.\ drives transient leakage to $\mathcal{P}_{2+}$. Because the inter-well dynamics is periodic, with oscillations $T_{10} = 2\pi/\omega_{10} \approx 612$ a.u., it is a legitimate question to ask whether initial times $T_{G}$ separated by a whole cycle $T_{10}$ will reproduce similar patterns. 

Fig. ~\ref{fig:resultpopulationpi}(c) shows that this is indeed the case: the pulse start time $T_{G}$ determines the phase 
$\phi = \omega_{10} T_{G}$ accumulated by the strong-field 
inter-well oscillation at the moment the weak control pulse is applied. An initial time $T_{G} = 1100$ a.u.\ produces the same leakage 
pattern as $T_{G} = 500$ a.u.: $\mathcal{P}_0+\mathcal{P}_1$ dips sharply 
at $t_\pi$ while $\mathcal{P}_{2+}$ rises transiently above the 
strong-field-only baseline. The repetition of the leaky 
behaviour confirms that gate 
quality is governed by the oscillation phase 
$\phi = \omega_{10}T_{G}$ at pulse start rather 
than by the total evolution time elapsed before 
the pulse or the degree of bound population 
stabilisation reached by that time. For clarity, Fig.~\ref{fig:resultpopulationpi}(f) shows that  
$T_{G} = 500$ and $1100$ a.u. place the qubit state at nearly the 
same phase point on the Bloch sphere equator at both 
pulse start times, while for $T_{G} = 800$ a.u. it exhibits a  phase shift of $\Delta\phi = \omega_{10}(800-500) \approx 
\pi/2$.  Finally, the absence of population exchange between the two eigenstates forming the qubit, in contrast to what is observed for $T_{G} = 500$ and $1100$ a.u., is evidence that 
$T_{G}$ is not merely a timing parameter, but also  
selects which axis of the Bloch sphere the qubit rotates 
around. A summary of the influence of $T_G$ for the $\pi$ pulse is provided in Table \ref{tab:T2one_axis_pi}.

\begin{table}[H]
\centering
\renewcommand{\arraystretch}{1.8}
\caption{ Influence of $T_{G}$ 
— $\pi$ pulse, $\varepsilon_{0w} = 0.0003$ a.u.}
\label{tab:T2one_axis_pi}
\resizebox{\columnwidth}{!}{%
\begin{tabular}{lp{4cm}p{3.2cm}}
\hline\hline
& $T_{G}=500$ a.u., $1100$ a.u. & $T_{G}=800$ a.u. \\
\hline
Rotation axis & 
$R_z(\pi)$ & 
$R_z(\pi)+R_x$ \\
$F_\varphi$ (late) & 
0.945 & 0.853 \\
$\mathcal{P}_0\leftrightarrow \mathcal{P}_1$ & 
\xmark & \checkmark \\
$\mathcal{P}_{2+}$ at $t_\pi$ & 
transient rise & near strong-only \\
Post-pulse & 
inverted oscillation & clean gate \\
\hline\hline
\\
\end{tabular}}
\end{table}

\subsubsection{$\pi/2$ pulse — S gate}
\label{sec:halfpi_results}

\begin{widetext}
\begin{figure}[H]
    \centering
    \includegraphics[width=\textwidth]{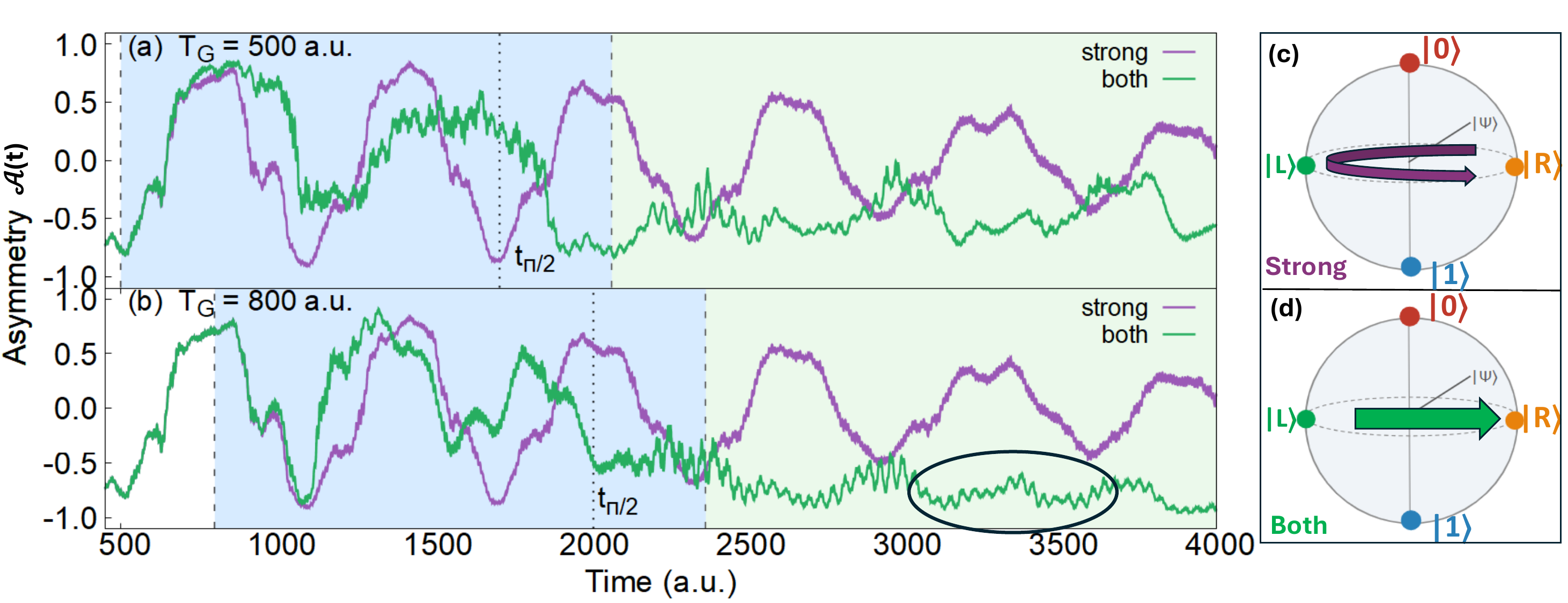}
    \begin{minipage}{\textwidth}
    
    \vspace*{0.8cm}
    \caption{Asymmetry $\mathcal{A}(t)$ for the $\pi/2$ pulse at 
$\varepsilon_{0w} = 0.0003$ a.u.\ for 
\textbf{(a)} $T_{G} = 500$ a.u.\ and 
\textbf{(b)} $T_{G} = 800$ a.u.: the both-fields 
result (green) is compared against the 
strong-field-only reference (purple). The blue 
shaded region marks the weak pulse window, the green 
region the post-pulse evolution under the strong 
field alone, and the dotted vertical line marks the completion 
time $t_\pi = \tau_2 + \pi/\Omega_R$ (at $T_{G} + 
t_\pi$ on the absolute axis).
At $T_{G} = 500$ a.u.\ the coherent oscillation 
persists post-pulse with a residual right-well 
preference but no clean locking, giving a late-pulse 
phase shift $\delta = -0.818\pi$ 
($F_\varphi = 0.364$). At $T_{G} = 800$ a.u.\ the 
inter-well oscillation ceases post-pulse and the 
asymmetry locks to a well-defined right-well bias 
$\mathcal{B} = -0.783$ (circled), giving a 
late-pulse phase shift $\delta = +0.535\pi$ 
($F_\varphi = 0.930$), close to the target $\pi/2$ 
and confirming a near-complete $S$ gate. This 
post-pulse localisation to a single well constitutes 
an effective qubit readout mechanism. The late-pulse 
phase shift is measured over the windows 
$[1500, 2069]$ a.u.\ for (a) and $[1800, 2369]$ 
a.u.\ for (b). 
\textbf{(c, d)} Bloch sphere representations showing 
the equatorial precession for the strong field only 
[(c)] and the $\pi/2$ phase rotation toward the 
right-well localised state $\ket{R}$ for both fields 
[(d)], the manifestation of the localisation locking 
seen in the circled region.}
    \label{fig:resultasymmetrypiover2}
     \end{minipage}
\end{figure}
\end{widetext}

Next, we use weak $\pi/2$ pulses to construct $S$ gates in the KH potential. 
Fig.~\ref{fig:resultasymmetrypiover2} shows the asymmetry $\mathcal{A}(t)$ 
for the $\pi/2$ pulse at $\varepsilon_{0w} = 0.0003$ a.u.\ for 
$T_{G} = 500$ a.u.\ [panel (a), green] and $T_{G} = 800$ a.u.\ 
[panel (b), green], compared against the strong-field-only 
reference (purple). For this gate, one aims for a $\pi/2$ phase shift in the time-dependent asymmetry, with regard to the case for which only the strong field is present (see Table \ref{tab:gates}). As for the $\pi$ pulse case, 
$t_{\pi/2}=\tau_2+\pi/(2\Omega_R)$ is used as a fixed 
reference (at $T_{G} + t_{\pi/2}$ on the 
absolute axis). 

The primary figure of merit for the S gate is the late-pulse 
phase fidelity, measured during the weak pulse window while the 
coherent oscillation is still present in both cases. For initial times 
$T_{G} = 800$ a.u.\ the late-pulse phase shift is 
$\delta = +0.535\pi$, giving $F_\varphi = 0.930$, close to 
the target $\delta = \pi/2$ and confirming a near-complete $\pi/2$ phase shift, the 
signature of an $S$ gate in the eigenstate basis. We verified that $\delta$ remains stable across 
the interval between the true pulse-area completion 
point and the marked $t_{\pi/2}$ (see discussion in Sec.~\ref{sec:gates}), confirming that this 
result is not sensitive to the marker convention.
For $T_{G} = 500$ 
a.u.\ the late-pulse phase shift is $\delta = -0.818\pi$, 
giving $F_\varphi = 0.364$, significantly below the target, 
indicating a partial and impure rotation at this $T_{G}$. 

Further to the phase shift, one is interested in localizing the wave
packet in one of the wells after the weak pulse is  switched off, as this constitutes an effective qubit readout mechanism. In addition, the right-well localisation 
bias post-pulse is a direct signature of the 
gate having acted, as   
the $\pi/2$ phase rotation shifts the mean of the 
inter-well oscillation toward the right well. 

Fig.~\ref{fig:resultasymmetrypiover2} shows a right-well 
localisation bias post-pulse, which is a direct signature 
of the $\pi/2$ gate having acted, but with 
markedly different character. For a pulse starting at $T_{G} = 500$ 
a.u.\ the coherent oscillation persists post-pulse 
with a residual right-well preference but no clean 
locking, while if it starts  at $T_{G} = 800$ a.u.\ the 
oscillation is suppressed and the asymmetry locks 
to a well-defined negative bias of 
$\mathcal{B} = -0.783$, indicating right-well 
localisation. 

The post-pulse fidelity is not the primary figure of merit here  
because for $T_{G} = 800$ a.u.\ the oscillation collapses 
post-pulse due to localisation locking and the phase $\delta$ 
cannot be reliably extracted from a locked asymmetry, giving 
the nominal $F_\varphi = 0.001$. The gate is therefore 
characterised by its late-pulse value $F_\varphi = 0.930$. 

However, fidelity and localisation appear to be inter-related. For example, for an initial time $T_{G} = 800$ a.u.\ the 
stronger gate fidelity ($F_\varphi = 0.930$) 
is associated with a well-defined bias $\mathcal{B} = -0.783$, 
with the oscillation largely suppressed. In contrast, for a weak pulse starting at 
$T_{G} = 500$ a.u.\ the weaker gate 
($F_\varphi = 0.364$) produces a smaller bias 
with residual oscillation persisting on top. The 
magnitude of the localisation bias therefore 
reflects the gate fidelity, making it a natural 
gate quality indicator: a stronger, cleaner gate 
produces a more pronounced and stable well 
preference. 

\begin{figure}[H]
  \centering
 \hspace*{-0.6cm} \includegraphics[width=0.55\textwidth]{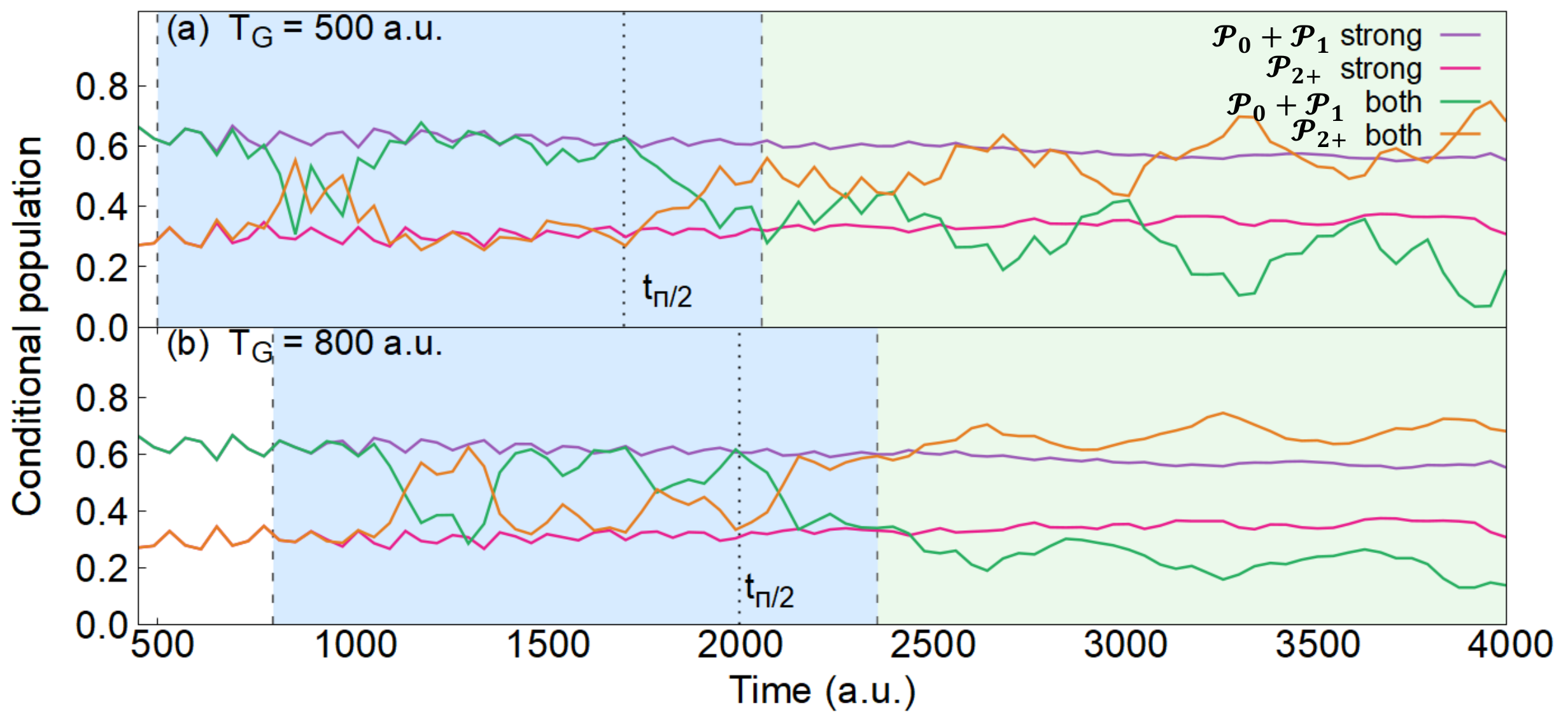}
    \begin{minipage}{0.5\textwidth}
    \caption{Qubit-subspace population $\mathcal{P}_0 + 
\mathcal{P}_1$ and higher-state leakage 
$\mathcal{P}_{2+}$ for the $\pi/2$ pulse at 
$\varepsilon_{0w} = 0.0003$ a.u., for 
\textbf{(a)} $T_{G} = 500$ a.u.\ and 
\textbf{(b)} $T_{G} = 800$ a.u. The 
strong-field-only reference is shown in purple 
($\mathcal{P}_0 + \mathcal{P}_1$) and magenta 
($\mathcal{P}_{2+}$), and the both-fields case in 
green ($\mathcal{P}_0 + \mathcal{P}_1$) and orange 
($\mathcal{P}_{2+}$). The blue shaded region marks 
the weak pulse window, the green region the 
post-pulse evolution under the strong field alone, 
and the dotted vertical line ... marks the completion time $t_{\pi/2} = \tau_2 + 
\pi/(2\Omega_R)$ (at $T_{G} + t_{\pi/2}$ on the 
absolute axis). In both 
cases $\mathcal{P}_0 + \mathcal{P}_1$ and 
$\mathcal{P}_{2+}$ remain near their 
strong-field-only values at $t_{\pi/2}$, confirming 
that the weak pulse causes no additional leakage at 
the gate completion point. After $t_{\pi/2}$, the 
continuing ramp-down drives population 
off-resonantly to higher KH eigenstates: 
$\mathcal{P}_{2+}$ rises above and $\mathcal{P}_0 + 
\mathcal{P}_1$ falls below their strong-field-only 
baselines, remaining so post-pulse. Despite this 
leakage, the right-well localisation persists and 
could in principle be read out. A summary is given 
in Table~\ref{tab:halfpi_pop_T2one}.}
    \label{fig:resultleakagepiover2}
      
    \end{minipage}
\end{figure}

Additionally to achieving localisation and a $\pi/2$ phase-shift in the asymmetry, to ensure that the S gate is successfully built, one must guarantee that the localized wave packet is within the qubit subspace for which it operates. This information is obtained by inspecting the population dynamics. 
Fig.~\ref{fig:resultleakagepiover2} shows the population $\mathcal{P}_0+\mathcal{P}_1$ in the qubit 
subspace, and the population $\mathcal{P}_{2+}$ leaking towards higher excited states. 
During the late 
pulse window both $T_{G}$ values show that $\mathcal{P}_0+\mathcal{P}_1$ in the subspace of interest and the population  $\mathcal{P}_{2+}$ leaked to the other KH eigenstates
remain near their strong-field-only value at the marked reference time $t_{\pi/2}$. This confirms that the weak pulse does not cause additional leakage at this moment. However, after $t_{\pi/2}$, 
$\mathcal{P}_{2+}$ begins to rise as the pulse ramp-down 
continues to drive the system beyond the gate 
completion point, coupling population off-resonantly 
to higher KH eigenstates before the field switches 
off. After $t_{\pi/2}$,  $\mathcal{P}_0+\mathcal{P}_1$ falls below the 
strong-field-only baseline while $\mathcal{P}_{2+}$ rises 
and remains elevated for both cases, mirroring $\mathcal{P}_0+\mathcal{P}_1$. Despite the leakage, the localization remains and could in principle be read out. 
A summary is provided in Table~\ref{tab:halfpi_pop_T2one}.

These results also show that the same selectivity with regard to the initial time $T_{G}$ as in the $\pi-$ pulse case applies. At $T_{G} = 800$ a.u.\ the late-pulse $F_\varphi = 0.930$ 
confirms a clean $R_z(\pi/2)$ S gate with $\mathcal{P}_0+\mathcal{P}_1$ near 
strong-only during the pulse, while at $T_{G} = 500$ a.u.\ 
only a partial rotation is achieved ($F_\varphi = 0.364$). 
Thus, $T_{G} = 800$ a.u. favours parity-selective intra-subspace rotation 
for both $\pi$ and $\pi/2$ pulses. For clarity, these parameters are given in Table \ref{tab:T2one_axis_halfpi}.

\begin{table}[H]
\centering
\small
\renewcommand{\arraystretch}{1.6}
\caption{$\pi/2$ pulse — gate characterisation, 
$\varepsilon_{0w} = 0.0003$ a.u., $\varepsilon_{0s}= 5$ a.u., 
$\omega_s = 0.7$ a.u.}
\label{tab:halfpi_pop_T2one}
\begin{tabular}{p{2.2cm}p{2.8cm}p{2.8cm}}
\hline\hline
& $T_{G} = 500$ a.u. & $T_{G} = 800$ a.u. \\
\hline
$t_{\pi/2,r} = \pi/(2\Omega_R)$ (a.u.) & 853 & 853 \\
$t_{\pi/2} = \tau_2 + \pi/(2\Omega_R)$ (a.u.) & 1211 & 1211 \\
$t_{\pi/2}$ (fs)              & 29   & 29   \\
$\delta$ (late pulse)         & $-0.818\pi$ & $+0.535\pi$ \\
$F_\varphi$ (late pulse)      & 0.364       & 0.930       \\
$\delta$ (post pulse)         & $+0.894\pi$ & $\sim 0$    \\
$F_\varphi$ (post pulse)      & 0.212       & —           \\
Localisation bias $\mathcal{B}$ & —         & $-0.783$    \\
$P_0+P_1$ during pulse &
Near strong-only &
Near strong-only \\
$P_0+P_1$ post pulse &
Falls below strong-only &
Falls below strong-only \\
$P_{2+}$ post pulse &
Rises above strong-only &
Rises above strong-only \\
$P_1$ post pulse &
Remains present &
Drops to near zero \\
Post-pulse character &
Phase-shifted oscillation persists &
Oscillation suppressed; right-well localisation locked \\
Overall assessment &
Partial rotation, $F_\varphi = 0.364$ &
Clean S gate, $F_\varphi = 0.930$ (late pulse) \\
\hline\hline
\end{tabular}
\end{table}

\begin{table}[H]
\centering
\renewcommand{\arraystretch}{1.8}
\caption{Influence of $T_{G}$  - $\pi/2$ pulse,  $\varepsilon_{0w} = 0.0003$ a.u.}
\label{tab:T2one_axis_halfpi}
\resizebox{\columnwidth}{!}{%
\begin{tabular}{lp{4cm}p{3.2cm}}
\hline\hline
& $T_{G}=500$ a.u., $1100$ a.u. & $T_{G}=800$ a.u. \\
\hline
Rotation axis & 
partial $R_z$ & 
$R_z(\pi/2)$ \\
$F_\varphi$ (late) & 
0.364 & 0.930 \\
$\mathcal{P}_0\leftrightarrow \mathcal{P}_1$ & 
\xmark & \xmark \\
$\mathcal{P}_{2+}$ at $t_{\pi/2}$ & 
near strong-only & near strong-only \\
$P_{2+}$ post $t_{\pi/2}$ & 
rises, partially elevated & rises, more elevated \\
Post-pulse & 
oscillation persists, & localisation locked, \\
& weak right-well bias & $\mathcal{B} = -0.783$ \\
\hline\hline
\end{tabular}}
\end{table}

\subsection{Comparison with time-averaged models}
\label{sec:ideal_comparison}
To establish the validity of the two-level description 
and contextualise the full TDSE results, we compare the 
gate dynamics against the ideal KH case: the 
time-averaged two-level system with dipole coupling and 
no leakage to higher eigenstates. The weak field coupling 
used here is given by the matrix element $\bra{\phi^{KH}_1}\widetilde{H}_{\text{coupl}}\ket{\phi^{KH}_0}$ and its complex conjugate, where the Hamiltonian is given by Eq.~\eqref{eq:Hmixed}. This matrix element retains only the first term of the coupling 
expression; the 
second term vanishes identically by orthogonality of the 
KH basis states. This coupling implies that the KH transformation was applied with regard to the strong-field only. The coupling employed the full transformation [Eq.~\eqref{eq:couplingKH}] had poorer agreement with the TDSE computations and for that reasons is not employed here (not shown). The main 
question is whether the idealised case gives a good 
description of the full TDSE gate dynamics, which we 
address below.

As in the full TDSE, population diagnostics and the 
asymmetry serve complementary roles: the asymmetry 
carries the phase information that identifies the gate, 
while the populations distinguish between gates sharing 
the same phase signature. In the ideal two-level model, 
however, there is no leakage to higher KH 
eigenstates, so the populations here isolate the gate's 
action on the qubit subspace alone, making the 
distinction between different gates easier.  In principle, the KH potential supports other, higher-lying bound states (for clarity see Table \ref{tab:energy_eigenvalues}), but applying a resonant weak pulse with the two bound states of interest ensures that there is practically no population transfer outside the qubit subspace.  In the results that follow, we have applied the weak pulses at $t=0$ as, for an idealized, time-independent dichotomous potential, there is no need to wait for the onset of stabilization.

\begin{figure}[H]
    \centering
    \includegraphics[width=0.9\linewidth]
    {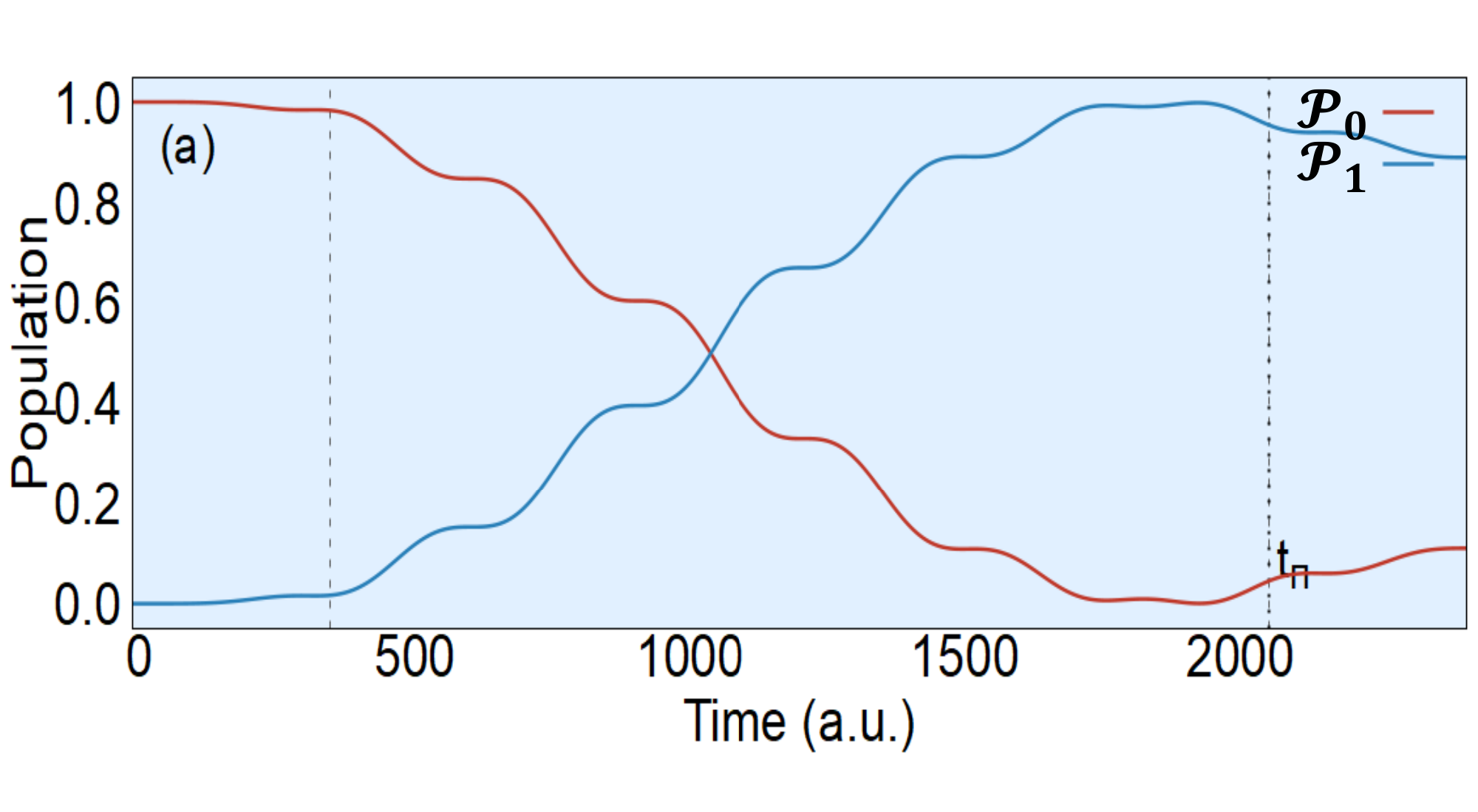}
    \caption{Ideal $X$ gate --- $\pi$ pulse at 
$\varepsilon_{0w} = 0.0003$ a.u.\ in the 
time-averaged two-level limit, with the KH ground 
state $\ket{\phi^{KH}_0}$ as the initial condition. 
The population inversion $\mathcal{P}_0 \to 0$, 
$\mathcal{P}_1 \to 1$ is the signature of an $X$ 
gate, achieved here with $\mathcal{P}_0 + 
\mathcal{P}_1 = 1$ maintained throughout, confirming 
no leakage in the ideal two-level case. The blue 
shaded background indicates that the weak pulse with a full trapezoidal envelope (ramp up, flat top, ramp down) is 
active from $t = 0$ to the end of the pulse; the 
dashed vertical line marks the end of the ramp-up 
(at $\tau_2$) and the dotted vertical line marks the 
completion time $t_\pi = \tau_2 + \pi/\Omega_R$. The 
population minimum occurs slightly before $t_\pi$, 
owing to the additional pulse area accumulated during 
the trapezoidal ramp-up, with a small partial 
back-transfer during the ramp-down.}
    \label{fig:ideal_Xgate}
\end{figure}

\subsubsection{Ideal gates and initial conditions}

Fig.~\ref{fig:ideal_Xgate} demonstrates a near-perfect X gate. 
 With the KH ground 
state $\ket{\phi^{KH}_0}$ as the initial condition, a $\pi$ 
pulse at $\varepsilon_{0w} = 0.0003$ a.u.\ drives a 
 complete population inversion 
$\mathcal{P}_0 \to 0$, $\mathcal{P}_1 \to 1$ with no leakage, i.e., 
$\mathcal{P}_0 + \mathcal{P}_1 = 1$ throughout. The population minimum 
occurs slightly before $t_\pi$ due to the additional 
pulse area accumulated during the trapezoidal ramp-up, 
consistent with the trapezoidal envelope correction discussed in Sec.~\ref{sec:gates}, with a small partial back-transfer 
during the ramp-down.

\begin{figure}[H]
    \centering
    \includegraphics[width=1\linewidth]
    {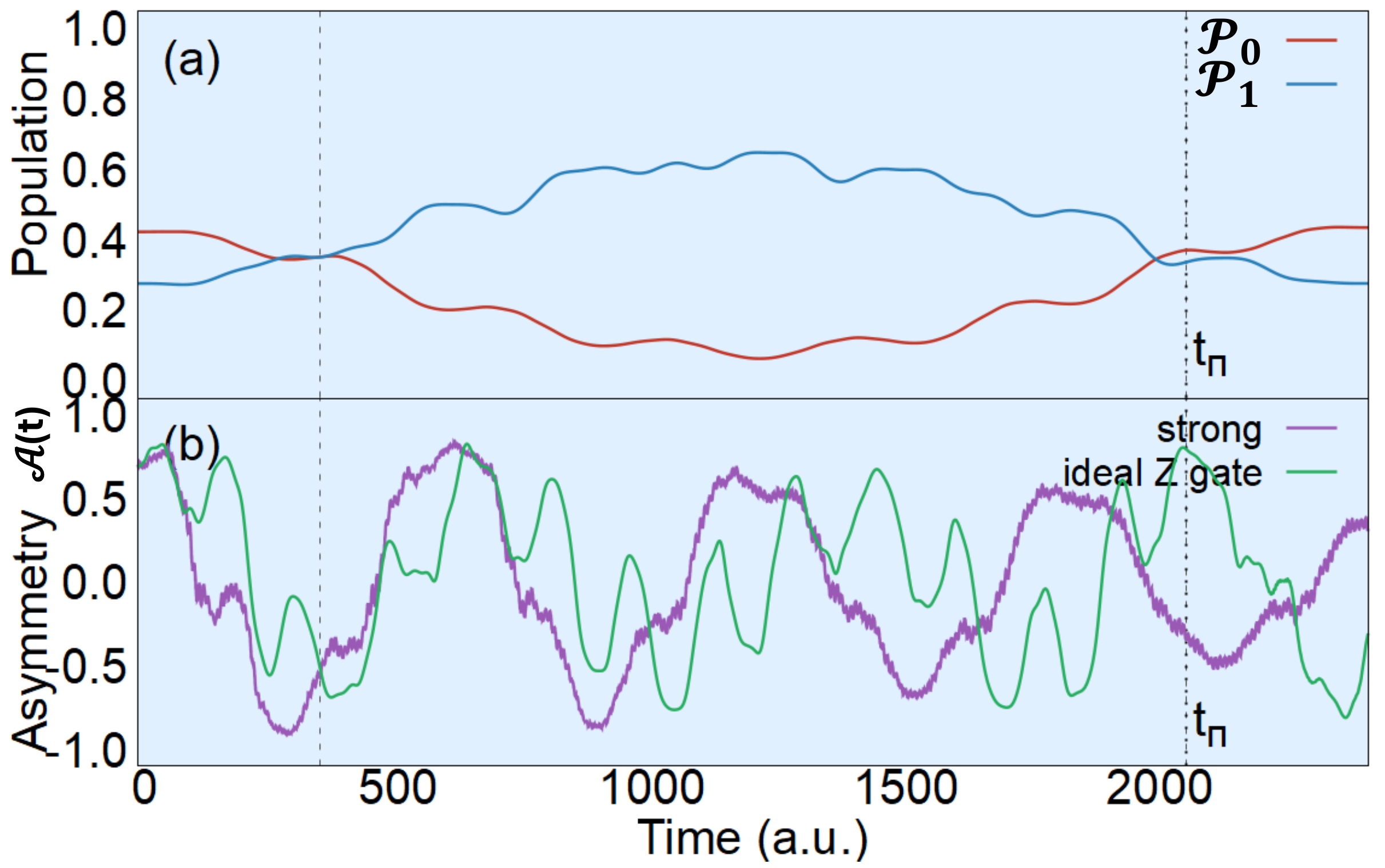}
    \caption{Ideal $Z$ gate --- $\pi$ pulse at 
$\varepsilon_{0w} = 0.0003$ a.u.\ in the 
time-averaged two-level limit, with a hybrid 
equatorial initial condition extracted from the full 
TDSE at $T_{G} = 800$ a.u.\ [Eq.~\eqref{eq:hybrid}]. 
\textbf{(a)} Populations $\mathcal{P}_0$ (red) and 
$\mathcal{P}_1$ (blue) remain close to their initial 
values throughout the pulse, as expected for a phase 
gate. The blue shaded background indicates that the weak pulse with a full trapezoidal envelope (ramp up, flat top, ramp down) is active from $t = 0$ to the end of the 
pulse; the dashed vertical line marks the end of the 
ramp-up (at $\tau_2$) and the dotted vertical line 
marks the completion time $t_\pi = \tau_2 + 
\pi/\Omega_R$. 
\textbf{(b)} Asymmetry $\mathcal{A}(t)$ for the ideal 
$Z$ gate (green) compared against the 
strong-field-only reference (purple), the latter 
time-shifted by $-800$ a.u.\ to align with the ideal 
pulse-on time at $t = 0$. A fit over the late-pulse 
window $[1700, 2200]$ a.u.\ gives a phase shift 
$\delta = -0.755\pi$, a near-complete $\pi$ phase 
inversion and the signature of a $Z$ gate, in fair 
agreement with the full TDSE result 
($\delta = -0.853\pi$ over the same window). The 
ideal case tracks the strong-field-only oscillation 
closely in the early pulse window, before 
$\mathcal{P}_{2+}$ leakage accumulates in the full 
case; this confirms that the two-level coupling 
[Eq.~\eqref{eq:Hmixed}] accurately describes the gate 
dynamics.}
    
    \label{fig:ideal_Zgate}
\end{figure}

In Fig.~\ref{fig:ideal_Zgate}, we illustrate that applying a $\pi$ pulse to a different initial state will lead to another gate.  For that purpose, we have extracted the KH eigenstate superposition from the full TDSE computation at  $T_{G} = 800$ a.u, and used it as an initial condition for the time-independent KH Hamiltonian. This state is approximately equatorial, and is described by Eq.~\eqref{eq:hybrid}. 
With this hybrid initial condition,
the ideal populations $\mathcal{P}_0$ and $\mathcal{P}_1$ 
remain close to their initial values throughout the 
pulse, as shown in Fig.~\ref{fig:ideal_Zgate}(a). 
The asymmetry, shown in Fig.~\ref{fig:ideal_Zgate}(b), is 
fitted over the same late-pulse window as the full TDSE 
case, $[1700, 2200]$ a.u., giving a phase shift of 
$\delta = -0.755\pi$, indicating a near-complete $\pi$ 
phase inversion (the signature of a $Z$ gate), in fair 
agreement with the full TDSE result 
$\delta = -0.853\pi$ (Section~\ref{sec:pi_results}). The 
strong-field-only reference is time-shifted by $-800$ 
a.u.\ to align with the ideal case pulse-on time at 
$t=0$, enabling direct comparison of the inter-well 
oscillation phase. This confirms that gate identity is  determined by the initial state position on 
the Bloch sphere, as predicted by the two-level 
formalism.

The hybrid initial condition provides a direct like-for-like 
comparison between the ideal and full TDSE cases at the same 
starting state. The Z gate asymmetry panel 
(Fig.~\ref{fig:ideal_Zgate}(b) shows that the ideal case 
tracks the strong-field-only oscillation pattern closely 
in the early pulse window before leakage to $P_{2+}$ has 
accumulated significantly, confirming that the coupling given by Eq.~\eqref{eq:Hmixed} yields a good description of the 
gate dynamics for the full time-dependent Hamiltonian. The divergence 
between ideal and full TDSE grows progressively during the 
pulse as $\mathcal{P}_{2+}$ leakage accumulates in the full case, 
broadening and damping the oscillation relative to the 
clean ideal oscillation. This confirms that the 
two-level description is accurate in the early gate window 
and that leakage is the primary mechanism degrading fidelity 
in the full TDSE, not a breakdown of the coupling itself.

\subsubsection{State initialization}

An important condition for a qubit is to be able to initialize it to a known state before the start of the gates \cite{DiVincenzo2000}. At the moment, the full TDSE allows access to approximate equatorial states. However, the 
equatorial state at the times $T_G$ cannot be cleanly rotated to a 
pure eigenstate pole; $P_{2+}$ leakage during the 
preparatory pulse dominates and prevents a clean 
$R_x(\pi/2)$ rotation toward $\ket{\phi^{KH}_0}$ or $\ket{\phi^{KH}_1}$. This hinders the construction of other gates, such as the X gate discussed here.  Therefore, one must devise a procedure to induce transitions to the pole states. The ideal case provides a 
testing ground for a route  to addressing the initialisation challenge identified in the DiVincenzo assessment.

\begin{figure}[H]
\centering
\begin{tikzpicture}[scale=0.6]
\draw[->, thick, black] (0,0) -- (14,0) 
    node[right, font=\footnotesize] {time};
\node[left, font=\footnotesize] at (0,0) {$t=0$};

\node[below, gray!70, font=\tiny] at (6.9,-0.65) 
    {};
\draw[blue!70, thick, fill=blue!15]
    (1.0, 0) -- (1.0, 0) 
    .. controls (1.3, 1.6) and (1.7, 1.6) .. 
    (2.0, 1.8)
    .. controls (2.3, 1.6) and (2.7, 1.6) ..
    (3.0, 0) -- cycle;
\node[blue!80, font=\footnotesize, align=center] at (2.0, 2.2) 
    {$\pi/2$ pulse};
\node[blue!60, font=\small, align=center] at (2.0, -0.35)
    {$T_{G_1}$};
\draw[blue!40, rounded corners, dashed] 
    (0.2, -2.8) rectangle (3.6, 2.6);
\node[blue!80, font=\small, align=center] at (2.0, 2.85)
    {\textbf{Stage 1}};
\node[blue!70, font=\footnotesize, align=center] at (2.0, -1.3)
    {Initialisation};
\node[blue!60, font=\footnotesize, align=center] at (2.0, -2)
    {$\ket{\Psi_\mathrm{eq}} \to \ket{\phi^{KH}_0}$ };
\begin{scope}[shift={(2.0, 4.8)}, scale=0.88]
    \draw[gray!50] (0,0) circle (1);
    \draw[gray!40, dashed] (1,0) arc (0:180:1 and 0.3);
    \draw[gray!60] (-1,0) arc (180:360:1 and 0.3);
    \draw[->, gray!60, thin] (0,-1.2) -- (0,1.3);
    \filldraw[blue!60] (0.9, 0.3) circle (2.5pt);
    \filldraw[red!70] (0,1) circle (2.5pt);
    \draw[->, blue!70, very thick] (0.9,0.3) -- (0.1,0.95);
    \node[red!70, font=\small] at (0.5, 1.6) {$\ket{\phi^{KH}_0}$};
\end{scope}
\draw[orange!80, thick, fill=orange!15]
    (5.0, 0) -- (5.0, 0)
    .. controls (5.4, 2.0) and (5.8, 2.0) ..
    (6.2, 2.4)
    .. controls (6.6, 2.0) and (7.0, 2.0) ..
    (7.4, 0) -- cycle;
\node[orange!90, font=\footnotesize, align=center] at (6.2, 2.85)
    {$\pi$ or $\pi/2$ pulse};
\node[orange!70, font=\small, align=center] at (6.2, -0.35)
    {$T_{G_2}$};
\draw[orange!50, rounded corners, dashed]
    (4.5, -2.8) rectangle (7.9, 3.2);
\node[orange!90, font=\small, align=center] at (6.2, 3.5)
    {\textbf{Stage 2}};
\node[orange!80, font=\footnotesize, align=center] at (6.2, -1.3)
    {Gate operation};
\node[orange!70, font=\footnotesize, align=center] at (6.2, -1.75)
    {};
\begin{scope}[shift={(6.2, 5.6)}, scale=0.88]
    \draw[gray!50] (0,0) circle (1);
    \draw[gray!40, dashed] (1,0) arc (0:180:1 and 0.3);
    \draw[gray!60] (-1,0) arc (180:360:1 and 0.3);
    \draw[->, gray!60, thin] (0,-1.2) -- (0,1.3);
    \filldraw[red!70] (0,1) circle (2.5pt);
    \draw[->, orange!80, very thick] 
        (0,1) arc (90:5:0.4 and 1);
    \node[orange!80, font=\small] at (1.0, 0.5) 
        {$R_x$};
\end{scope}
\draw[green!60!black, thick, fill=green!10]
    (9.5, 0) -- (9.5, 0)
    .. controls (9.8, 1.2) and (10.2, 1.2) ..
    (10.5, 1.4)
    .. controls (10.8, 1.2) and (11.2, 1.2) ..
    (11.5, 0) -- cycle;
\node[green!60!black, font=\footnotesize, align=center] 
    at (10.5, 1.7) {$\pi/2$ pulse};
\draw[purple!70, thick, fill=purple!10]
    (11.8, 0) -- (11.8, 0)
    .. controls (11.9, 0.8) and (12.1, 0.8) ..
    (12.2, 1.0)
    .. controls (12.3, 0.8) and (12.5, 0.8) ..
    (12.6, 0) -- cycle;
\node[purple!70, font=\footnotesize, align=center] 
    at (12.2, 1.35) {XUV};
\draw[green!40!black, rounded corners, dashed]
    (9.0, -2.8) rectangle (13.2, 2.2);
\node[green!60!black, font=\small, align=center] 
    at (11.1, 2.55) {\textbf{Stage 3}};
\node[green!60!black, font=\footnotesize, align=center] 
    at (11.1, -1.3) {Readout};
\node[green!50!black, font=\footnotesize, align=center] 
    at (11.1, -2.0) {PMD $\to$ $\ket{L}$/$\ket{R}$};
\begin{scope}[shift={(11.1, 4.4)}, scale=0.88]
    \draw[gray!50] (0,0) circle (1);
    \draw[gray!40, dashed] (1,0) arc (0:180:1 and 0.3);
    \draw[gray!60] (-1,0) arc (180:360:1 and 0.3);
    \draw[->, gray!60, thin] (0,-1.2) -- (0,1.3);
    \draw[->, gray!60, thin] (-1.2,0) -- (1.3,0);
    \filldraw[green!60!black] (1,0) circle (2.5pt);
    \node[green!60!black, font=\small] at (1.55, 0.15) 
        {$\ket{R}$};
    \filldraw[green!40!black] (-1,0) circle (2.5pt);
    \node[green!40!black, font=\small] at (-1.6, 0.15) 
        {$\ket{L}$};
\end{scope}
\draw[->, thick, gray!60] (3.5, 0.8) -- (4.4, 0.8);
\draw[->, thick, gray!60] (8.0, 0.8) -- (8.9, 0.8);
\node[gray!60, font=\tiny, align=center] at (2.0, -2.35)
    {};
\node[gray!60, font=\tiny, align=center] at (6.2, -2.35)
    {};
\node[gray!60, font=\tiny, align=center] at (11.1, -2.35)
    {};
\end{tikzpicture}
\caption{Schematic pulse sequence for KH qubit operation, 
shown with the three stages applied while the strong 
driving field remains active, preserving the KH 
double-well basis throughout. 
\textbf{Stage 1 (initialisation):} at a time 
$T_{G_1}$ after the onset of stabilisation, a $\pi/2$ 
pulse with carrier phase $\varphi = \pi$ rotates the 
equatorial initial state $\ket{\Psi_{\mathrm{eq}}}$ 
toward the eigenstate pole $\ket{\phi^{KH}_0}$, 
demonstrated in the ideal two-level case 
(Section~\ref{sec:ideal_comparison}). 
\textbf{Stage 2 (gate):} a second resonant pulse at 
$T_{G_2}$ drives the required gate operation. 
\textbf{Stage 3 (readout):} a further $\pi/2$ pulse 
maps the qubit state onto the spatially localised 
well states $\ket{L}$ or $\ket{R}$, converting the 
phase information generated by the gate into a 
directly observable population imbalance. This 
right-well localisation locking is already observed 
in the full TDSE $S$ gate results 
(Section~\ref{sec:halfpi_results}) and may be read 
out by time-delayed XUV photoelectron 
momentum-distribution imaging~\cite{Morales2011, 
He2020, Ivanov2022}.}
\label{fig:pulse_sequence}
\end{figure}
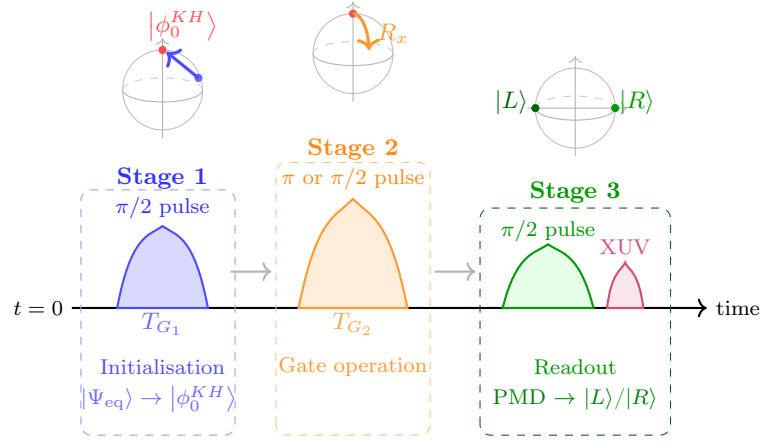

The full KH-qubit protocol is illustrated in 
Fig.~\ref{fig:pulse_sequence}, including initial-state preparation, and applies a a sequence of resonant pulses at frequency $\omega_{10}$.  The proposed KH-qubit protocol consists of three consecutive stages while the strong driving field remains active, thereby preserving the KH double-well basis throughout. First, the qubit is initialised at a time $T_{G_1}$ after the onset of stabilization, by applying a $\pi/2$ pulse with carrier phase $\phi=\pi$, which rotates the initial equatorial state towards the KH eigenstate $\lvert\phi^{KH}_0\rangle$. A second resonant pulse, applied at a later time $T_{G_2}$ (such as those chosen in the TDSE computations), performs the desired gate operation. Finally, a further $\pi/2$ pulse maps the resulting qubit state onto the spatially localised left- and right-well states, $\lvert L\rangle$ and $\lvert R\rangle$. This converts the phase information generated by the gate into a directly observable population imbalance, which may be measured, e.g., using time-delayed XUV photoelectron momentum-distribution imaging~\cite{Morales2011,He2020,Ivanov2022}.

\begin{figure}[H]
    \centering
    \includegraphics[width=1\linewidth]
    {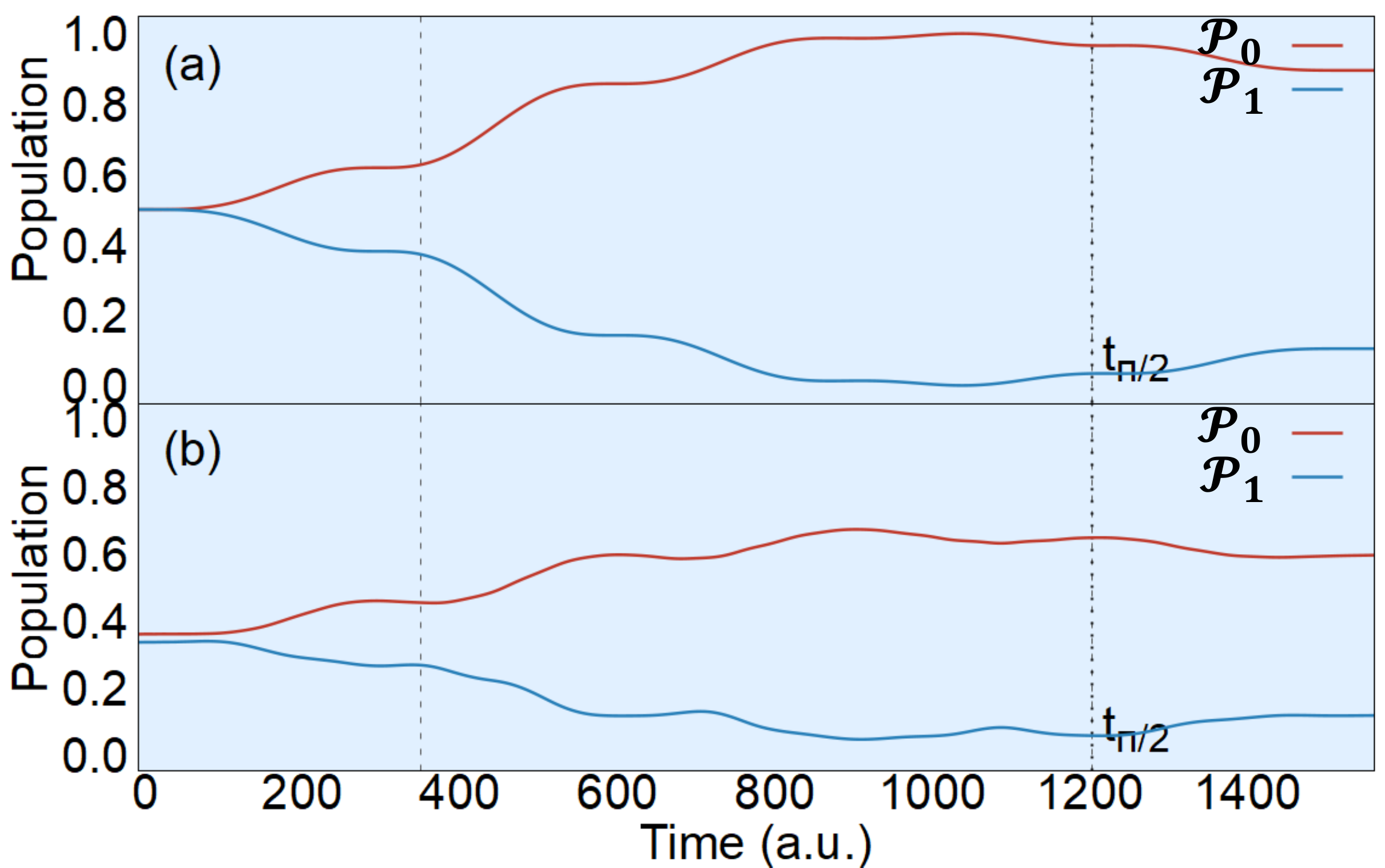}
    \caption{Ideal initialisation test --- $\pi/2$ pulse with 
carrier phase $\varphi = \pi$ at 
$\varepsilon_{0w} = 0.0003$ a.u.\ in the 
time-averaged two-level limit. The signature of 
successful initialisation is the transfer of 
population to a single eigenstate pole, 
$\mathcal{P}_0 \to 1$ and $\mathcal{P}_1 \to 0$, 
corresponding to an $R_x(\pi/2)$ rotation of the 
equatorial state up to $\ket{\phi^{KH}_0}$. 
\textbf{(a)} Perfect coherent superposition initial 
condition $(\ket{\phi^{KH}_0} + 
\ket{\phi^{KH}_1})/\sqrt{2}$: the rotation drives 
$\mathcal{P}_0 \to 1$, $\mathcal{P}_1 \to 0$ 
post-pulse (red and blue), demonstrating clean 
eigenstate initialisation in the two-level limit. 
\textbf{(b)} Hybrid TDSE initial state at 
$T_{G} = 500$ a.u., carrying $\mathcal{P}_{2+}$ 
contamination and unequal $\mathcal{P}_0/\mathcal{P}_1$ 
populations: $\mathcal{P}_0$ increases and 
$\mathcal{P}_1$ decreases but neither reaches its 
target value, reflecting the impurity of the initial 
state rather than a breakdown of the rotation 
mechanism. In both panels the blue shaded background indicates that the weak pulse with a full trapezoidal envelope (ramp up, flat top, ramp down) is active from $t = 0$ 
to the end of the pulse; the dashed vertical line 
marks the end of the ramp-up (at $\tau_2$) and the 
dotted vertical line marks the completion time 
$t_{\pi/2} = \tau_2 + \pi/(2\Omega_R)$.}
    \label{fig:ideal_initialisation}
\end{figure}

This state initialization is depicted in Fig.~\ref{fig:ideal_initialisation}, which shows the population 
dynamics for a $\pi/2$ pulse with carrier phase 
$\varphi = \pi$ under two initial conditions. With a 
perfect coherent superposition 
$(\ket{\phi^{KH}_0} + \ket{\phi^{KH}_1})/\sqrt{2}$ as the initial 
state [Fig.~\ref{fig:ideal_initialisation}(a)], the $R_x(\pi/2)$ rotation drives $\mathcal{P}_0 \to 1$, 
$\mathcal{P}_1 \to 0$ post-pulse, demonstrating clean eigenstate 
initialisation in the two-level limit. With the approximate  
TDSE initial state at $T_{G} = 500$ a.u., which carries 
$\mathcal{P}_{2+}$ contamination and unequal $\mathcal{P}_0/\mathcal{P}_1$ populations, 
the initialisation is partial: $\mathcal{P}_0$ increases and $\mathcal{P}_1$ 
decreases but neither reaches its target value, reflecting 
the impurity of the initial state rather than a breakdown 
of the rotation mechanism. Note that $T_{G} = 500$ a.u.\ is used here because the initialisation test requires 
only a hybrid initial state while exhibiting less 
$\mathcal{P}_{2+}$ leakage over the gate duration of interest than $T_{G} = 800$ a.u.\ 
[see Fig.~\ref{fig:resultleakagepiover2}]. 


This comparison establishes two points: eigenstate 
initialisation is achievable in principle within the 
two-level description, and the deviation in the hybrid 
case arises from initial state impurity rather than a 
breakdown of the $R_x(\pi/2)$ rotation mechanism itself. 
Initialisation is therefore an engineering constraint, 
suppressible through leakage reduction via higher frequency 
operation or pulse shaping, rather than a fundamental 
barrier to the KH qubit.

\subsubsection{Complete single-qubit gate set — ideal case}

\begin{figure}[H]
    \centering
    \includegraphics[width=1\linewidth]
    {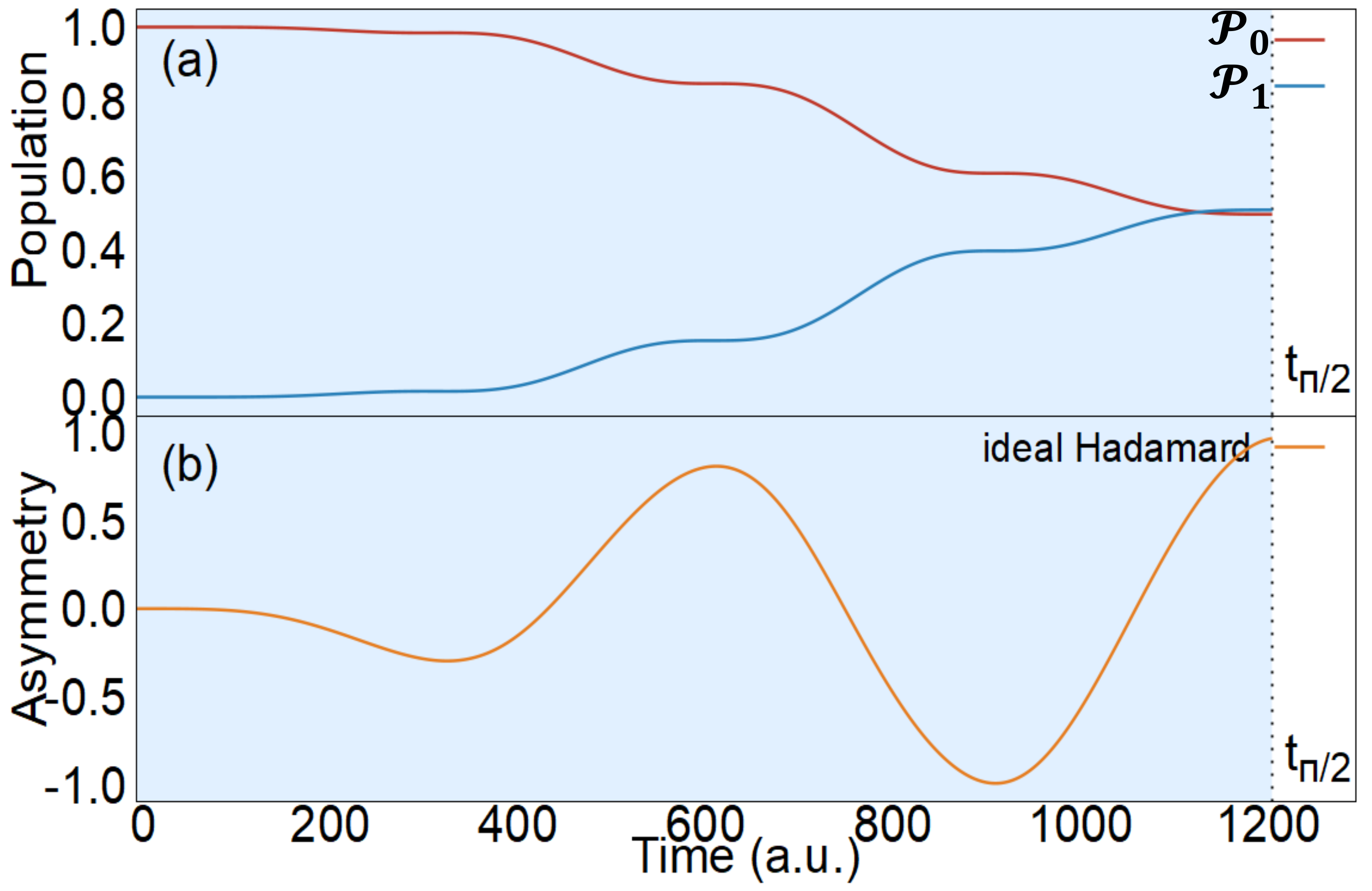}
    \caption{Ideal Hadamard gate --- $\pi/2$ pulse, carrier phase 
$\varphi = \pi$, KH ground state $\ket{\psi_0}$ 
initial condition, $\varepsilon_{0w} = 0.0003$ a.u.\ 
in the time-averaged two-level limit, using the 
shortened flat-top construction so that the full 
trapezoidal envelope delivers the target $\pi/2$ 
rotation at the marked completion time. 
\textbf{(a)} Populations: $\mathcal{P}_0$ (red) 
decreases from 1 toward 0.5 and $\mathcal{P}_1$ 
(blue) increases from 0 toward 0.5, reaching equal 
superposition at $t_{\pi/2}$. 
\textbf{(b)} Asymmetry: the oscillation builds from 
zero during the pulse and reaches full amplitude at 
$\omega_{10}$ after $t_{\pi/2}$, confirming coherent 
superposition creation rather than an incoherent 
mixture. The blue shaded background indicates that the weak pulse with a full trapezoidal envelope (ramp up, flat top, ramp down) is active up to $t_{\pi/2}$; the dotted 
vertical line marks the completion time $t_{\pi/2}$.}
    \label{fig:ideal_Hadamard}
\end{figure}

Beyond the $Z$ and $X$ gates demonstrated above, the 
ideal two-level system supports the complete set of 
standard single-qubit gates. The Hadamard gate is 
realised by applying a $\pi/2$ pulse to the KH ground 
state $\ket{\psi_0}$ initial condition with carrier 
phase $\varphi = \pi$; the $R_x(\pi/2)$ rotation maps 
the north pole to the equator, producing an equal 
superposition $\mathcal{P}_0 = \mathcal{P}_1 = 0.5$ 
post-pulse. The asymmetry builds from zero during the 
pulse as the coherent superposition develops and 
reaches full amplitude oscillation at $\omega_{10}$ 
after $t_{\pi/2}$, confirming coherent superposition 
creation rather than an incoherent mixture, as shown 
in Fig.~\ref{fig:ideal_Hadamard}. Here the shortened 
flat-top construction is used (see 
Sec.~\ref{sec:gates}), so that the full trapezoidal 
envelope delivers exactly the target $\pi/2$ rotation 
at the marked completion time $t_{\pi/2}$, achieving 
maximal gate fidelity at that point.

\begin{figure}[H]
    \centering
    \includegraphics[width=1\linewidth]
    {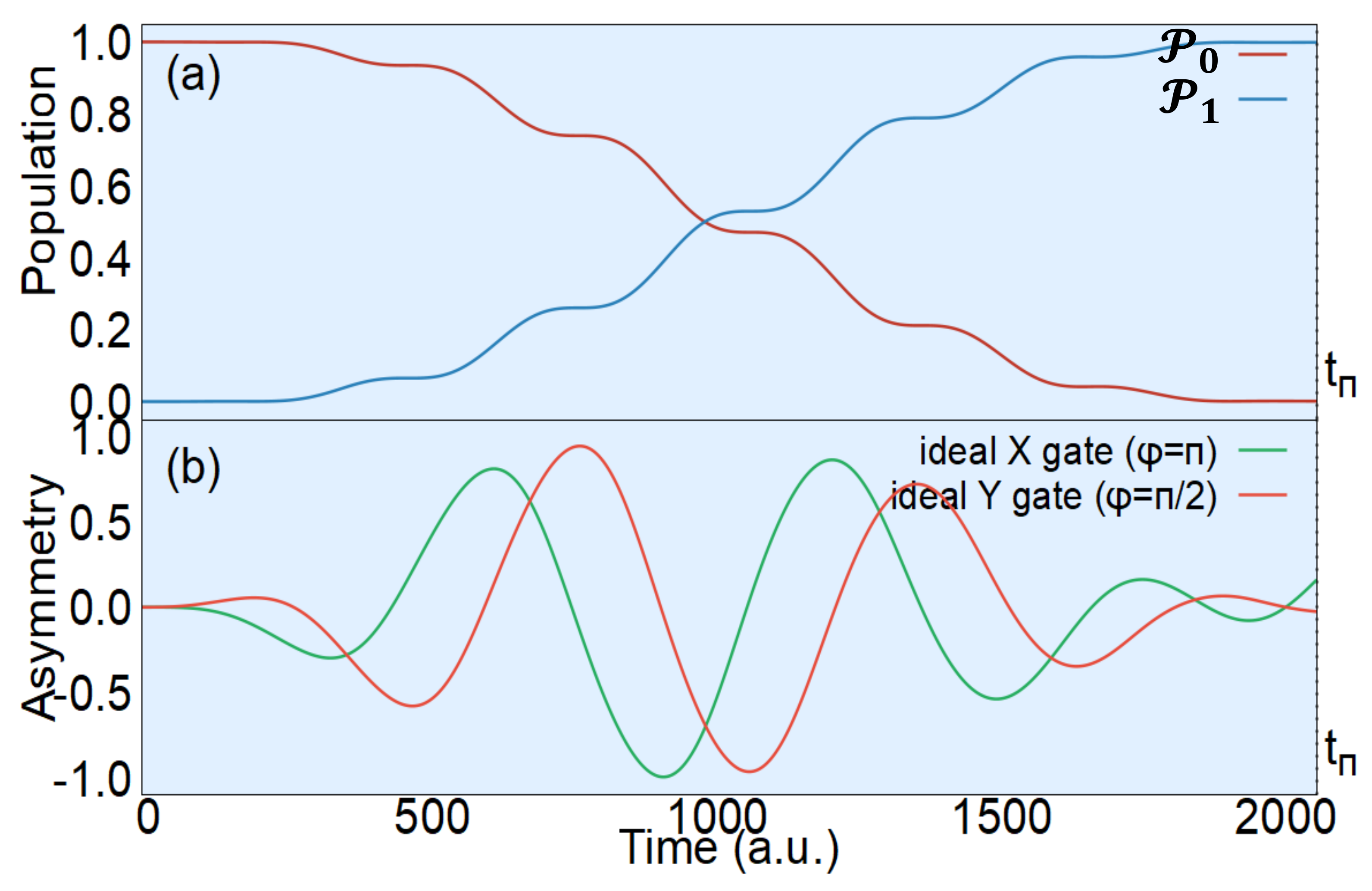}
    \caption{Ideal $Y$ gate --- $\pi$ pulse, carrier phase 
$\varphi = \pi/2$, KH ground state $\ket{\psi_0}$ 
initial condition, $\varepsilon_{0w} = 0.0003$ a.u.\ 
in the time-averaged two-level limit, using the 
shortened flat-top construction. 
\textbf{(a)} Populations: identical to the $X$ gate 
with $\mathcal{P}_0 \to 0$ (red), $\mathcal{P}_1 \to 
1$ (blue), confirming population inversion 
independent of rotation axis. 
\textbf{(b)} Asymmetry: the $Y$ gate (red) is shifted 
by $87.4^\circ$ relative to the $X$ gate (green), 
consistent with the expected $\pi/2$ phase shift from 
the carrier phase difference $\Delta\varphi = \pi/2$, 
confirming an $R_y$ rather than $R_x$ rotation. The 
blue shaded background indicates that the weak pulse with a full trapezoidal envelope (ramp up, flat top, ramp down) is 
active up to $t_\pi$; the dotted vertical line marks 
the completion time $t_\pi$.}
    \label{fig:ideal_Ygate}
\end{figure}

The $Y$ gate is realised by a $\pi$ pulse with carrier 
phase $\varphi = \pi/2$, rotating the north pole about 
the $y$-axis of the Bloch sphere. The population 
signature is identical to the $X$ gate 
($\mathcal{P}_0 \to 0$, $\mathcal{P}_1 \to 1$) but with 
a characteristic $\pi/2$ phase shift in the asymmetry 
relative to the $X$ gate, as shown in 
Fig.~\ref{fig:ideal_Ygate}. The phase shift is 
confirmed quantitatively by sinusoidal fitting of both 
asymmetry curves in the flat-top region, giving a 
phase difference of $87.4^\circ$ relative to the $X$ 
gate, consistent with the expected $90^\circ$ shift 
arising from the carrier phase difference 
$\Delta\varphi = \pi/2$, confirming an $R_y$ rather 
than $R_x$ rotation.

\subsubsection{Composite phase gates}

\begin{widetext}
\begin{figure}[H]
 \centering
   \includegraphics[width=0.9\textwidth]{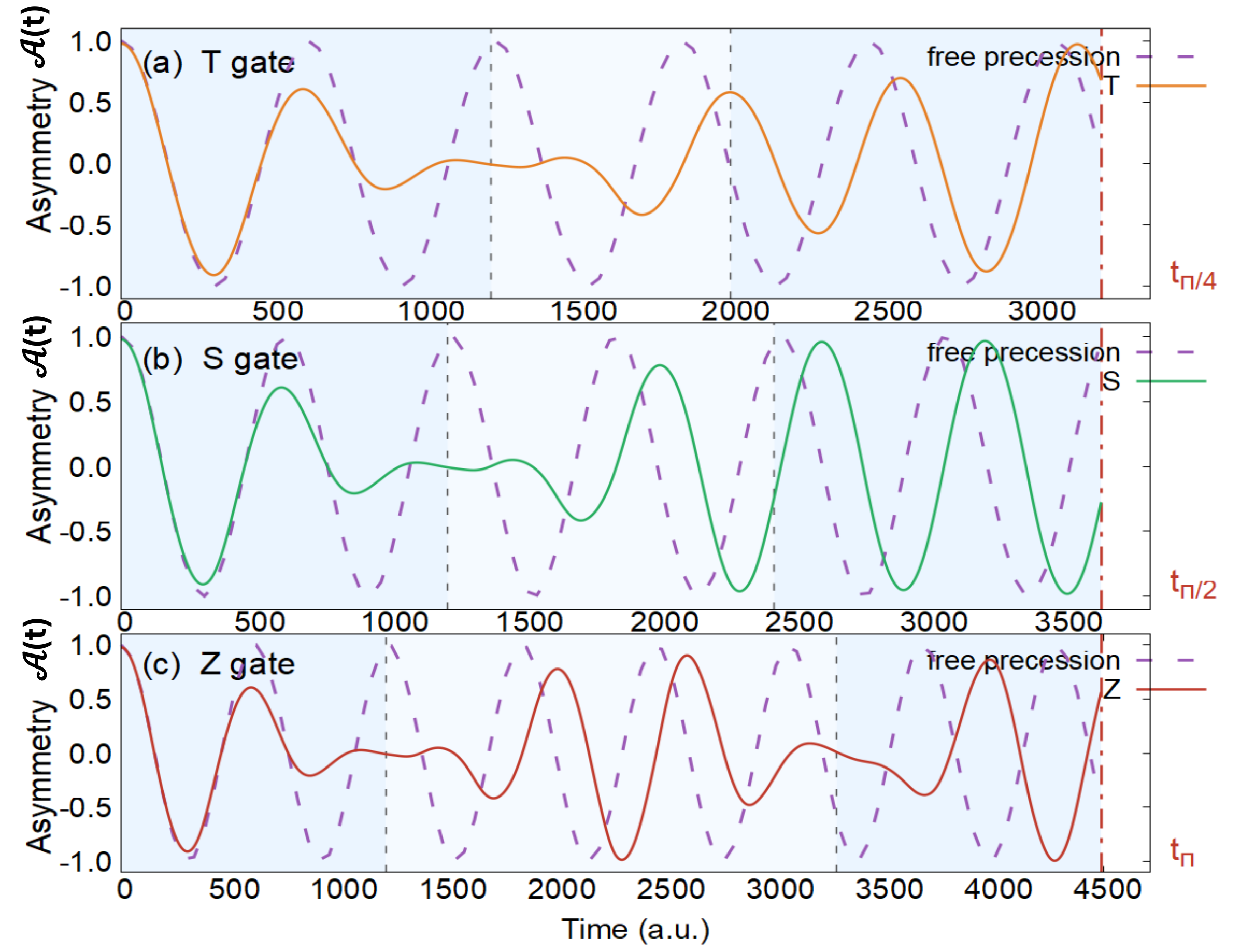}
    \begin{minipage}{0.95\textwidth}
    \caption{Composite $R_z(\theta)$ sequences in the 
time-averaged two-level limit, 
$\varepsilon_{0w} = 0.0003$ a.u., starting from the 
coherent superposition $(\ket{\phi^{KH}_0} + 
\ket{\phi^{KH}_1})/\sqrt{2}$. 
\textbf{(a)} $T$ gate ($\theta = \pi/4$), 
\textbf{(b)} $S$ gate ($\theta = \pi/2$), and 
\textbf{(c)} $Z$ gate ($\theta = \pi$), each realised 
as a three-pulse sequence $R_z(\theta) = 
R_x(\pi/2)\,R_y(\theta)\,R_x(-\pi/2)$ sharing 
identical outer $R_x(\pm\pi/2)$ pulses and differing 
only in the middle pulse's area. In each panel the 
gate asymmetry (solid) is compared against the 
field-free precession reference (purple, dashed), 
which oscillates at $\omega_{10}$ for the same 
initial state~\cite{Aynul2025}. The internal dashed 
vertical lines mark the pulse boundaries within the 
sequence, and the dash-dotted vertical line marks the 
sequence completion time ($t_{\pi/4}$, $t_{\pi/2}$, 
$t_\pi$). All three curves track the free-precession 
reference through the shared first pulse and 
increasingly lag it as the sequence progresses, the 
accumulated phase shift at completion growing with 
$\theta$. Fitting the asymmetry over a window ending 
at completion --- $[2600, 3207]$ a.u.\ for (a), 
$[3000, 3633]$ a.u.\ for (b), and $[3900, 4486]$ 
a.u.\ for (c) --- gives phase shifts $\delta_T = 
0.257\pi$, $\delta_S = 0.469\pi$, and $\delta_Z = 
0.973\pi$, close to their targets $\pi/4$, $\pi/2$, 
$\pi$ (fidelities $F_T = 0.972$, $F_S = 0.938$, 
$F_Z = 0.973$), with the $2\!:\!1$ ratio 
$\delta_S/\delta_T = 1.83$ confirming that a single 
composite skeleton reproduces all three phase gates 
as genuine $z$-axis rotations. The blue shaded background indicates the weak pulses 
are active throughout the sequence, with the lighter 
central region marking the middle $R_y(\theta)$ pulse 
and the outer regions the $R_x(\pm\pi/2)$ pulses.}
    \label{fig:ideal_STZ}
    \end{minipage}
\end{figure}
\end{widetext}

As discussed in Section~\ref{sec:gates}, the $S$, $T$, 
and $Z$ gates are additionally demonstrated here as 
composite three-pulse sequences, $R_z(\theta) = 
R_x(\pi/2)\,R_y(\theta)\,R_x(-\pi/2)$, starting from a localized 
coherent superposition $(\ket{\phi^{KH}_0}+\ket{\phi^{KH}_1})/
\sqrt{2}$ and using the shortened-flat envelope so that 
the target angle $\theta$ is delivered exactly at the 
marked completion time of each pulse and fidelity is maximized. Unlike the 
single-pulse $Z$ gate above, this construction realises 
a genuine $R_z(\theta)$ rotation within the ideal 
two-level model, built entirely from the $R_x$ and $R_y$ 
rotations directly accessible on resonance. Fig.~\ref{fig:ideal_STZ} shows the resulting asymmetry 
for $\theta=\pi/4$ ($T$), $\pi/2$ ($S$), and $\pi$ 
($Z$), each built from identical outer pulses with only 
the middle pulse's area differing. For reference, we plot the idealized asymmetry parameter calculated in the absence of the additional pulses (see dashed lines). For an initial coherent superposition of KH eigenstates, this parameter oscillates with the frequency $\omega_{10}$ \cite{Aynul2025}.

All three curves 
track the field free precession reference closely through the 
shared first pulse, then increasingly lag it as the 
sequence progresses, with the accumulated phase shift 
at completion growing with the target angle $\theta$. 
Measured phase shifts $\delta_T = 0.257\pi$, 
$\delta_S = 0.469\pi$, $\delta_Z = 0.973\pi$ 
are close to the target values $\pi/4$, $\pi/2$, $\pi$ 
with fidelities $F_T = 0.972$, $F_S = 0.938$, $F_Z = 0.973$, 
and the $2\!:\!1$ ratio between $\delta_T$ and $\delta_S$ 
($\delta_S/\delta_T = 1.83$, close to the expected $2.00$) 
confirms that a single composite skeleton reproduces 
all three phase gates as genuine $z$-axis rotations.
Together with the $X$, $Y$ and $H$ gates 
demonstrated above, this establishes all six standard 
single-qubit gates in the ideal two-level limit: $X$ 
($R_x(\pi)$), $Y$ ($R_y(\pi)$), $H$ 
($R_x(\pi/2)$), $Z$ ($R_z(\pi)$), $S$ ($R_z(\pi/2)$), and $T$ 
($R_z(\pi/4)$), the latter three realised via the 
composite sequence above, confirming that the KH qubit 
coupling supports a complete single-qubit gate set. The 
limitation to the full TDSE is leakage rather than the 
coupling mechanism, as established by the like-for-like 
comparison above.

The ideal case results collectively confirm three points: 
the dipole coupling provides an accurate description 
of the gate dynamics before leakage dominates; gate 
identity is determined by initial state as predicted by 
the two-level formalism; and the fidelity limitations of 
the full TDSE arise from leakage rather than a breakdown 
of the coupling mechanism. These results establish a clear 
optimisation target: reducing $P_{2+}$ leakage through 
higher frequency operation or pulse shaping would bring 
the full TDSE results into close agreement with the ideal 
two-level predictions, unlocking both higher gate fidelity 
and eigenstate initialisation.

\subsection{Effect of the field strength $\varepsilon_{0w}$}
\label{sec:E1_results}
The weak field amplitude $\varepsilon_{0w}$ controls the Rabi 
frequency $\Omega_R = \varepsilon_{0w}\mu/\hbar$ and therefore 
the gate time $t_\mathrm{gate} \propto 1/\varepsilon_{0w}$. Thus, a  
smaller $\varepsilon_{0w}$ gives a slower gate, with more 
oscillation cycles elapsing during the pulse and more phase 
decay accumulating from decoherence, while a larger 
$\varepsilon_{0w}$ gives a faster gate but risks disturbing the 
KH potential. Thus, one must assess how the field strength $\varepsilon_{0w}$ affects the proposed gates.

\begin{widetext}
\begin{figure}[H]
\hspace*{1.8cm}      \includegraphics[width=0.7\textwidth]{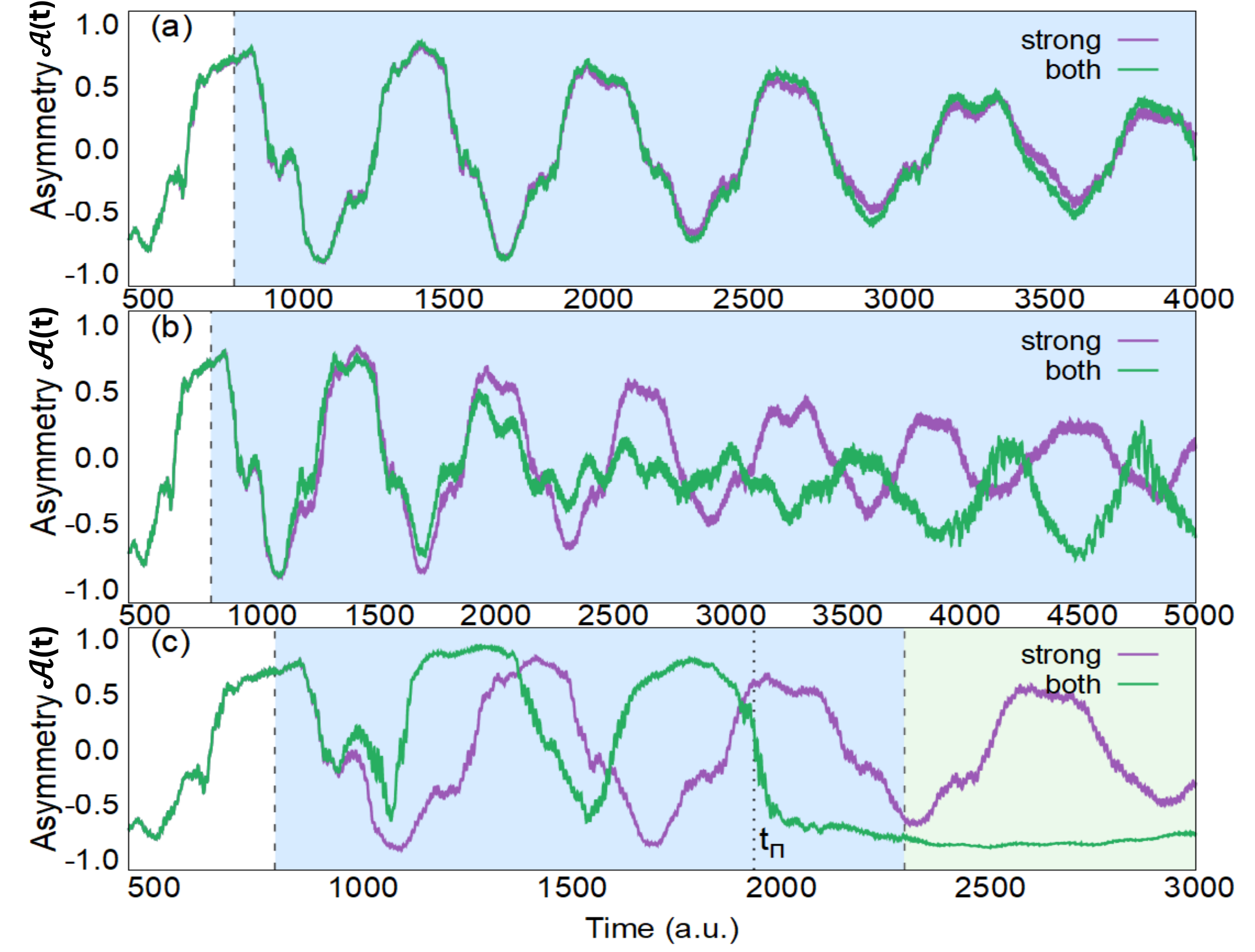}
    \begin{minipage}{0.95\textwidth}
    \caption{Asymmetry $\mathcal{A}(t)$ for the $\pi$ pulse at 
three weak-field amplitudes, with $T_{G} = 800$ a.u.\ 
throughout: \textbf{(a)} $\varepsilon_{0w} = 0.00003$ 
a.u., \textbf{(b)} $\varepsilon_{0w} = 0.0001$ a.u., 
and \textbf{(c)} $\varepsilon_{0w} = 0.00065$ a.u. In 
each panel the both-fields result (green) is compared 
against the strong-field-only reference (purple), and 
the dashed vertical line marks the start of the weak 
pulse. The blue shaded region marks the weak pulse 
window; in (a) and (b) the plotted range lies entirely 
within this window, so no post-pulse region is shown, 
while in (c) the green region marks the post-pulse 
evolution under the strong field alone and the dotted 
vertical line marks the completion time $t_\pi$ (at $T_{G} + 
t_\pi$ on the absolute axis). In 
(a) the both-fields curve overlaps almost exactly with 
the strong-field-only reference, with no measurable 
phase shift -- an effective identity gate, confirming 
that below a threshold amplitude the weak field is too 
weak to drive coherent Rabi oscillations within the 
coherence window. In (b) a clear late-pulse phase 
shift $\delta = -0.912\pi$ ($F_\varphi = 0.912$), 
measured over the window $[5700, 6275]$ a.u., confirms 
$Z$-gate action, with the longer gate time 
($t_\pi = 132$ fs) giving a higher fidelity than at 
$\varepsilon_{0w} = 0.0003$ a.u.\ owing to reduced 
off-resonant coupling. In (c), the fastest gate 
($t_\pi = 28$ fs), the both-fields curve shows a clean 
large-amplitude oscillation diverging strongly from 
the reference during the pulse and locks to a strong 
right-well bias $\mathcal{B} \approx -0.9$ post-pulse.}
    \label{fig:three-panels}
    \end{minipage}
\end{figure}
\end{widetext}

Fig.~\ref{fig:three-panels} shows the asymmetry parameter obtained from the full TDSE computation, for $\pi$ pulses starting at $T_{G} = 800$ a.u, but with three different field strengths. At $\varepsilon_{0w} = 0.00003$ a.u.\ [Fig.~\ref{fig:three-panels}(a)], the Rabi frequency is so 
small that the gate time far exceeds the coherence window. The 
both-fields asymmetry traces almost exactly the 
strong-field-only reference throughout, with no measurable 
phase shift imprinted and no gate action occurring. This 
is effectively an identity gate, confirming that below a 
threshold $\varepsilon_{0w}$ the weak field is too weak to 
drive coherent Rabi oscillations within the available 
coherence time. This result also serves as a useful negative 
control: the absence of any effect at 
$\varepsilon_{0w} = 0.00003$ a.u.\ confirms that the phase 
shifts observed at higher $\varepsilon_{0w}$ are genuine 
gate-driven effects rather than artefacts of the simulation. 
If the field is stronger ($\varepsilon_{0w} = 0.0001$ a.u.), Fig.~ \ref{fig:three-panels}(b) shows a clear phase 
shift of $\delta = -0.912\pi$ in the late-pulse window. This shows that the $Z$ gate has been initialized, although the gate has increased from $t_\pi = 50$ fs to 
$t_\pi = 132$ fs with regard to that shown in  Fig.~\ref{fig:resultasymmetrypi}, for $\varepsilon_{0w}=0.0003$, which is three times as strong. Despite the longer gate, $F_\varphi = 0.912$ is actually higher than the 
$\varepsilon_{0w} = 0.0003$ value of $0.853$. This is associated with a reduced off-resonant coupling to the higher-lying KH eigenstates. 
At $\varepsilon_{0w} = 0.00065$ a.u.\ [Fig. \ref{fig:three-panels}(c)], the gate time reduces 
to $t_\pi = 1145$ a.u.\ $= 28$ fs, less than two 
inter-well oscillation cycles. This  gives the fastest gate achieved 
in this study, with $N_\mathrm{gates} = T_{\mathcal{A}}/t_\pi = 6$ 
within the coherence window. There is  a clean 
large-amplitude oscillation of the both-fields asymmetry 
during the pulse, diverging strongly from the 
strong-field-only reference. Post-pulse the asymmetry 
locks to a strong right-well bias of 
$\mathcal{B} \approx -0.9$, more pronounced than at 
$\varepsilon_{0w} = 0.0003$.   This bias is due the post-pulse rise of the bound-state populations $\mathcal{P}_{2+}$,  while $\mathcal{P}_0$ and $\mathcal{P}_1$ both 
fall. However, we have verified that there is no sharp increase ionization, so that the KH atom is not destroyed.  For detailed population dynamics, see Appendix \ref{app:populations}. 

Fig.~\ref{fig:three-panels} illustrates a clear 
trade-off: increasing $\varepsilon_{0w}$ shortens the gate 
time and increases $N_\mathrm{gates}$, but also raises the ratio 
$\alpha_{0w}/\alpha_{0s}$ and the disturbance to the KH 
potential.  For weak fields, such as  $\varepsilon_{0w} = 0.0001$ a.u.\, the gate is 
slow but clean, with minimal leakage. As the field increases, the gate becomes faster but the population dynamics becomes more complex. An example is  $\varepsilon_{0w} = 0.00065$ 
a.u.\, for which the gate is fastest and clean just after $t_\pi$, 
but there is significant 
redistribution to higher KH eigenstates post-pulse. In practice, this leakage also limits the number of gates one can execute in practice, although in principle the coherence time $T_{\mathcal{A}}$ would allow for more gate operations. For the idealized, time-averaged scenario we have verified the same pattern, but the absence of leakage would ensure a higher number of gates as $\alpha_{0w}$ increases.

\section{Conclusions}
\label{sec:discussion}

The results presented here demonstrate coherent quantum 
gate operations in a qubit whose computational basis is 
not merely compatible with the strong driving field but 
is created and maintained by it. The gates are built employing coherent superpositions of the two most deeply bound eigenstates of the Kramers-Henneberger (KH) potential, which is a light-induced dichotomous potential created with a strong laser field. The required coherent superpositions are induced by additional, much weaker low-frequency pulses resonant with the energy gap between the KH eigenstates. The remaining weak pulse  parameters are chosen depending on the gate to be induced. For an idealised two-level system built from a time-averaged KH potential, we have demonstrated all six 
standard single-qubit gates (X, Y, Z, H, S, T), while for full time-dependent Schr\"odinger equation (TDSE) computations, we have confirmed the Z, 
S, and identity gates. In the full time-dependent case, the gates generally exhibit phase fidelities $F_\varphi$ over $90\%$, and are thus robust 
against wavepacket excursions to higher KH states and 
fast oscillations at the strong-field frequency $\omega_{s}$.  The strong 
field pre-applies a Hadamard gate, placing the system at 
the equator rather than the pole of the Bloch sphere. This limits the gates that can be built in for the full time dependent case and 
is a consequence of the time evolution of the system. Initially, the model atom is in the field-free ground state, and is subsequently driven by the strong field in a way that, after stabilization sets in, the KH potential is induced. However, eigenstate 
initialisation via a preparatory $R_x(\pi/2)$ pulse 
is demonstrated in the ideal two-level limit 
(Section~\ref{sec:ideal_comparison}), establishing 
this as an engineering constraint rather than a 
fundamental barrier. 
Taken together, these results show that the qubit model 
satisfies most of the DiVincenzo 
criteria~\cite{DiVincenzo2000}: the qubit is well 
characterised, a universal single-qubit gate set has 
been demonstrated, coherence times exceed the gate 
operation time, and initialisation has been shown in the 
ideal two-level limit, achievable in the full TDSE once 
leakage is sufficiently suppressed through finding optimal 
frequency operation or pulse shaping. Potentially, the 
scheme also satisfies the measurement criterion: the 
right-well localisation locking observed post-pulse in 
the full TDSE $S$ gate results provides a direct spatial 
signature of the gate operation, though the proposed 
readout via time-delayed photoelectron momentum 
distribution imaging of the KH 
eigenstates~\cite{Morales2011, Ivanov2022, He2020} has 
not itself been computationally demonstrated. 
Scalability and a two-qubit entangling gate remain open 
questions, identified as key directions for future work.

This light-induced setup represents a 
qualitative departure from conventional qubit 
architectures such as superconducting circuits~\cite{Kjaergaard2020}, trapped 
ions~\cite{Bruzewicz2019}, and spin 
qubits~\cite{Loss1998, Hanson2007}, where strong driving is expected to be 
a source of error that would need to be minimised. In the KH qubit the 
strong field is the computational resource: it engineers 
the double-well potential that defines the qubit basis, 
stabilises the bound population against ionisation 
through KH trapping, and counterintuitively provides 
stronger protection at higher intensity. This may be 
described as \textit{strong-field Hamiltonian 
engineering}: the computational structure itself is 
field-created and exists only in the presence of the 
driving field.  Furthermore, the initial time $T_{G}$ also controls the 
rotation axis on the Bloch sphere and population conservation character, while  the field strength 
$\varepsilon_{0w}$ controls gate time. 
Both are continuously tuneable laser parameters requiring 
no hardware changes.

A further distinction from conventional architectures 
concerns the nature of the dominant error channel. In 
solid-state qubits the primary limitation is stochastic 
environmental decoherence from phonon baths ~\cite{Breuer2002, Borsch2023, Kjaergaard2020}, charge 
noise ~\cite{Paladino2014, Kjaergaard2020}, and magnetic fluctuations ~\cite{Hanson2007, 
Kjaergaard2020}, which is largely 
unstructured and difficult to suppress by control 
engineering alone. In the KH qubit the dominant error 
is population leakage to higher KH eigenstates $\mathcal{P}_{2+}$, 
driven by off-resonant dipole coupling during the additional, gate 
pulse. This leakage is Hamiltonian-derived and therefore 
structured, with a well-defined dependence on the initial pulse time $T_G$, its frequency, strength and shape 
as 
demonstrated throughout the results. Structured errors 
are in principle suppressible through control 
engineering: pulse shaping, optimal $T_{G}$ selection, 
and composite pulse sequences are all viable routes to 
leakage suppression, none of which require changes to 
the physical system. 

Moreover, the strong-field environment provides natural decoherence protection, as KH 
stabilisation suppresses ionisation by trapping the 
electron in the time-averaged double-well potential. Higher field intensity gives deeper trapping and better 
protection — the opposite of conventional qubits, where 
stronger driving increases errors. The qubit is 
stabilised by the field, not despite it — the same field 
that ionises ordinary atoms creates and maintains this 
qubit. There is no conflict between the computational 
resource and the qubit stability. This counterintuitive 
inversion is a unique feature of the strong-field 
architecture.

One should also note that the gates reported in this work are around tens of femtoseconds ($\tau_\mathrm{gate} \sim 10^{-14}$ s) 
, i.e.,  six orders of magnitude 
faster than physically driven superconducting qubit gates ($\sim $ ns) \cite{Werninghaus2021}\footnote{In superconducting processors, Z and S gates are virtual Z rotations, implemented by updating the phase reference of subsequent microwave pulses. Therefore, they have effectively no physical duration. The microwave phase reference is updated in software rather than applying a physical pulse. Such a gate is described as having effectively zero physical duration. However, our KH Z and S gates are conceptually different: the relative phase is produced by actual driven evolution, with accompanying population dynamics and possible leakage. Therefore, we compare with finite-duration, resonantly driven $\pi/2$ rotations, $\pm X/2$ and $\pm Y/2$, in a superconducting transmon, which are used to construct single-qubit Clifford gates \cite{Werninghaus2021}. The shortest optimised control pulse therein has a duration of $4.16~\mathrm{ns}$. Although these rotations differ from the phase gates considered here, they provide a more meaningful comparison because both operations arise from actual driven quantum evolution. }. 
The femtosecond gate regime 
places this work 
in a fundamentally different computational regime from 
all existing qubit platforms. At these timescales, many 
environmental decoherence channels including phonon 
scattering, charge rearrangement, and nuclear spin 
dynamics are simply too slow to act during a single 
gate operation. Thus, the gates may in principle operate at room temperature, without the need to resort to cryogenics. Even if the absolute coherence time 
$T_{\mathcal{A}} = 168$ fs were modest by the standards of slower 
qubit platforms, the ratio $T_{\mathcal{A}}/\tau_\mathrm{gate} = 3.4$ for $\varepsilon_{0w} = 0.0003$ a.u.
establishes proof-of-concept viability, and at 
$\varepsilon_{0w} = 0.00065$ a.u.\ this rises to $6$. 
The upper bound on useful gate operations is ultimately 
set by the KH stabilisation lifetime; whether 
$T_{\mathcal{P}_{2+}} = 373$ fs represents a fundamental limit 
of the KH trapping mechanism or an engineering 
constraint reducible through laser parameter 
optimisation remains an open question and a key 
direction for future work. Preliminary estimates 
suggest that maintaining stabilisation on picosecond 
timescales is not a fundamental barrier, with the 
primary experimental challenge being the pulse 
turn-on transient rather than the stabilised dynamics 
thereafter; this remains an open question for future 
experimental investigation. This has significant 
implications for the gate budget: if stabilisation 
persists on picosecond timescales while gate times 
are reduced toward the femtosecond and ultimately 
attosecond regime through higher laser frequency or 
Rydberg parameter optimisation, the ratio 
$N_\mathrm{gates} = T_{\mathcal{A}}/\tau_\mathrm{gate}$ could 
increase by several orders of magnitude from the 
proof-of-concept value of $3.4$ demonstrated here, 
placing the KH qubit on good footing with 
established platforms \cite{Richter_private}.

Nonetheless, the proposed scheme exhibits some limitations. First, the qubit subspace is small. The bound-state populations  of this subspace constitute $\mathcal{P}_0+\mathcal{P}_1 \approx 
0.59$ of the surviving bound-state population, which itself 
is only $\sim$10--15\% of the original population. 
Still, the small qubit subspace fraction ($\mathcal{P}_0+\mathcal{P}_1 \approx 
0.59$) and leakage population are significant concerns 
but do not invalidate the proof-of-concept demonstration. 
Historical precedent shows that early qubit 
implementations with small usable fractions can yield 
meaningful gate operations and establish platforms for 
future optimisation.
Early NV centre experiments~\cite{Jelezko2004} could 
only use a small fraction of NV centres, as most lacked 
the right orientation, strain, or photonic coupling, 
yet gate fidelities of $\sim$75\% on the usable subset 
were strong early results that established the platform. 
Similarly, early cavity QED experiments~\cite{Raimond2001} 
had atom-cavity coupling of only a few percent of the 
cavity decay rate, with effective qubit participation 
well below theoretical maxima. In both cases the key 
metric was the coherence and fidelity of operations 
on the accessible subspace, precisely what the present 
work addresses through $F_\varphi$. Additionally, experimental work demonstrates 
that such surviving stabilised populations are 
detectable~\cite{vanDruten1997}. 

An important open question is related to more 
realistic models. The present work is meant as a proof-of concept rather than a concrete road map towards experimental realization. Nonetheless, the qubit framework and gate scheme presented here 
are general and apply to any parameter regime or atomic model 
supporting KH stabilisation. The primary parameter regime explored here uses 
$\varepsilon_{0s}= 5$ a.u.\ and $\omega_s = 0.7$ a.u., where the 
intensity $I \approx 8.8 \times 10^{17}$ W/cm$^2$ 
approaches relativistic scales. While preliminary 
calculations at $\varepsilon_{0s} = 8$ a.u., $\omega_s = 2.0$ a.u.\ 
show an improved ionisation plateau of $\sim$31\% 
compared to $\sim$87\% at primary parameters, 
further increases in intensity become experimentally 
and theoretically challenging as relativistic 
corrections grow. Relativistic 
corrections
have not been considered here, consistent with 
standard practice in theoretical KH stabilisation 
studies at comparable intensities~\cite{Gavrila2002}. 

Potentially, an  experimentally accessible route is through 
Rydberg excited states, where the reduced binding 
energy allows KH stabilisation at intensities of 
$\sim 10^{13}$ W/cm$^2$~\cite{Eichmann2014}, 
several orders of magnitude more experimentally 
accessible. Population loading into the KH 
eigenstates is also significantly more efficient 
in the Rydberg regime. The interference 
stabilisation mechanism suppresses ionisation 
during the pulse turn-on, allowing the atom to 
enter the KH stabilised regime with minimal 
population loss~\cite{Fedorov1988, Fedorov1989}. 
The larger Rydberg transition dipole moments 
($\mu \sim n^2 a_0$) additionally increase the 
Rabi frequency $\Omega_R = \varepsilon_{0w}\mu/\hbar$ 
at fixed $\varepsilon_{0w}$, opening a direct pathway 
toward attosecond-scale gate operations. Higher laser frequency also preserves 
the KH stabilisation mechanism while reducing 
wavepacket excursions to higher KH states, 
consistent with the expectation that deeper into 
the high-frequency KH regime ($\omega \gg |E_g|$) 
the time-averaged double-well is better defined. Although experiments revealing the existence of the KH atom remain challenging, evidence of stabilisation has been reported in both ground-state and Rydberg atoms 
surviving intense laser fields~\cite{vanDruten1997, 
Eichmann2009, Eichmann2014}, establishing the physical 
basis on which the qubit platform rests.

\vspace*{0.5cm}
\noindent\textbf{Acknowledgements:} We are greatly indebted to Maria Richter and Gopal Dixit for useful discussions. This research has been funded by the UK Engineering and Physical Sciences Research Council (EPSRC) (grant No.EP/T019530/1), by UCL and by the Institute of Physics (IoP) Bell Burnell fund. The authors acknowledge the use of the UCL Myriad High Performance Computing Facility (Myriad@UCL), and associated support services, in the completion of this work.



\appendix
\section*{Appendix}

\section{Stability of the Kramers Henneberger potential}
\label{app:stability}

Because the KH potential is light induced, a key question is how a second wave affects its dynamics and the parameter ranges that can be applied. Indicators that the wave packet is trapped in the KH potential are: (a) its width, which should remain bounded with time; (b) the electron displacement, which should be oscillatory and not increase in amplitude; (c) the time-dependent wave packet, which should exhibit a dichotomous structure. Further evidence that the atom remains in the stabilization regime are the populations of the KH eigenstates, but these will be discussed in the main body of the paper as they will be manipulated with specific purposes. These observables have been employed in our previous publication \cite{Aynul2025} and in the seminal paper \cite{Gavrila2002}.

The probability density is defined as  
\begin{equation}
    \biggl| \psi_{KH}(x,t)\biggr|^2=\biggl| \psi_{L}(x+\alpha(t),t)\biggr|^2,
\end{equation}
where the subscripts $L$ and $KH$ indicate the length gauge and the Kramers-Henneberger frame, respectively, and the transformation causes a shift in the argument. When integrated over space this leads to a gauge-invariant unitary probability, as expected. The width is defined as twice the standard deviation $\sigma_x$ of the distribution of points within this spatial region. We use a spatial filter and consider that for $-60 \hspace{0.1cm}\mathrm{a.u.}\leq x\leq 60 \hspace*{0.1cm}\mathrm{a.u.}$ the wave packet is trapped, and compute the width in this region. These observables were calculated using the TDSE. 

\begin{figure}[H]
    \centering
    \includegraphics[width=1\linewidth]{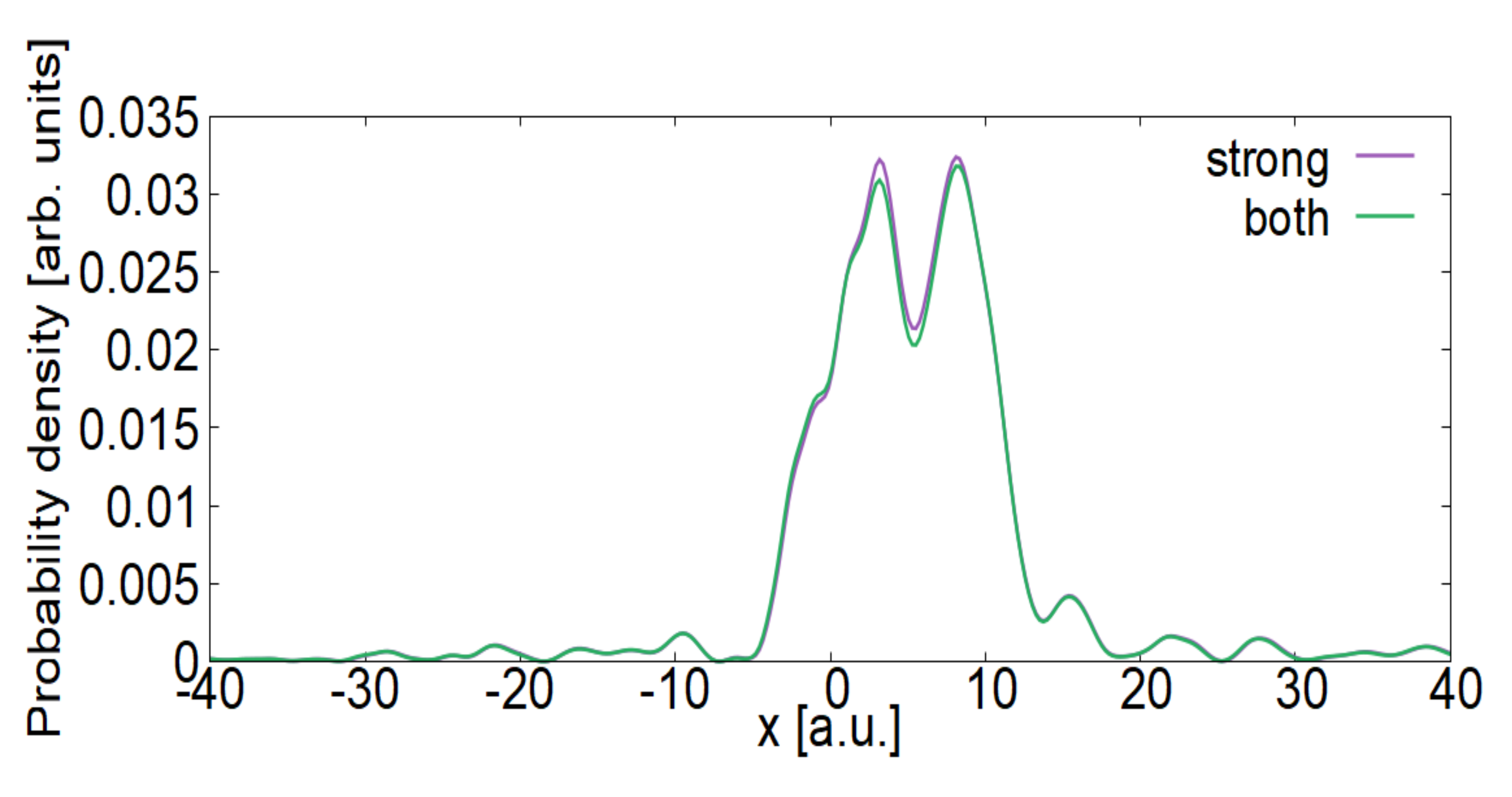}
    \caption{Probability density $|\psi_{KH}(x,t)|^2$ in the Kramers-Henneberger frame at $t = 513$ a.u., slightly after the onset of stabilisation, for the strong-field-only case (purple) and the bichromatic field with an additional $\pi$ pulse of amplitude $\varepsilon_{0w} = 0.0003$ a.u.\ starting at $T_{G} = 500$ a.u.\ with total pulse duration $2422$ a.u.\ (ramp-up $\tau_2 = 358$ a.u.), shown in green. The clear dichotomous (double-peaked) structure is evidence that the KH atom has formed, with the electron density localised in the two wells of the light-induced double-well potential. The near-perfect overlap of the two curves shows that the weak $\pi$ pulse practically does not alter the KH potential or its bound density, confirming that the qubit basis is preserved during gate operation.}
    \label{fig:2appendix}
\end{figure}

Fig.~\ref{fig:2appendix} shows the probability density in the Kramers-Henneberger frame at $t=513$ a.u., which is slightly after the onset of stabilization, considering the strong-field only and a bichromatic field with an additional  $\pi$ pulse of strength $\varepsilon_{0w}=0.0003$ a.u. starting at $T_{G}=500$ a.u. The figure exhibits a dichotomous structure, which is evidence that the KH atom has formed and is practically not altered by the second field. 

\begin{figure}[H]
    \centering
    \includegraphics[width=1\linewidth]{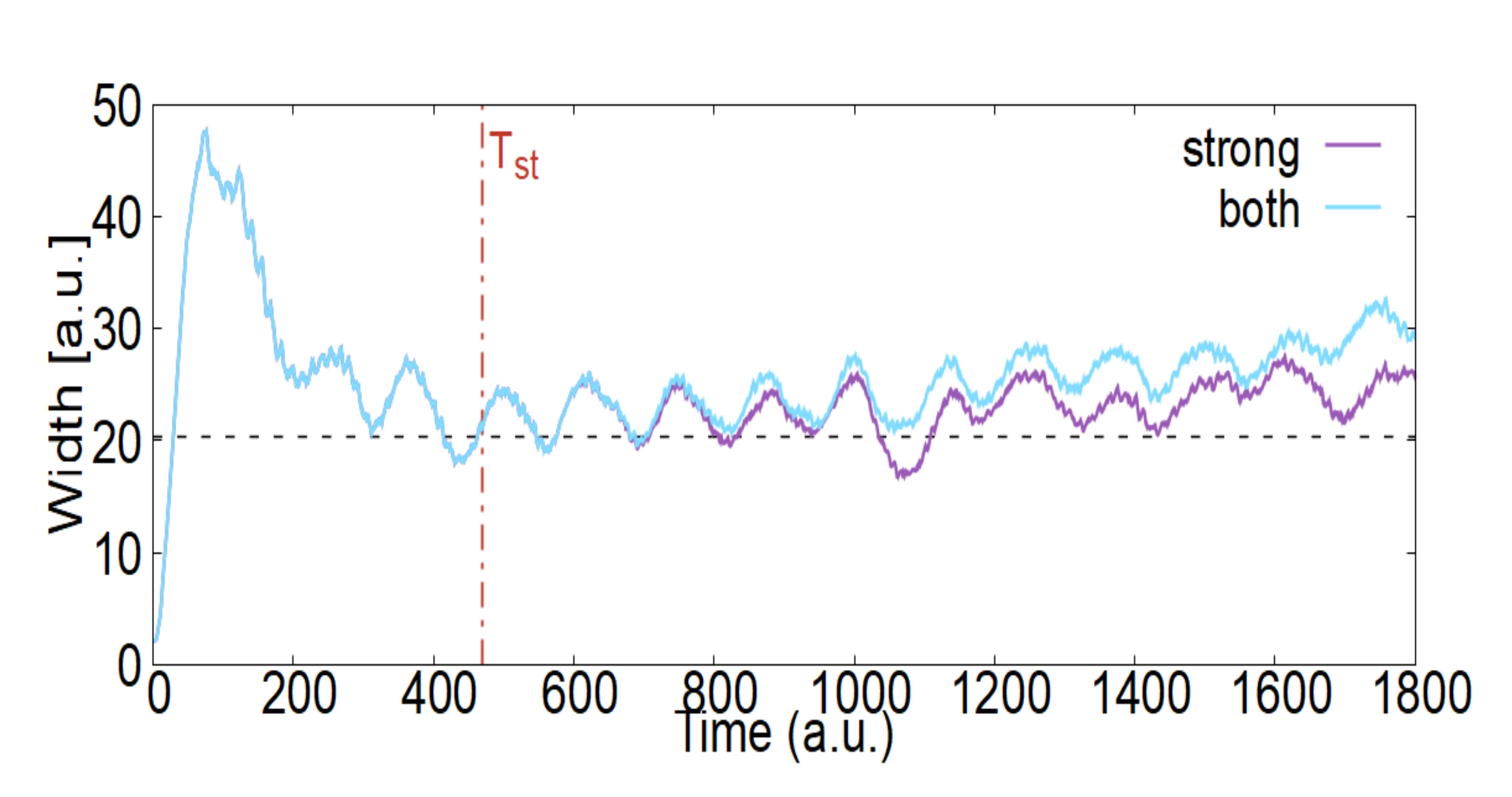}
    \caption{Time-dependent width (twice the standard deviation $\sigma_x$) of the wave packet in the Kramers-Henneberger frame for the strong-field-only case (purple) and the bichromatic field with an additional $\pi$ pulse of amplitude $\varepsilon_{0w} = 0.0003$ a.u.\ starting at $T_{G} = 500$ a.u.\ (cyan), with the same parameters as Fig.~\ref{fig:2appendix}. After an initial transient, the width remains bounded and oscillatory rather than growing with time, confirming that the wave packet stays trapped in the light-induced double-well potential. The red dash-dotted line marks the onset of stabilisation $T_{\mathrm{st}} = 470$ a.u.; any time after this point may be chosen as the stabilisation reference. The weak $\pi$ pulse produces only a small increase in the width relative to the strong-field-only case, confirming that the second field does not significantly disturb the KH atom.}
    \label{fig:1appendix}
\end{figure}

Fig.~\ref{fig:1appendix} confirms this trend over a longer period of time, showing that the width of the time-dependent wave packet is not significantly altered by the weaker field.

\section{Population dynamics for varying $\varepsilon_{0w}$}
\label{app:populations}

In this appendix, we provide details on how the populations $\mathcal{P}_0$, $\mathcal{P}_1$ and $\mathcal{P}_{2+}$ evolve for the gates considered in Sec. \ref{sec:E1_results}, and discuss the influence of the field strength $\varepsilon_{0w}$ on these dynamics. 

\begin{figure}[H]
    \centering
    \includegraphics[width=0.5\textwidth]{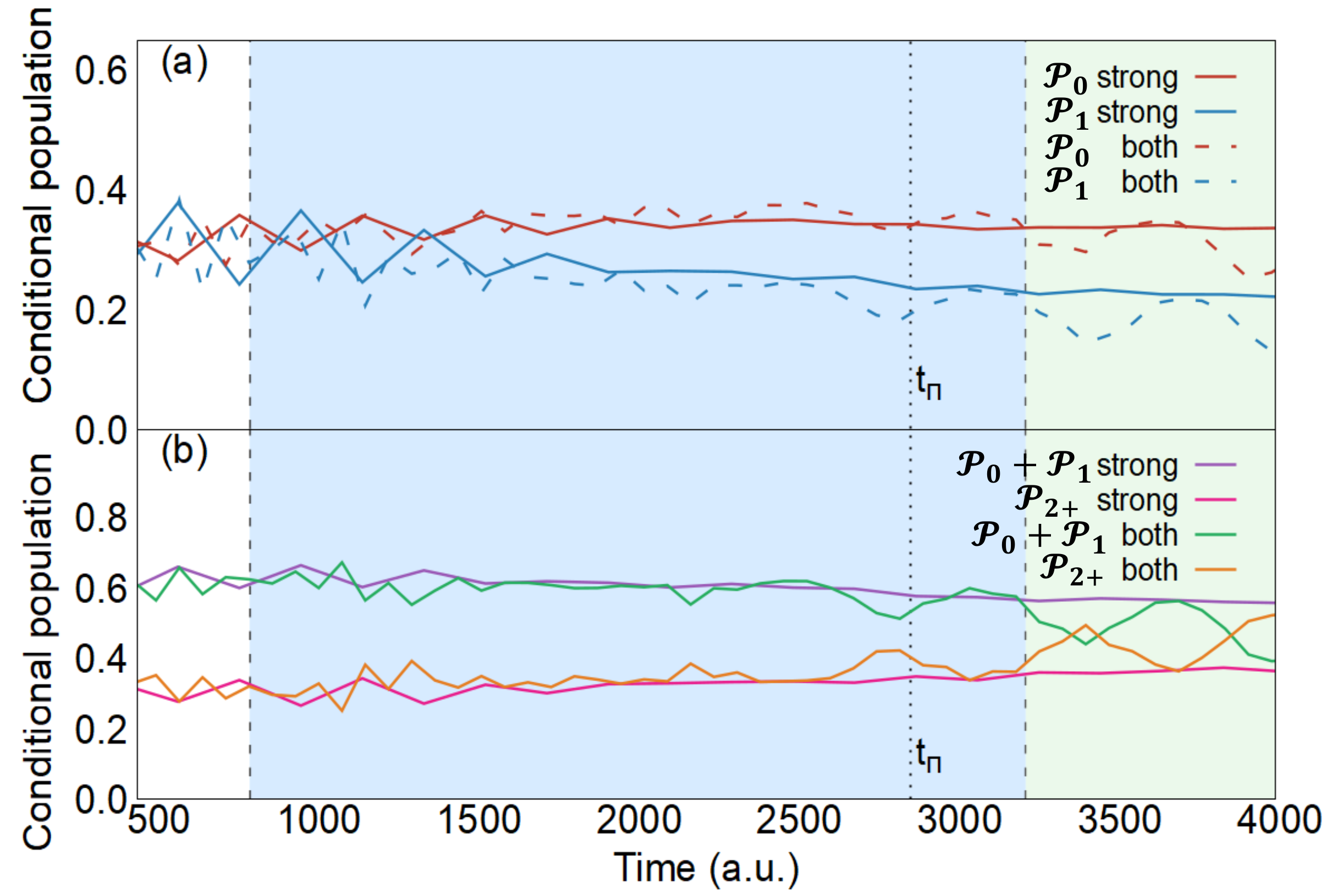}
    \begin{minipage}{0.5\textwidth}
    \caption{Population diagnostics for the $\pi$ pulse at $\varepsilon_{0w} = 0.0001$ a.u.\ and $T_{G} = 800$ a.u. \textbf{(a)} Individual eigenstate populations $\mathcal{P}_0$ (red) and $\mathcal{P}_1$ (blue) for the strong-field-only reference (solid) and both fields (dashed). \textbf{(b)} Qubit-subspace population $\mathcal{P}_0 + \mathcal{P}_1$ and higher-state leakage $\mathcal{P}_{2+}$, with the strong-field-only reference in purple and magenta and the both-fields case in green and orange. In both panels the blue shaded region marks the weak pulse window, the green region the post-pulse evolution under the strong field alone, and the dotted vertical line the marked completion time $t_\pi$ (at $T_{G} + 
t_\pi$ on the absolute axis). Throughout the pulse, $\mathcal{P}_0 + \mathcal{P}_1$ remains near the strong-field-only baseline and $\mathcal{P}_{2+}$ stays near its strong-only value, confirming a clean intra-subspace rotation with minimal gate-induced leakage --- less than at $\varepsilon_{0w} = 0.0003$ a.u.\ (Fig.~\ref{fig:resultpopulationpi}), consistent with the weaker perturbation of the KH potential at lower $\varepsilon_{0w}$.}
    \label{fig:resultleakagepi0001}
    \end{minipage}
\end{figure}

Fig.~\ref{fig:resultleakagepi0001} shows how these populations evolve in time for a field strength $\varepsilon_{0w} = 0.0001$, which gives a slow and clean gate. Both the qubit subspace population 
that $\mathcal{P}_0+\mathcal{P}_1$ and the leaked population $\mathcal{P}_{2+}$ remain near the strong-field-only baseline 
throughout the pulse, with less gate-induced 
leakage than at $\varepsilon_{0w} = 0.0003$ [see Fig.~\ref{fig:resultpopulationpi} in the main part of the paper for comparison]. This is consistent 
with the weaker perturbation of the KH potential at lower 
$\varepsilon_{0w}$ ($\alpha_{0w}/\alpha_{0s} = 0.093$ vs $0.28$), 
reducing off-resonant coupling to $\mathcal{P}_{2+}$. The post-pulse 
window is only 548 a.u.\ $< T_{10}$, insufficient for a 
reliable phase fit, so the late-pulse $F_\varphi = 0.912$ 
is the primary figure of merit.

\begin{figure}[H]
    \centering
 \includegraphics[width=0.5\textwidth]{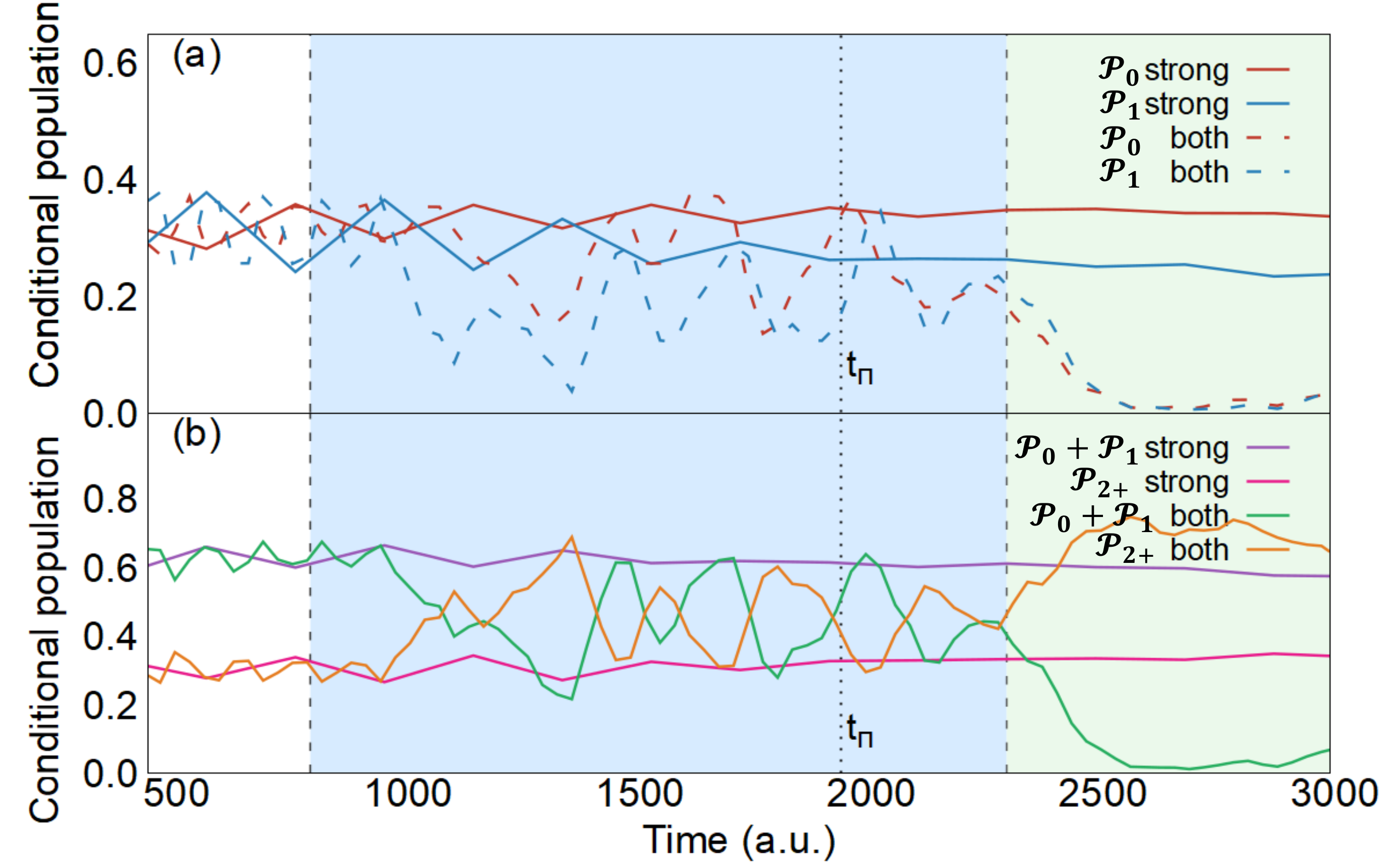}
    \begin{minipage}{0.5\textwidth}
    \caption{Population diagnostics for the $\pi$ pulse at $\varepsilon_{0w} = 0.00065$ a.u.\ and $T_{G} = 800$ a.u. \textbf{(a)} Individual eigenstate populations $\mathcal{P}_0$ (red) and $\mathcal{P}_1$ (blue) for the strong-field-only reference (solid) and both fields (dashed). \textbf{(b)} Qubit-subspace population $\mathcal{P}_0 + \mathcal{P}_1$ and higher-state leakage $\mathcal{P}_{2+}$, with the strong-field-only reference in purple and magenta and the both-fields case in green and orange. In both panels the blue shaded region marks the weak pulse window, the green region the post-pulse evolution under the strong field alone, and the dotted vertical line the marked completion time $t_\pi$ (at $T_{G} + 
t_\pi$ on the absolute axis). Just after $t_\pi$ all populations lie close to their strong-field-only values, with $\mathcal{P}_{2+}$ essentially at baseline, the clearest confirmation of the $Z$ gate at the ideal moment. Subsequently, the continued action of the weak field beyond gate completion at this amplitude ($\alpha_{0w}/\alpha_{0s} = 0.60$) drives further oscillation, and after the pulse ends $\mathcal{P}_{2+}$ rises while $\mathcal{P}_0$ and $\mathcal{P}_1$ fall, indicating redistribution into higher KH bound states.}
    \label{fig:resultleakageE00065}
    \end{minipage}
\end{figure}

In contrast, the population diagnostics in
Fig.~\ref{fig:resultleakageE00065} reveal a
multi-stage evolution. At $t_\pi$, the populations
$\mathcal{P}_0$, $\mathcal{P}_1$, and $\mathcal{P}_{2+}$
all move towards their strong-field-only baselines,
although none has fully reached it. Shortly after
$t_\pi$, all three populations lie close to their
strong-field-only values, with $\mathcal{P}_{2+}$
essentially at its baseline. This corresponds to the
clearest confirmation of the $Z$ gate, where the
$\pi$ pulse induces the expected phase shift rather
than population inversion, consistent with the hybrid
initial state. Within the remainder of the pulse, the populations continue to oscillate: $\mathcal{P}_1$ transiently exceeds its strong-field-only value while $\mathcal{P}_0$ decreases, before both partially recover. This behaviour reflects the continued action of the weak driving field beyond the optimal gate completion time for the chosen field amplitude ($\alpha_{0w}/\alpha_{0s} = 0.60$). After the pulse, $\mathcal{P}_{2+}$ increases significantly, accompanied by a decrease in both $\mathcal{P}_0$ and $\mathcal{P}_1$, indicating a redistribution of population from the qubit subspace into higher KH bound states. Despite this redistribution, the total ionisation remains close to the strong-field-only baseline. This is evidence that the KH atom has not been destroyed. 

\vspace*{0.5cm}

\end{document}